\documentclass[%
 reprint,
 amsmath,amssymb,
 aps,
]{revtex4-2}

\usepackage{graphicx}
\usepackage{dcolumn}
\usepackage{bm}
\usepackage{braket}
\usepackage{siunitx}
\usepackage{soul}

\begin{document}

\preprint{APS/123-QED}

\title{Nonlinear Multi-Resonant Cavity~Quantum Photonics~Gyroscopes\\ for
Quantum Light Navigation}

\author{Mengdi Sun}
\thanks{mengdis@vt.edu}%
\author{Vassilios Kovanis}%
\affiliation{Bradley Department of Electrical and Computer Engineering, Virginia Tech, Arlington, VA, USA}%


\author{Marko Lon\v{c}ar}
\affiliation{
John A. Paulson School of Engineering and Applied Sciences, Harvard University, Cambridge, MA, USA}%

\author{Zin Lin}%
\affiliation{Bradley Department of Electrical and Computer Engineering, Virginia Tech, Blacksburg, VA, USA}%



\date{\today}

\begin{abstract}
We propose an on-chip all-optical gyroscope based on nonlinear multi-resonant cavity quantum photonics in thin film $\chi^{(2)}$ resonators---Quantum-Optic Nonlinear Gyro or QONG in short. The key feature of our gyroscope is {\it co-arisal and co-accumulation} of quantum correlations, nonlinear wave mixing and non-inertial signals, all inside the same sensor-resonator. We theoretically analyze the Fisher Information of our QONGs under fundamental quantum noise conditions. Using Bayesian optimization, we maximize the Fisher Information and show that $\sim 900\times$ improvement is possible over the shot-noise limited linear gyroscope with the same footprint, intrinsic quality factors and power budget.
\end{abstract}

\maketitle


\section{Introduction}
Gyroscopes are critical components of an inertial navigation system for augmenting the GPS guidance or salvaging GPS-denied operational environments \citep{doddoc}. In an optical gyroscope, the rotation rate is measured through the phase shift between two counter-propagating beams in an optical loop. This approach was first proposed by Sagnac in 1913 \citep{Sagnac,post1967sagnac} and soon different types of optical gyroscopes were developed \citep{Arditty,Chow}. Careful studies have been performed on the sensitivity and the quantum noise of these gyroscopes \citep{Cresser,MATSKO20182289}, and remarkable levels of sensitivity ($<0.001^\circ$/h) have been achieved in state-of-the-art discrete component optical gyroscopes, including fiber optic gyroscopes (FOG) \citep{Wang:14,Ma,Sanders}, ring laser gyroscopes (RLG) \citep{Korth}, atom-laser gyroscope \citep{Dowling} and optical cavity gyroscopes \citep{Liang:17,Zhang:17,Khial2018}. However, bulky components and relatively high power consumption remain major roadblocks to further exploiting discrete component optical gyroscopes. On the other hand, on-chip optical gyroscopes \citep{vahala,Ge:15,Scheuer:07,Cimi,MatthewGrant} exhibit great potential for fully integrated inertial navigation platforms (free of fragile moving parts) and can outperform their discrete component counterparts in size, weight, power consumption, maneuverability, manufacturing scalability, robustness and the ability to operate in harsh environments. However, on-chip gyros are yet to reach sensitivity levels smaller than $1^\circ$/h. This is due to fundamentally limited optical path lengths even in ultra-high quality factor resonantors \citep{vahala}, leaving dubious prospects for further improvements via increasing resonator size or quality factors. To address the challenge of this seemingly intrinsic trade-off between sensitivity and compactness, novel physics and designs have been investigated, including exceptional point sensing \citep{Ren:17,Kononchuk2022}, slow light \citep{Yariv,Anping}, dispersive enhancements \citep{Smith08,Shahriar}, dynamic thermal drift cancellation \citep{Khial2018,Digonnet} and nuclear magnetic resonance \citep{Larsen}.

Meanwhile, driven by the emerging trend of quantum technologies \citep{Toth_2014,Cappellaro,CRC}, quantum light sensors have been identified as a \textit{promising} option that can extend the fundamental sensitivity limits beyond the shot noise regime \citep{Caves,Saikat,Fink_2019}. These ideas were reinforced by decades of development and analysis that lead to the construction of very large laser interferometers with extreme sensitivity that is capable of detecting gravitational waves from remote cosmological events. Recently, squeezed light was used in the LIGO in the US and the VIRGO in Italy to substantially improve the sensitivity of the observing runs that happened late in April 2019 \citep{PRL2019,PRL2}. On the other hand, recent advances in nanofabrication, integration and packaging of ultra-coherent laser sources \citep{Jin2021}, low-loss photonic circuits \citep{Krasnokutska:18,Liu:22} and highly efficient photo-detectors \citep{Li:18} have opened up exciting opportunities for realizing fully on-chip quantum devices. Along this trend, we identify on-chip quantum light gyroscopes, which combine high sensitivities, low power consumption, and small form factors, as promising candidates for next-generation rotation sensing.

In this paper, we theoretically introduce a new type of on-chip quantum light gyroscope that exploits nonlinear multi-resonant cavity quantum photonics in integrated thin film resonators with strong quadratic $\chi^{(2)}$ nonlinearities. We call our gyroscope Quantum-Optic Nonlinear Gyro or QONG in short. Instead of externally injecting quantum states of light into the gyroscope \citep{Saikat}, one of the distinguishing features of our gyroscope is that it \textit{fuses} quantum-coherent nonlinear interactions, quantum light generation and non-inertial signal accumulation inside the same sensor-resonator, enabling $\gtrsim 900\times$ improvements in gyroscopic sensitivity over the linear shot noise limit. In our scheme, classical laser light (coherent state) is injected into a doubly-resonant $\chi^{(2)}$ cavity, and output light is measured at the fundamental ($\omega_1$) and second-harmonic frequencies ($\omega_2=2\omega_1$). The sensitivity of the gyroscope is evaluated by Fisher information (FI) \citep{Zhang2019,Anderson2023}, and the latter is maximized by Bayesian optimization \citep{Shahriari}. Various parameter regimes associated with both fundamental and second harmonic injection schemes were investigated, which reveal correlated noise suppression and sensitivity enhancements via parametric oscillations and critically sensitive three wave mixing dynamics. We predict that, under quantum noise conditions, a minimum detectable rotation rate (MDR) of $<0.01$ $^\circ$/h can be achieved using a thin film lithium niobate (TFLN) ring resonator with a diameter of 20 mm, intrinsic quality factors $Q_{i2}=10^6$ at the second harmonic wavelength (795 nm), $Q_{i1}=10^7$ at the fundamental wavelength (1590 nm). We discuss the scope, validity and implications of our approach and results, while the key sensitivity enhancement factors are summarized in Table~\ref{tab:1} of Section~\ref{sec:dtab}.
\section{Gyroscopic model}
\subsection{Linear resonant gyroscope as a baseline}
\label{sec:Linearmodel}
\begin{figure}[htbp]
    \centering
    \includegraphics[width=\linewidth]{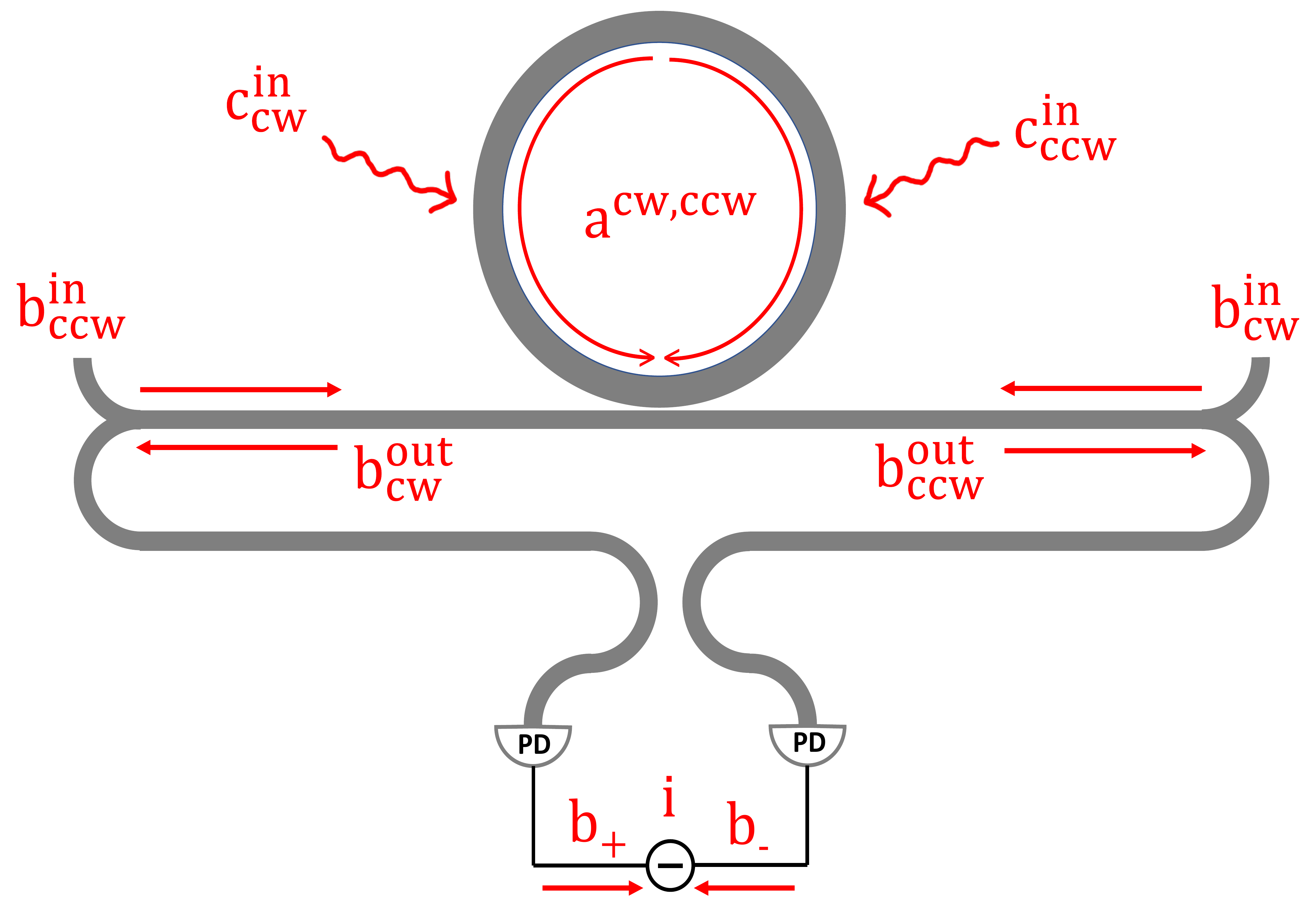}
    \caption{The schematic of the linear micro-ring gyroscope. The input and output light of the micro-ring cavity is injected at two waveguide ports $b_1^{\text{in}}$ (CW) / $b_1^{\text{in}}$ (CCW) and $b_1^{\text{out}}$ (CW) / $b_1^{\text{out}}$ (CCW). The radiation losses are expressed by the ficticious radiation channel $c_1^{\text{in}}$ (CW) / $c_1^{\text{in}}$ (CCW). The output light is measured by homodyne detection.}
    \label{fig:1}
\end{figure}

We first review a basic interferometric scheme probing the gyroscopic shift of a linear resonant cavity, as outlined in Fig.~\ref{fig:1}. We perform a quantum noise analysis similar to Ref.~\citep{zhang2019quantum} or Section 4 of Ref.~\citep{MATSKO20182289}. Two identical counter-propagating probes (seeded from the same on-chip laser) are injected into the clockwise (CW) and the counterclockwise (CCW) modes of a ring resonator; at the exit, the two probe fields are set to interfere via balanced homodyne detection \citep{Lee2013}. In the absence of rotation, the CW and the CCW modes are degenerate and the exiting fields register a vanishing {\it differential} photocurrent signal at the detection setup \citep{MATSKO20182289}. Rotational motion induces a frequency splitting proportional to the rotation rate $\Omega$, which in turn induces a phase difference between the outgoing CW and CCW probes. Subsequently, interference of the two probe fields gives rise to a non-zero differential signal and the underlying $\Omega$ can be measured. For conceptual simplicity, we assume that the frequency of the probe laser is always locked to the degenerate frequency of the unperturbed gyro \citep{MATSKO20182289}. In principle, this can be achieved by self-injection locking the laser to an independent rotation-insensitive cavity (such as a high-Q spiral resonator \citep{Ciminelli}) having the exact same frequency as the unperturbed gyro ring. It has been demonstrated \citep{Kondratiev2023} that self-injection locking to a high-Q cavity can produce an ultra-coherent \textit{integrated} laser with a sub-Hertz linewidth; therefore, we can readily approximate the laser state as a quantum-mechanical coherent state. In the Heisenberg picture, the ring resonator gyro obeys the Heisenberg-Langevin equations \citep{Drummond2004}:

\begin{align}
    \frac{d \hat{a}_\text{cw}}{dt} &= \left(-\frac{\kappa}{2} - \frac{\gamma}{2} + i\delta \right)\hat{a}_\text{cw} + i\beta \hat{a}_\text{ccw} + \sqrt{\kappa}\hat{b}^\text{in}_\text{cw} + \sqrt{\gamma}\hat{c}^\text{in}_\text{cw} \label{eq:a1}\\
    \frac{d \hat{a}_\text{ccw}}{dt} &= \left(-\frac{\kappa}{2} - \frac{\gamma}{2} - i\delta \right)\hat{a}_\text{ccw} + i\beta \hat{a}_\text{cw} + \sqrt{\kappa}\hat{b}^\text{in}_\text{ccw} + \sqrt{\gamma}\hat{c}^\text{in}_\text{ccw} \label{eq:a2}
\end{align}

where $\hat{a}_\text{cw}$ and $\hat{a}_\text{ccw}$ are the annihilation operators for the cavity CW and CCW modes excited by the injections $\hat{b}^\text{in}_\text{cw}$ and $\hat{b}^\text{in}_\text{ccw}$ respectively. $\hat{c}^\text{in}_\text{cw}$ and $\hat{c}^\text{in}_\text{ccw}$ represent intrinsic loss channels (such as radiative losses). $\kappa$ and $\gamma$ are the decay rates for the coupling and the intrinsic losses. We approximate Rayleigh-type back-scattering as a linear (conservative) coupling $\beta$ between CW and CCW modes inside the cavity \citep{vahala}. $\delta$ is the rotation-induced resonant frequency shift due to the Sagnac effect. Here we have assumed single-photon normalization for each eigenmode so that $\hat{a}^\dagger \hat{a}$, for example, represents the photon number operator inside the cavity.

We denote $\langle \hat{A} \rangle = \bra{\psi} \hat{A} \ket{\psi} $ as the usual notation for computing the expectation value of a physical observable $\hat{A}$ with respect to the quantum state $\ket{\psi}$. In the linear problem, we will consider coherent states of the same amplitude $b^\text{in}$ in the input waveguides and vacuum states in the intrinsic loss channels for both CW and CCW light \citep{MATSKO20182289}. The input quantum state of the gyro is then given by $\ket{\psi} = \ket{b^\text{in}}_\text{cw} \ket{b^\text{in}}_\text{ccw }\ket{0}_\text{cw} \ket{0}_\text{ccw}$. The classical counterpart of the input operator $\hat{b}^\text{in}$ is the input amplitude of a coherent state in the feeder waveguide, and can be related to the input power $P$ by the formula:

\begin{align}
    |b^\text{in}|^2 = \langle \hat{b}^{\text{in}\dagger} \hat{b}^\text{in} \rangle = \frac{P}{\hbar\omega} \label{eq:a3}
\end{align}

The output operators in the waveguides are given by \citep{MATSKO20182289}:

\begin{align}
    \hat{b}^\text{out}_\text{cw} &=\hat{b}^\text{in}_\text{cw}-\sqrt{\kappa_1}\hat{a}_\text{cw} \label{eq:a4}\\
    \hat{b}^\text{out}_\text{ccw} &=\hat{b}^\text{in}_\text{ccw}-\sqrt{\kappa_1}\hat{a}_\text{ccw} \label{eq:a5}
\end{align}

The clockwise and counterclockwise signals are set to interfere through a directional coupler/beam splitter with a controllable phase shift $\phi$, followed by photodetection. The signal incident on the photodetectors and the photocurrent operators are then given by:

\begin{align}
    \hat{b}_+ &=\left(\hat{b}^\text{out}_\text{cw} e^{i\phi/2}+i\hat{b}^\text{out}_\text{ccw} e^{-i\phi/2}\right)/\sqrt{2} \label{eq:a6} \\
    \hat{b}_- &= \left(i \hat{b}^\text{out}_\text{cw} e^{i\phi/2}+\hat{b}^\text{out}_\text{ccw} e^{-i\phi/2}\right)/\sqrt{2} \label{eq:a7} \\
    \hat{i}_+ &= \hat{b}_+^\dagger \hat{b}_+ \label{eq:a8} \\
    \hat{i}_- &= \hat{b}_-^\dagger \hat{b}_- \label{eq:a9}
\end{align}

We measure the differential current signal:

\begin{align}
    \hat{i} = \hat{i}_+ - \hat{i}_- \label{eq:a10}
\end{align}

As a figure of merit, we will investigate the minimum detectable frequency shift by calculating the ratio between the standard variation of the measured differential current and the derivative of the mean value of the current over the rotation-induced frequency shift, as reported by Dowling in 1998 \citep{Dowling}:

\begin{align}
    \delta_\text{min} = \frac{\sqrt{\langle \hat{i}^2 \rangle - \langle \hat{i} \rangle^2}}{\left\lvert \frac{\partial \langle \hat{i} \rangle}{\partial \delta} \right\rvert}\Bigg|_{\delta=0} \label{eq:a12}
\end{align}

Since the resonant frequency shift due to the Sagnac effect is given by $\delta = \frac{2\pi r \Omega}{\lambda n_0}$ \citep{MATSKO20182289}, the minimum detectable rotation rate (MDR) is given by:

\begin{align}
    \Omega_\text{min} = \frac{\lambda n_0}{2\pi R} \frac{\sqrt{\langle \hat{i}^2 \rangle - \langle \hat{i} \rangle^2}}{\left\lvert \frac{\partial \langle \hat{i} \rangle}{\partial \delta} \right\rvert}\Bigg|_{\delta=0} \label{eq:a13}
\end{align}

where R and $n_0$ are the radius and the refractive index of the micro-ring. $\lambda$ is the wavelength of the input light. We emphasize that $\Omega_\text{min}$ is a holistic measure that considers the \textit{deterministic} sensitivity of the noise-averaged photocurrent with respect to $\Omega$ as well as the variance of the measured current signals \textit{due to noise} (Both are critical to correctly characterizing the overall sensitivity of the gyro; it has been pointed out~\citep{Wang2020} that an analysis only of the deterministic sensitivity could often lead to misleading conclusions).

If we ignore Rayleigh back-scattering $\beta=0$, we can derive a simple closed-form expression for MDR in the linear gyroscope (LG):

\begin{align}
    \Omega_\text{min}^\text{LG} = \frac{\sqrt{2} \lambda n_0 (\kappa + \gamma)^2}{32 \pi R \kappa \sqrt{N}} \label{eq:a14}
\end{align}

where $N = \frac{P}{\hbar \omega}$ is the incident number of photons per unit time. $\Omega_\text{min}^\text{LG}$ is minimized at $\kappa = \gamma$, yielding $\text{MDR}^\text{LG}_\text{min} = \frac{\sqrt{2} c n_0}{4 R \sqrt{N} Q_i}$, where the intrinsic quality factor is defined by $Q_i = \frac{\omega}{\gamma}$. Note that this is only an example to illustrate the sensitivity dependence of the simplest linear gyroscope without considering Rayleigh back-scattering, which will be taken into account in the following discussions. We note that, in our analysis, we only consider fundamental quantum noise: without loss of generality, we have assumed perfect beam splitters and detectors external to the resonator, while we do consider realistic losses inside the resonator. This linear quantum result will serve as a baseline comparison for our later analysis of a new mode of gyroscope that relies on nonlinear quantum optical effects. Note that the scaling $\text{MDR}^\text{LG}_\text{min} \sim \frac{1}{\sqrt{N} Q_i}$ recovers the familiar shot noise limit or the standard quantum limit \citep{Dowling}. In addition, the $\frac{1}{R}$ dependence in Equation \eqref{eq:a14} indicates that a larger ring radius $R$ offers better sensitivity, which is one of the common control knobs of classical linear optical gyroscope.

\subsection{Nonlinear Multi-Resonant Cavity Quantum Photonics Gyro} \label{sec:Noninearmodel}
\begin{figure}[htbp]
    \centering
    \includegraphics[width=\linewidth]{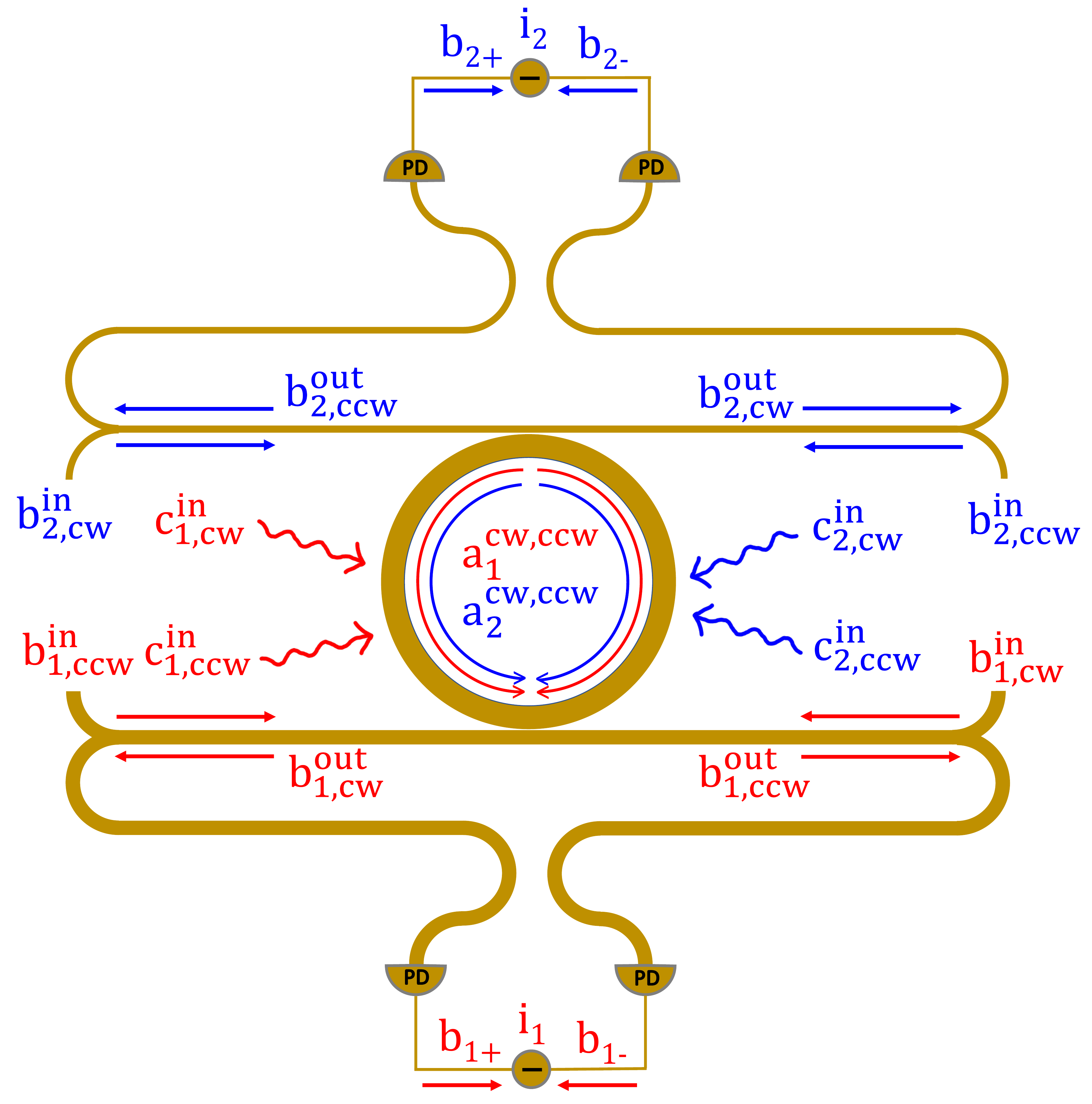}
    \caption{The schematic of the nonlinear micro-ring gyroscope. The input light is injected at four waveguide ports $b_1^{\text{in}}$ (CW) / $b_1^{\text{in}}$ (CCW) and $b_2^{\text{in}}$ (CW) / $b_2^{\text{in}}$ (CCW). The input/output and loss channels for the second harmonic light are expressed by blue arrows. The output light is measured by a joint measurement using the differential currents at both output ports ($b_1^{\text{out}}$ and $b_2^{\text{out}}$). The accuracy of the measurement is determined by evaluating the quantum Fisher information of the output light.}
    \label{fig:2}
\end{figure}

We now consider the gyroscopic operation of a doubly resonant ring resonator with quadratic $\chi^{(2)}$ nonlinearities (Fig. \ref{fig:2}). Quadratic nonlinearities are well-known generators of quantum-coherent correlations such as squeezing and entanglement \citep{GaetaandLipson,Masada2015}. Our Nonlinear Multi-resonant Cavity Quantum Photonics Gyro, or Quantum-Optic Nonlinear Gyro (QONG) in short, \textit{fuses} nonlinear dynamics, quantum correlations, and non-inertial Sagnac effects in the same sensor-resonator, to maximally leverage any possible nonlinear quantum-optical effects for gyroscopic sensitivity. Specifically, we investigate the following Heisenberg-Langevin equations:

\begin{align}
\begin{split}
    \frac{d \hat{a}_{1,\text{cw}}}{dt} &= -\left(\frac{\kappa_1}{2} + \frac{\gamma_1}{2} - i\delta_1\right)\hat{a}_{1,\text{cw}} + i\beta_1 \hat{a}_{1,\text{ccw}} \\
    & + \chi \hat{a}_{1,\text{cw}}^\dagger \hat{a}_{2,\text{cw}} + \sqrt{\kappa_1}\hat{b}^\text{in}_{1,\text{cw}} + \sqrt{\gamma_1}\hat{c}^\text{in}_{1,\text{cw}}
\end{split}\label{eq:a15} \\
\begin{split}
    \frac{d \hat{a}_{1,\text{ccw}}}{dt} &= -\left(\frac{\kappa_1}{2} + \frac{\gamma_1}{2} + i\delta_1\right)\hat{a}_{1,\text{ccw}} + i\beta_1 \hat{a}_{1,\text{cw}} \\
    & + \chi \hat{a}_{1,\text{ccw}}^\dagger \hat{a}_{2,\text{ccw}} + \sqrt{\kappa_1}\hat{b}^\text{in}_{1,\text{ccw}} + \sqrt{\gamma_1}\hat{c}^\text{in}_{1,\text{ccw}}
\end{split}\label{eq:a16} \\
\begin{split}
    \frac{d \hat{a}_{2,\text{cw}}}{dt} &= -\left(\frac{\kappa_2}{2} + \frac{\gamma_2}{2} - i\delta_2\right)\hat{a}_{2,\text{cw}} + i\beta_2 \hat{a}_{2,\text{ccw}} \\
    & - \frac{1}{2} \chi \hat{a}_{1,\text{cw}}^2 + \sqrt{\kappa_2}\hat{b}^\text{in}_{2,\text{cw}} + \sqrt{\gamma_2}\hat{c}^\text{in}_{2,\text{cw}}
\end{split}\label{eq:a17} \\
\begin{split}
    \frac{d \hat{a}_{2,\text{ccw}}}{dt} &= -\left(\frac{\kappa_2}{2} + \frac{\gamma_2}{2} + i\delta_2\right)\hat{a}_{2,\text{ccw}} + i\beta_2 \hat{a}_{2,\text{cw}} \\
    & - \frac{1}{2} \chi \hat{a}_{1,\text{ccw}}^2 + \sqrt{\kappa_2}\hat{b}^\text{in}_{2,\text{ccw}} + \sqrt{\gamma_2}\hat{c}^\text{in}_{2,\text{ccw}}
\end{split}\label{eq:a18}
\end{align}

In these equations, the index $j=1,2$ in the field operators ($\hat{a}_j$, $\hat{b}_j$, $\hat{c}_j$) stands for the fundamental $\omega_1$ and second harmonic $\omega_2 = 2\omega_1$ resonances. $\kappa_j$ and $\gamma_j$ are the decay rates of the coupling and the intrinsic loss channels. Rayleigh scattering rate between CW and CCW modes at each resonance $j$ is again characterized by $\beta_j$, \hl{which is inversely proportional to the intrisic quality factor at each resonance $j$ ($Q_{ij}$)}. The nonlinear coupling terms $\chi \hat{a}_\text{1}^\dagger \hat{a}_\text{2}$ indicate a multi-photon process in which one incident photon with $\omega_2$ breaks down into two photons of half the frequency $\omega_1 = \omega_2/2$ (parametric down conversion \citep{Couteau}), or its reverse, ${1 \over 2} \chi \hat{a}_\text{1}^2$, indicating that two photons with $\omega_1$ combine into one photon with the double frequency $\omega_2 = 2\omega_1$ (second harmonic generation \citep{Kleinman}). These are quantum-coherent, energy-conserving, three-wave mixing processes, which preserve the fundamental commutation relations \citep{Breunig}. The rotation-induced frequency shifts ($\delta_1$, $\delta_2$) are different for each resonance, have opposite polarity between CW and CCW modes, and can be approximated by $\delta_2 = 2\delta_1$ (since $\delta = {2\pi r \Omega \over \lambda n_0}$ \citep{MATSKO20182289}). \hl{Note that here the material dispersion of the lithium niobate is neglected because the index difference (n=2.21 at 1590 nm and n=2.25 at 795 nm) can be compensated by engineering the geometry dispersion} \citep{Sarma:15}. In order to improve the accuracy of the model, however, the dispersion effect should be taken into account in future exploration. The key parameter in this model is the nonlinear modal coupling strength \citep{Drummond2004,Lu:19}:

\begin{align}
    \chi = {\epsilon_0 \over \hbar} \iiint {3 \chi^{\left( 2 \right)} (r) \over 4\sqrt{2}} u_{1}^*(z,r,\theta)^2 u_{2}(z,r,\theta) r dr d\theta dz \label{eq:a19}
\end{align}

where $u_{1}^*(z,r,\theta)$ and $u_{2}(z,r,\theta)$ are the electric field profiles (in polar coordinates) of the fundamental and the second harmonic eigenmodes of the gyroscopic resonator. Given that our sensor is a ring resonator of radius $R$, it is instructive to decompose $\chi$ into cross-sectional modal overlap $\zeta$ and the remaining contributions. Following \citep{Guo:16}, we approximate:

\begin{align}
    \chi &\approx \sqrt{\hbar \omega_1^2 \omega_2 \over \epsilon_0 2\pi R} {\zeta \over \epsilon_1 \sqrt{\epsilon_2}} {3 \chi^{(2)} \over 4 \sqrt{2}} \label{eq:a20}\\
    \zeta &= {\iint u_{1}^*(z,r)^2 u_{2}(z,r) dr dz \over \iint |u_{1}^*(z,r)|^2 dr dz \sqrt{\iint |u_{2}(z,r)|^2 dr dz}} \label{eq:a21}
\end{align}

It is important to realize that the nonlinear Langevin equations \citep{Drummond2004,Drummond} encode the time evolution of \textit{four coupled infinite-dimensional} quantum operators; as such, it is very challenging to obtain an exact solution either analytically or numerically (we note that straightforward numerical methods using a truncated Fock basis \citep{johansson2012qutip} are not feasible because our system typically involves milli-watts of optical power amounting to $\sim 10^{16}$ photons). However, at milli-watt injection powers, quantum fluctuations can be considered ``small signals'' compared to much stronger average field intensities at steady state, so that each operator can be decomposed into a classical scalar-valued amplitude and a quantum fluctuation operator, e.g. $\hat{a} = \alpha + \hat{\delta a}$. The details of calculating steady state solutions are included in Appendix \ref{sec:appendix1}. The classical amplitude represents a steady-state solution to the mean-field averaged Langevin equations at the classical (large photon number) limit while the ``small-signal'' fluctuation operator approximately obeys the linearized Langevin equations in the vicinity of the steady-state mean-field solution. Linearizing a nonlinear steady state to study the fluctuations in its vicinity is commonly known as small-signal modeling in electronics engineering \citep{Tang}. In a similar spirit, the small-signal treatment of quantum fluctuation operators in the Heisenberg-Langevin picture is a simple but effective approach widely accepted for steady-state noise analysis in laser and nonlinear quantum optics literature with experimental support \citep{MATSKO20182289,Gillner,Yamamoto1,chembo2016quantum,pontula2022strong}. Theoretically, it is important to note that such an approach is justified as long as the steady state we consider is a hyperbolic fixed point whose neighborhood is a topologically stable manifold that ensures small fluctuations (Hartman-Grobman theorem \citep{Costa2021}). On the other hand, a more sophisticated phase-space formalism, which employs quasi-probability distributions, Fokker-Planck equations and stochastic calculus, can be used to study more complicated dynamics such as large fluctuations at non-hyperbolic critical points and self-pulsing (limit-cycle) solutions \citep{Drummond_1980}.
Using the small-signal approximation, we can compute the differential photocurrent signals at both the fundamental and the second harmonic resonances (see also Fig. \ref{fig:2}):

\begin{align}
    \hat{i}_1 = i A_1 \left( \hat{b}^{\text{out} \dagger}_\text{1,cw} \hat{b}^\text{out}_\text{1,ccw} e^{-i\phi_1}-\hat{b}^{\text{out} \dagger}_\text{1,ccw} \hat{b}^\text{out}_\text{1,cw} e^{i\phi_1}\right) \label{eq:a22}\\
    \hat{i}_2 = i A_2 \left( \hat{b}^{\text{out} \dagger}_\text{2,cw} \hat{b}^\text{out}_\text{2,ccw} e^{-i\phi_2}-\hat{b}^{\text{out} \dagger}_\text{2,ccw} \hat{b}^\text{out}_\text{2,cw} e^{i\phi_2}\right) \label{eq:a23}
\end{align}

Here $A_1$ and $A_2$ are constant factors determined by the frequencies of the light and the responsivity of the photodetectors. $e^{i\phi_1}$ and $e^{i\phi_2}$ are the propagation phase shifts that each output light experiences, \hl{which are determined by the propagation distances after the output CW and CCW waves are mixed. Therefore, by controlling the lengths of the output waveguides, $e^{i\phi_1}$ and $e^{i\phi_2}$ are set to one here.} Here we measure both $\hat{i}_1$ and $\hat{i}_2$ to extract maximal information out of the nonlinear wave-mixing gyro.

The output of our quantum-optic nonlinear gyro (QONG) is now characterized by a mean vector $\langle \mathbf{i} \rangle$ and a covariance matrix $\langle \Delta \mathbf{i}^2 \rangle$:

\begin{align}
    {\langle \mathbf{i} \rangle} &= \begin{pmatrix}
    \langle \hat{i}_1 \rangle \\
    \langle \hat{i}_2 \rangle
    \end{pmatrix} \label{eq:a24}\\
    {\langle \Delta \mathbf{i}^2 \rangle} &= \begin{pmatrix}
    \langle \hat{i}_1^2 \rangle - {\langle \hat{i}_1 \rangle}^2 & {\langle \hat{i}_1 \hat{i}_2 \rangle + \langle \hat{i}_2 \hat{i}_1 \rangle \over 2} - \langle \hat{i}_1 \rangle \langle \hat{i}_2 \rangle \\
   {\langle \hat{i}_1 \hat{i}_2 \rangle + \langle \hat{i}_2 \hat{i}_1 \rangle \over 2} - \langle \hat{i}_2 \rangle \langle \hat{i}_1 \rangle & \langle \hat{i}_2^2 \rangle - {\langle \hat{i}_2 \rangle}^2
    \end{pmatrix} \label{eq:a25}
\end{align}

Assuming that the joint probability distribution of the measured photocurrents follow a bi-variate Gaussian, we can express the Fisher information~\citep{Anderson2023} of our QONG:

\begin{align}
    I(\delta) = ({{d \langle \mathbf{i} \rangle} \over {d \delta}})^T {\langle \Delta \mathbf{i}^2 \rangle}^{-1} ({{d \langle \mathbf{i} \rangle} \over {d \delta}}) \label{eq:a26}
\end{align}

The details of the statistial analysis of the differential currents are included in Appendix \ref{sec:appendix2}. Then the sensitivity is determined by Cramer-Rao bound \citep{Braunstein}:

\begin{align}
    \delta_\text{min} = {1 \over {\sqrt{I(\delta)}}} \label{eq:a27}
\end{align}
Similar to the linear gyroscope, here MDR is given by:
\begin{align}
    \Omega_\text{min} = {\lambda n_0 \over 2 \pi R} \delta_\text{min} \label{eq:a28}
\end{align}

Before we provide further estimation for particular QONG implementation, we want to offer a few remarks of our modeling approach:

\begin{itemize}
    \item In this paper, we have stuck to a Langevin description of our quantum gyro, which takes into account quantum noise through the Langevin fluctuation operator $\hat{b}$ or $\hat{c}$ in each coupling or dissipation channel (with the rates determined by the fluctuation-dissipation theorem), preserving the fundamental commutation relations \citep{Drummond2004}.

We note that while the Langevin form is widely utilized in many experimental situations \citep{Guo:16}, more sophisticated theoretical analysis, delineating the open-system quantum dynamics \citep{Kapral_2015}, can be performed using the density operator formalism and the Master equation, which will be the subject of future investigations. In particular, our simple perturbative approach restricts our solution to examine the quantum fluctuations around a stable hyperbolic fixed point. On the other hand, non-hyperbolic fixed points and non-steady state attractors (such as limit cycles) require more sophisticated non-perturbative treatment (while their implications for quantum correlations and sensing remain unexplored). One such treatment involves expanding the density operator in a non-diagonal coherent state basis (so-called positive P representation), deriving a Fokker-Planck equivalent of the Lindblad Master equation and simulating the associated stochastic dynamics \citep{Drummond2004}. However, to the best of our knowledge, Fokker-Planck equations corresponding to more than two bosonic operators \citep{Drummond_1980,Drummond1} have not been well studied; our nonlinear multi-resonant cavity quantum photonic gyro is described by 4 coupled Langevin equations and will lead to an 8+1 dimensional Fokker Planck equation, which requires substantial computational resources and will be the subject of future investigations.

    \item In our approach, we have assumed idealized sources and detectors in order to simplify our gyroscopic model to physically most crucial components, and thereby to unveil the fundamental information-theoretic limits (in the same spirit as the analysis presented in Ref.~\citep{zhang2019quantum} or Section 4 of Ref.~\citep{MATSKO20182289}). Future works will develop more detailed models that can compute commonly accepted experimental metrics such as the integration-time dependent Allan deviation curve \citep{Giglio}, for example, by incorporating the quantum theory of photodetection \citep{Shapiro,Carmichael1999}, which can explicitly take into account photo-electron generation rates and detector integration times.
    
    \item Last but not least, we note that our present model focuses on $\chi^{(2)}$ processes to delineate their effects on the gyroscopic sensitivity. A more thorough gyro model may also consider $\chi^{(3)}$ (Kerr-type self modulation) nonlinearities, which may come into effect at ultra-high quality factors and are found to limit the sensitivity of the (otherwise) \textit{linear} gyroscope~\cite{MATSKO20182289,vahala}. While we shall take into account $\chi^{(3)}$ processes in detailed comprehensive models in the future (see Section \ref{sec:s&o}), we note that Eqs.~\ref{eq:a15}-\ref{eq:a18} are fully applicable to material platforms, such as thin film lithium niobate \citep{Zhu:21,Lu:19}, which possess prominent $\chi^{(2)}$. Furthermore, unlike their linear counterparts, resonators with $\chi^{(2)}$ can be engineered to exhibit negative Kerr shifts via cascaded second-order effects \citep{Cui2022}, which can mitigate the intrinsic positive Kerr shift; we shall investigate such cancellation schemes in our future works. On the other hand, we would like to emphasize that $\chi^{(3)}$ processes, including even the Kerr shift, need not be treated as a nuisance, but as extra complexities and additional degrees of freedom that can be optimized to our advantage (see Section \ref{sec:s&o}). For example, it has been recently reported that the bistability effects associated with the Kerr-shift self-modulation can even enhance sensitivities under appropriate sensing schemes~\cite{silver2021nonlinear,peters2022exceptional}.
\end{itemize}

\subsection{Thin film lithium niobate as an implementation platform}
Our nonlinear multi-resonant cavity quantum photonic gyro (or quantum-optic nonlinear gyro QONG in short) can be implemented in any thin film material platform, including LiNbO3 \citep{Zhu:21}, AlN \citep{ALN}, SiC \citep{Sic}, GaAs \citep{GaAs}, etc, which has prominent $\chi^{(2)}$. In this work we consider thin film lithium niobate (TFLN) as a particularly promising platform, as it has gained widespread popularity for realizing quantum-grade ultra-low loss photonic integrated circuits \citep{Alireza,McCutcheon:09}. Indeed, lithium niobate has been traditionally employed in quantum optics applications as a nonlinear medium for generating squeezed light and entangled photon states \citep{Chen:22,Zhao}. However, traditional LN crystals are bulky and suffer from relatively limited strength of light-matter interactions (leading to very weak nonlinear coupling $\chi \sim 10^3~\mathrm{Hz}$ in Eqs.~\ref{eq:a20}). Only recently, high quality wafer-scale TFLN becomes widely available for realizing integrated photonic circuits with nonlinear and electro-optic functionalities~\citep{Zhu:21}. Associated with large $\chi^{(2)}$, low optical loss and strong nanophotonic confinement \citep{MianZhang}, TFLN devices offer orders of magnitude enhancements in nonlinear coupling $\chi \sim 10^6 \mathrm{Hz}$ \citep{Lu:19}.

\begin{figure}[htbp]
    \centering
    \includegraphics[width=\linewidth]{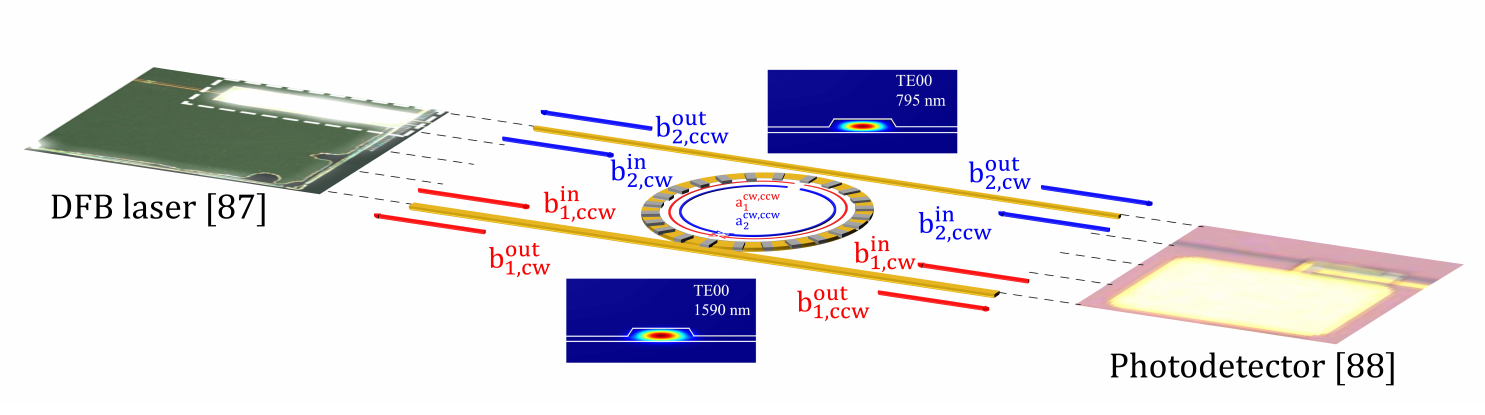}
    \caption{The 3D schematic (not drawn to scale) of quasi-phase-matching (QPM) achieved by periodic poling. Poled rings are used to form quasi-phase matched structures. Semiconductor lasers \citep{Okhotnikov} and photodetectors \citep{Guo:22} are also integrated on the chip. The numerically simulated field profiles of both the fundamental frequency (1590 nm)  and the second harmonic (795 nm) cavity modes are shown in the inset.}
    \label{fig:3}
\end{figure}

Fig.~\ref{fig:3} shows the design of a TFLN ring resonator which can used as an QONG. Note that feeder waveguides of different dimensions, frequency cutoffs, and dispersion characteristics, can be designed to selectively couple to the fundamental (1590nm) and the second harmonic (795nm) modes \citep{Bi:12}, and their coupling rates can be further tuned by TFLN electro-optics~\citep{Shams-Ansari2022}. To realize strong nonlinear coupling $\chi$ between the two resonances, two zeroth-order transverse electric eigenmodes (TE00) can be (quasi-)phase-matched \citep{Lu:19} via periodic poling \citep{Fejer} that achieves crystal domain inversion, leading to periodically varying nonlinear susceptibility $\chi^{(2)}$ which compensates wave vector mismatch between the fundamental and the second harmonic modes $k_\chi = k(\omega_2) - 2 k(\omega_1)$ \citep{Lu:19}. Apart from the phase-matched resonator itself, a fully integrated QONG can be implemented in TFLN, incorporating flip-chip bonded semiconductor lasers \citep{Okhotnikov} and heterogeneously integrated uni-travelling carrier photodetectors \citep{Guo:22}. Furthermore, we note that TFLN comes with unique electro-optic control capabilities \citep{Zhu:21} which can be used for tuning resonator parameters such as the coupling rates to the waveguides \citep{Guarino2007}, managing long term temperature stability, cancelling thermal drifts and electronic noise \citep{Khial2018,Wang:22}, and performing signal processing \citep{8979176}.

In our gyroscopic model, intrinsic losses $\gamma$ and back-scattering rates $\beta$ should be treated as ``fixed'' parameters which depend on the experimentally feasible characteristics of a particular implementation platform, such as residual material losses and surface roughness due to fabrication imperfections. In thin film lithium niobate, intrinsic quality factors reaching $\sim 10^8$ have been demonstrated \citep{Tobias}, and we expect proportionate back-scattering rates with the same order of magnitude. It is important to realize that, apart from $\gamma$ and $\beta$, almost all other parameters can be designed, engineered and optimized, including injection powers, coupling rates, resonator radius, resonator waveguide cross section and dispersion as well as quasi-phase matching processes. We will utilize these parameters as degrees of freedom (DoF) in optimizing the Fisher Information and hence the minimum detectable rotation (MDR) of our QONG (as compared to the linear gyro). While any optimization algorithm can be employed, gradient-free global optimization methods are most suitable for a relatively low-dimensional problem like our two-resonance gyro (where about 5--10 DoFs can be optimized). We will use Bayesian optimization, a simple but powerful machine-learning-based optimization algorithm which requires relatively few function evaluations (as compared to other heuristic methods such as simulated annealing and evolutionary algorithms \citep{SimulatedAnnealing,Eiben}) and has been observed to be particularly effective for optimizing $\sim 20$ DoFs \citep{Shahriari}.

\section{Results}
\label{sec:r&d}
As noted above, the Fisher Information and the Minimum Detectable Rotation (MDR) of the gyroscope is determined by various parameters and can be optimized by judiciously adjusting their values. In our design, the operational wavelengths (of the input/output light) are fixed at $\lambda_1$ = 1590 nm and at $\lambda_2$ =795 nm. The material property of TFLN is taken from the literature \citep{MianZhang}: the refractive index is $n=2.2$ while the second order nonlinear susceptibility is $\chi^{(2)}$ = 30 pm/V. The intrinsic quality factors are fixed at $Q_{i1} = 10^7$ for the 1590 nm and $Q_{i2} = 10^6$ for the 795 nm. The Rayleigh scattering rates for both cavity modes $\beta_1=5.4 \times 10^4$ Hz and $\beta_2=5.4 \times 10^5$ Hz are inferred by adjusting the literature-reported values~\citep{vahala} to the quality factors $Q_i$ of our TFLN platform. We have also fixed the radius of the resonator at $R =20$ mm as well as the cross-sectional dimensions of the resonator waveguide ($1.2~\si{\micro\meter}$ width $\times 0.6~\si{\micro\meter}$ thickness) and the fabrication side wall angle of $75 \si{\degree}$, leading to a cross-sectional area of $0.8~\si{\micro\meter}^2$. The two TE00 modes have phase mismatch of $1.354~\si{\micro\meter}^{-1}$, which can be compensated by poling with a period of $4.64~\si{\micro\meter}$. Based on Eqs.~\ref{eq:a20}, the quasi-phase matched $\chi^{(2)}$ \citep{Lu:19}, and the numerically simulated modal overlapping factor $\zeta = 1.18~/\si{\micro\meter}$, the nonlinear coupling strength is $\chi = 1.26 \times 10^6$ Hz, which is independent of the injection schemes. The rest of the parameters remain to be determined, including the injection power at the fundamental and the second harmonics $P_1$ and $P_2$, the quality factors due to coupling to the waveguides for each cavity mode: $Q_{c1}$ and $Q_{c2}$. These four parameters will be determined by Bayesian optimization. We will investigate gyroscopic performance under different injection schemes including (1) coherent state input at the second harmonic ($\lambda_2$), (2) coherent state input at the sub-harmonic ($\lambda_1$), and (3) coherent state inputs at both second and sub-harmonics ($\lambda_1$ and $\lambda_2$).

\subsection{Optical parametric oscillator gyro (coherent injection at second harmonic)}\label{Sec:para}
\begin{figure}[htbp]
    \centering
    \includegraphics[width=\linewidth]{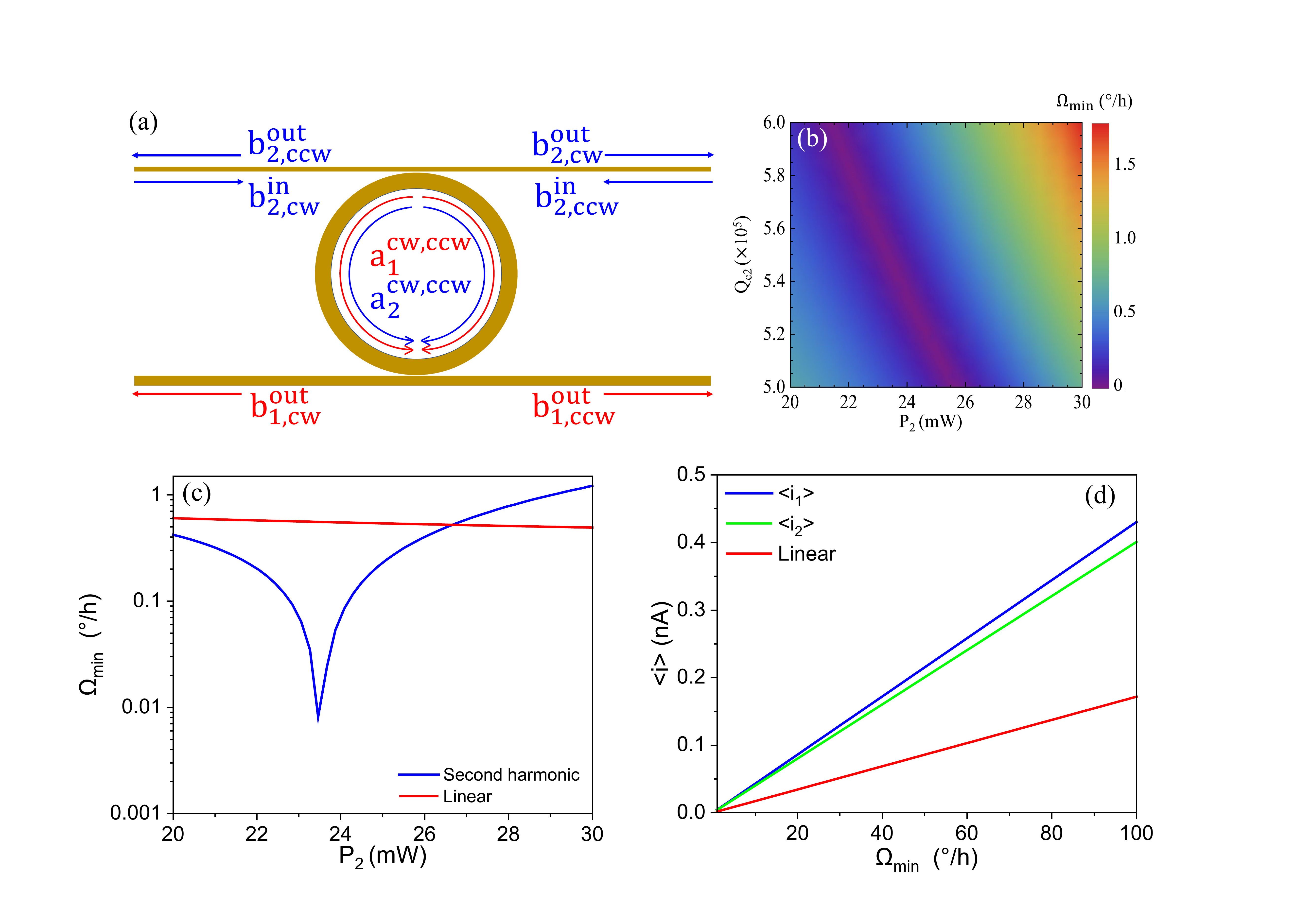}
    \caption{(a) The schematic of the second harmonic injection scheme. The input light is only injected at the second harmonic (blue arrow). (b) 2D density plot of the sensitivity at second harmonic injection. The sensitivity (MDR) is expressed as a function of the input power $P_2$ and the quality factor due to coupling loss of the second harmonic cavity mode $Q_{c2}$. (c) 1D plot of the sensitivity of the second harmonic injection scheme (blue) and the standard linear gyroscope (red) at optimal $Q_c$ in terms of the input power. (d) 1D linear plot of the mean differential current of the output light at fundamental frequency (blue), second harmonic (green) and the standard linear gyroscope (red).}
    \label{fig:4}
\end{figure}

First, we study the performance of an optical parametric oscillator gyroscope under the coherent injection at the second harmonic frequency. As shown in Fig. \ref{fig:4}(a), classical laser light with a wavelength $\lambda_2$ = 795 nm is injected from opposite directions from the waveguide ports $b^{\text{in}}_\text{2,cw}$ and $b^{\text{in}}_\text{2,ccw}$, while no light is injected from the waveguide port $b^{\text{in}}_\text{1,cw}$ or $b^{\text{in}}_\text{1,cw}$, i.e., $P_1=0$. The carefully phase-matched fundamental and second harmonic modes facilitate parametric down conversion, in which one photon with higher frequency (shorter wavelength $\lambda_2$ breaks down into two photons with half the frequency (longer wavelength $\lambda_1$), generating phase-squeezed signals~\citep{Drummond2004}. First we investigate the mininum detectable rotation (MDR) as a function of the input power $P_2$ and the coupling factor $Q_{c2}$. Using Bayesian optimization, we identify a high sensitivity regime, that is, low MDR, as shown in the 2D density plot of MDR in terms of $P_2$ and $Q_{c2}$ (Fig.~\ref{fig:4}b). \hl{Here $P_1$, $P_2$, $Q_{c1}$ and $Q_{c2}$ are subject to Bayesian optimization. Initial scanning range is required for each parameter to start the optimization. For example, the optimization of $P_1$ and $P_2$ starts from 0.1 $\mu$W to 100 mW since integrated photonic devices don't support high power input, while $Q_{c1}$ and $Q_{c2}$ range from $10^5$ to $10^8$, covering most of the thin film lithium niobate micro-ring resonators.} We mapped over the parameter space where $Q_{c2}$ ranges from $5 \times 10^5$ to $6 \times 10^5$ and $P_2$ ranges from 20 mW to 30 mW. At the second harmonic injection, the selection of $P_2$ is determined by the critical power $P_c$, below which the steady-state solutions of the system become unstable. This phenomenon has been discussed in details by Drummond \citep{Drummond}. Here $P_c$ = 14.05 mW such that $P_2$  should be larger than this value. Within this region, small MDR (0 - 1.7 $^\circ$/h) is observed (indicated by rainbow colors), showing that high sensitivity is achieved for a sizable parameter range such that the enhanced sensitivity is not an isolated singularity. The lowest MDR ($\Omega_\text{min} < 0.25$ $^\circ$/h) appears in the narrow purple band. Outside this region, MDR gradually increases as the color becomes more greenish and reddish, indicating reduced sensitivity. In principle, as the rotation rate increases, the difference between the CW and CCW mode becomes more significant. To benchmark the gyroscope performance, we compared the sensitivity of our optical parametric oscillator (OPO) gyro under the second harmonic injection (solid blue line) with the sensitivity of a standard linear gyroscope (solid red line) in Fig. \ref{fig:4}(c). The optimal coupling factors for the OPO gyro are $Q_{c1}=1.018 \times 10^5$ and $Q_{c2} = 5.462 \times 10^5$, both discovered by Bayesian optimization. Meanwhile, the highest sensitivity of the linear gyroscope ($\Omega_\text{min} = 4.465 \times 10^{-3}$ $^\circ$/h) is found at $Q_{c1} = 9.58 \times 10^6 \approx 10^7$. Note that, in the limit of vanishing $\beta_1$, the lowest linear gyro MDR is achieved when $\kappa_1 = \gamma_1$ (see Eq.~\ref{eq:a14}). Since $Q_{i1}$ is fixed at $10^7$, the optimal $Q_{c1}$ for the linear gyro is expected to be close but not exactly equal to this value, considering the influence of the small but non-zero back-scattering. As the input power is increased from 20 mW to 30 mW, the MDR of the OPO gyro drops from $\sim$ 0.42 $^\circ$/h down to near zero and rises back to $\sim$ 1.21 $^\circ$/h with a local minimum at 23.507 mW, corresponding to the optimal sensitivity point. We also found that the optimal-sensitivity point is associated with 9.9 dB phase squeezing at an experimentally feasible value on a TFLN platform \citep{Alireza}. Meanwhile, the MDR of the linear gyroscope remains $> 0.49~^\circ$/h, which demonstrates that the OPO gyro is $\sim$ 124.4 $\times$ more sensitive than the linear gyro under the same injection power, resonator size and intrinsic quality factors. In order to visualize this effect, we investigated the mean current values as a function of the rotation rate $\Omega$ at different frequencies in Fig. \ref{fig:4}(d). The mean values of the output differential currents at the subharomonic ($<i_1>$) and the second harmonic ($<i_2>$) are expressed by the solid blue and red line respectively, while that of the linear gyroscope at the same power consumption is expressed by the solid green line for reference. It is shown that as $\Omega$ increases from 0 - 100 $^\circ$/h, $<i_1>$ increases from 0 - 0.43 nA, $<i_2>$ increases from 0 - 0.4 nA while the output current of the linear gyroscope increases from 0 - 0.17 nA. This result shows that the nonlinear eigenmode dispersion, as discussed in Section \ref{sec:Noninearmodel}, produces stronger output signals (differential currents) at both wavelengths compared to the linear gyroscope.

\subsection{Coherent injection at the fundamental frequency}
\begin{figure}[htbp]
    \centering
    \includegraphics[width=\linewidth]{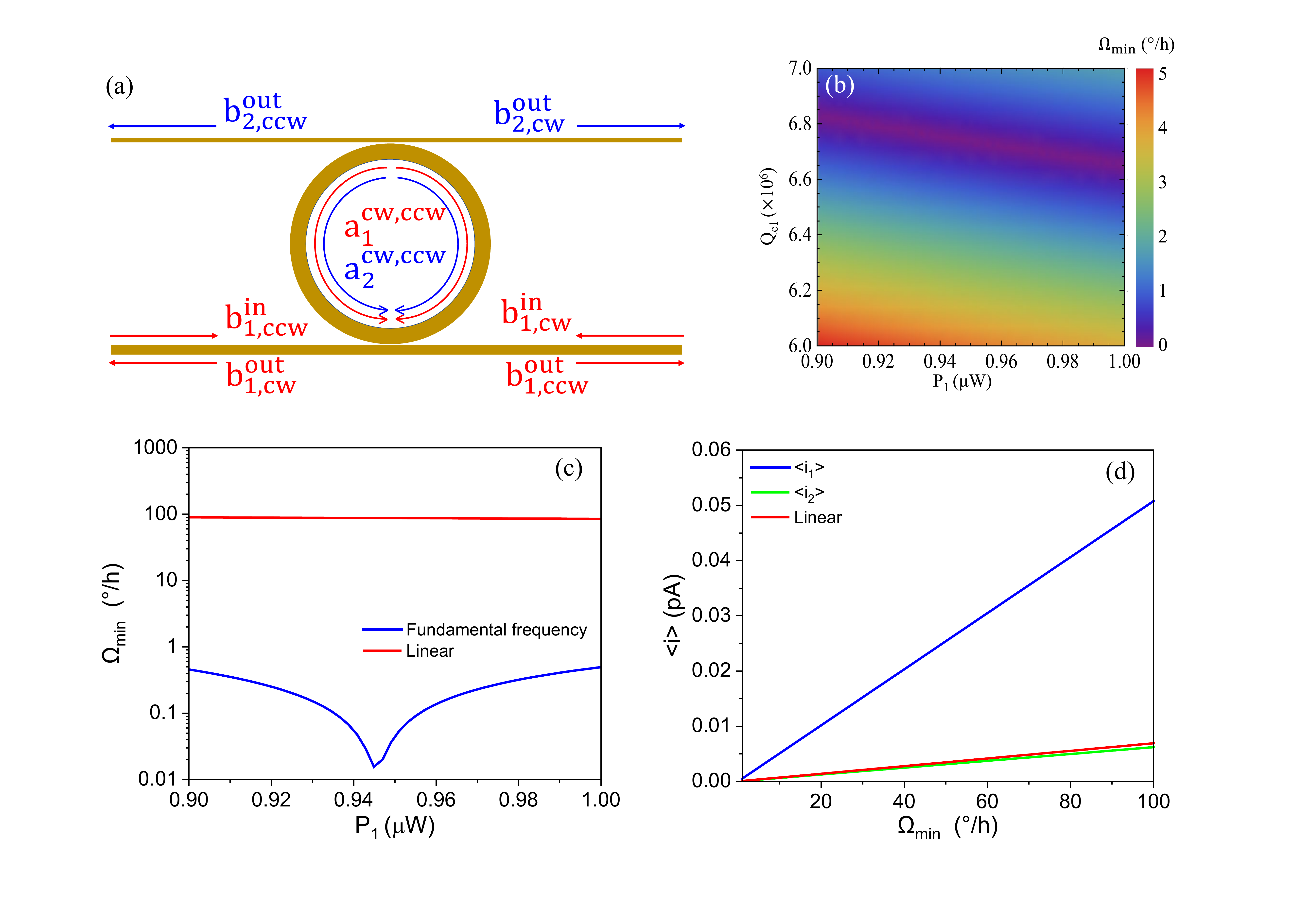}
    \caption{(a) The schematic of the fundamental frequency (subharmonic) injection scheme. The input light is only injected at the fundamental frequency (red arrow). (b) 2D density plot of the sensitivity at the fundamental frequency injection scheme. The sensitivity (MDR) is expressed as a function of the input power $P_1$ and the quality factor due to coupling loss of the cavity mode $Q_{c1}$. (c) 1D plot of the sensitivity of the fundamental frequency injection scheme (blue) and the standard linear gyroscope (red) at optimal $Q_c$ in terms of the input power. The inset shows the magnified plot of the critical region. (d) 1D linear plot of the mean differential current of the output light at fundamental frequency (blue), second harmonic (green) and the standard linear gyroscope (red).}
    \label{fig:5}
\end{figure}

Aside from injecting light at the second harmonic, we also investigated the scheme of the fundamental frequency (subharmonic) injection. As shown in Fig. \ref{fig:5}(a), the input light at the wavelength $\lambda_1$ = 1590 nm is injected from the waveguide ports $b^{\text{in}}_\text{1,cw}$ and $b^{\text{in}}_\text{1,cw}$ while $P_2=0$, stimulating intra-cavity up-conversion (two photons of lower energy are combined to one photon of higher energy). We study the sensitivity (MDR) in terms of $P_1$ and $Q_{c1}$, as indicated by the 2D density plot Fig.~\ref{fig:5}(b), in which variations in MDR again show up in rainbow colors.  In Fig. \ref{fig:5}(b) we mapped over the parameter space where $Q_{c1}$ ranges from $6 \times 10^6$ to $7 \times 10^6$ and $P_1$ ranges from $0.9~\mu W$ to $1~\mu W$. Note that much lower power injection is required at this injection scheme compared to coherent injection at the second harmonic, related to the fact that the steady-state solutions of the cavity modes are stable only when $P_1$ is below the critical power when the second harmonic injection is absent \citep{Drummond}. Here the critical power is $P_c = 3.24$ mW. Only solutions of $P_1$ smaller than this value are stable, resulting in orders of magnitude lower power consumption. In Fig. \ref{fig:5}(b), MDR varies from 0 - 5 $^\circ$/h across the entire parameter space. The lowest MDR is found in a narrow band-like region. The mean values of the output currents are also studied here. To better evaluate the gyroscope performance, we compared it with the linear gyroscope at the same power consumption. Figure \ref{fig:5}(c) shows the sensitivity of the fundamental frequency injection (solid blue line) and the linear gyroscope (solid red line) in terms of the injection power at fixed optimal Q factors ($Q_{c1}=6.747 \times 10^6, Q_{c2} = 6.675 \times 10^7$), which are determined by Bayesian optimization, and optimal $Q_c = 9.58 \times 10^6$ in the linear gyro. As $P_1$ increases from $0.9~\mu W$ to $1~\mu W$, MDR drops from $\sim$ 0.46 $^\circ$/h down to near zero at $0.945~\mu W$, then rises back to $\sim$ 0.49 $^\circ$/h. Meanwhile, the sensitivity of the linear gyroscope monotonically but slowly decreases from $\sim$ 89.78 $^\circ$/h to $\sim$ 85.17 $^\circ$/h. The optimal sensitivity $\Omega_\text{min} = 0.093$ $^\circ$/h is found at $P_1=0.945$ $\mu W$. At the optimal point, a surprisingly high sensitivity improvement of $\sim$ 942.5 $\times$ is observed (0.093 $^\circ$/h vs 87.62 $^\circ$/h). Most importantly, the fundamental frequency injection scheme merges high sensitivity and low power consumption together in a compact form, which shows great potential in practical applications. Similar to the analysis in Section \ref{Sec:para}, the mean differential currents ($<i_1>$, $<i_2>$ and linear) as a function of the rotation rate $\Omega$ are shown in Fig. \ref{fig:5}(d) in solid blue, red and green lines. As $\Omega$ increases from 0 - 100 $^\circ$/h, $<i_1>$ increases from  0 - 0.05 pA, $<i_2>$ increases from 0 - 0.0062 pA and the output differential current of the linear gyroscope increases from 0 - 0.0069 pA. The differential currents are in the pA range, not the nA range, again because of much weaker input power. It is shown that the sensitivity of $<i_1>$ is almost $10 \times$ larger than that of $<i_2>$ and that of the output current in the linear gyroscope, suggesting that, even though very little second harmonic power is ultimately extracted, the presence of a non-linearly interacting second harmonic mode critically enhances the sensitivity of the fundamental mode.

\subsection{Dual frequency injection}
\begin{figure}[htbp]
    \centering
    \includegraphics[width=\linewidth]{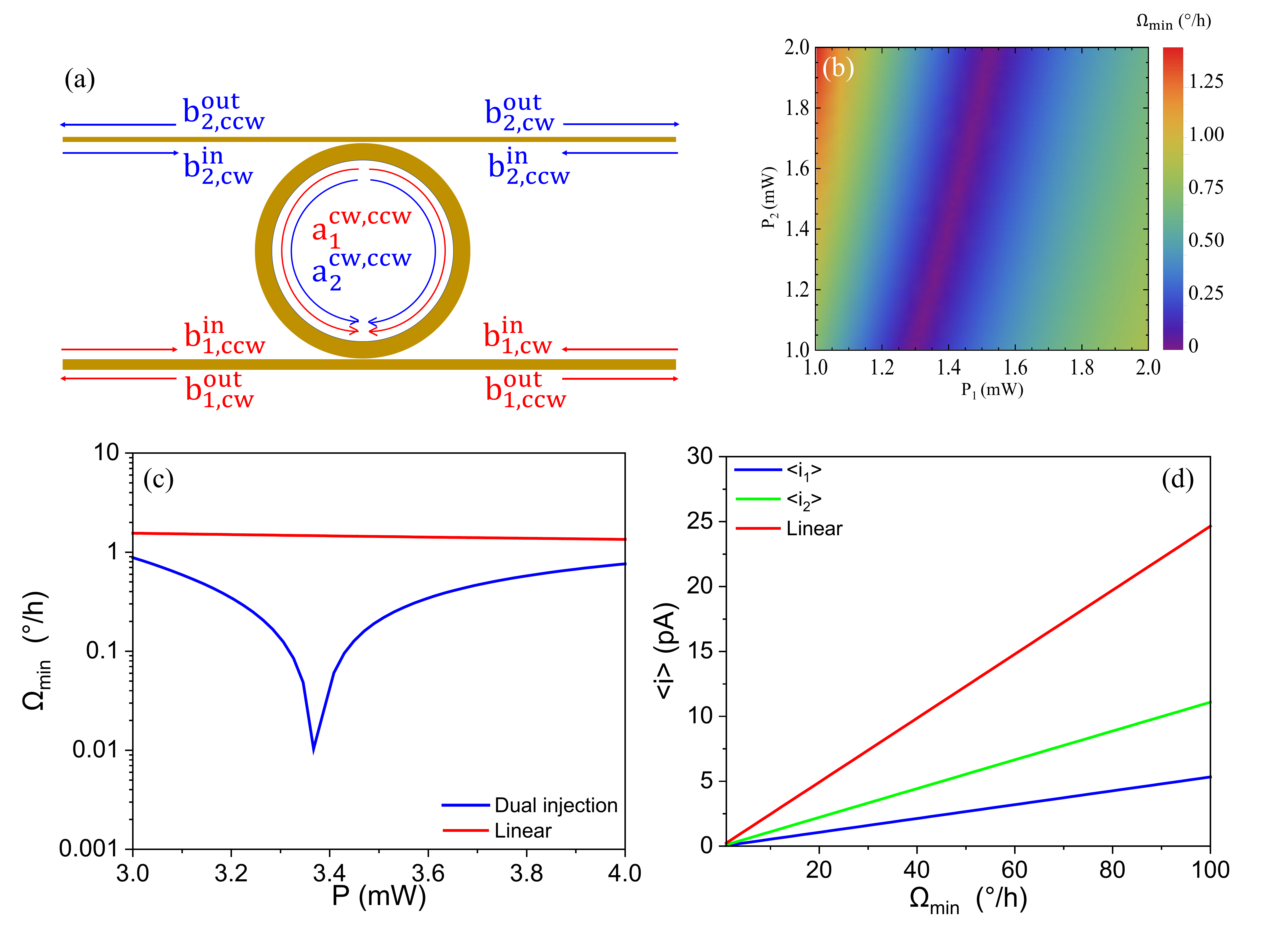}
    \caption{(a) The schematic of the dual (sub-second harmonic) injection scheme. The input light is only injected at both the second (blue arrow) and the fundamental frequency (red arrow). (b) 2D density plot of the sensitivity at the dual injection scheme. The sensitivity (MDR) is expressed as a function of the input power $P_1$ and $P_2$. (c) 1D plot of the sensitivity of the dual injection scheme (blue) and the standard linear gyroscope (red) at optimal $Q_c$ in terms of the input power. The inset shows the magnified plot of the critical region. (d) 1D linear plot of the mean differential current of the output light at the fundamental frequency (blue), second harmonic (green) and the standard linear gyroscope (red).}
    \label{fig:6}
\end{figure}

For the sake of completeness, we also studied the dual injection scheme (Fig.~\ref{fig:6}a) where coherent light is injected at both fundamental and second harmonics ($\lambda_1$ = 1590 nm and $\lambda_2$ = 795 nm). Here, we optimize four independent parameters: $P_1$, $P_2$, $Q_{c1}$, and $Q_{c2}$. The 2D density plot of the sensitivity (MDR) in terms of $P_1$ and $P_2$ is shown in Fig. \ref{fig:6}(b). Here $Q_{c1}$ and $Q_{c2}$ are fixed at the optimal values of $4.353 \times 10^5$ and $8.769 \times 10^6$, again discovered by Bayesian optimization. In Fig. \ref{fig:6}(b), $P_1$ ranges from 1 mW to 2 mW and $P_2$ ranges from 1 mW to 2 mW. Similar to the scheme of fundamental frequency injection, we found that low input power is also necessary for high sensitivity, which is beneficial for integrated optical gyroscopes. As shown in the figure, MDR ranges from 0 - 1.4 $^\circ$/h across the entire parameter space. The region of the lowest MDR is expressed by a narrow purple band in the figure while outside this region MDR gradually increases. We also compare the MDR of the dual injection scheme (solid blue line) with the linear gyroscope (solid red line), as shown in Fig. \ref{fig:6}(c). Note that in the figure the x-axis denotes the total input power $P$, which equals to $P_1+P_2$ for the dual frequency injection and $P_1$ for the linear gyroscope. \hl{Here $P_2$ is fixed at 1.873 mW, which is the optimal value discovered by Bayesian optimization}. It is shown that the sensitivity of dual frequency injection drops from $\sim 0.88 ^\circ$/h down to near zero as $P$ increases from 3 mW to 3.373 mW, then rises back to $\sim$ 0.762 $ ^\circ$/h as $P$ increases to 4 mW while sensitivity of the linear gyroscope drops from 1.555 $^\circ$/h to 1.347 $^\circ$/h within the same range of injection power. At the optimal power (3.373 mW), a high sensitivity of 0.013 $^\circ$/h is observed, leading to a substantial sensitivity improvement of $\sim$ 113.1 $\times$ over the linear gyroscope (0.013 $^\circ$/h vs 1.47 $^\circ$/h). To better understand the influence of rotation, the mean output currents are shown in Fig. \ref{fig:6}(d). As the rotation rate $\Omega$ increases from 0 - 100 $ ^\circ$/h, $<i_1>$, $<i_2>$ and the mean differential current of the linear gyroscope monotonically increases from 0 to 5.3 pA, 11.1 pA and 24.7 pA respectively. Note that here neither the sub nor the second harmonic mode shows stronger output current compared to the linear gyroscope. This is because the injection power of the linear gyroscope equals to the summation of both the sub and the second harmonic injection. Nonetheless, significant sensitivity improvement is still observed in this case due to the combination of the nonlinear coupling and the generation of the phase-squeezed photons.

\subsection{Discussion}
\label{sec:dtab}
Table \ref{tab:1} summarizes the maximal sensitivity enhancement factors (over the linear baseline) that can obtained in multiple operational regimes over a wide range of critical power requirements. Under the optimal second-harmonic injection at $\approx 23.5$ mW, a 9.9 dB quadrature squeezing is predicted, where our nonlinear multi-resonant cavity quantum photonics gyro (or quantum-optic nonlinear gyro QONG in short) can be nearly $124.4 \times$ more sensitive than an optimized linear gyro with the same radius, intrinsic quality factor and power budget, allowing for a minimum detectable rotation (MDR) as small as 0.0044 $^\circ$/h. Alternatively, even larger enhancement factors can be obtained at lower powers under the fundamental and the dual frequency injection schemes. The dual frequency injection scheme achieves near 7 $\times$ sensitivity improvement over the fundamental frequency injection scheme (0.013 $^\circ$/h vs 0.093 $^\circ$/h), which could be the result of the extra squeezed photons generated by the process of parametric down conversion. \hl{In fact, these two latter scenarios do not only promote squeezing (4.8 dB in fundamental frequency injection and 5.09 dB in dual injection) but also rely on nonlinear coupling, which will be detailly discussed in Appendix} \ref{sec:appendix3}. In either of our fundamental or second-harmonic injection scheme, we measured the output signals at both the fundamental and the second harmonic frequencies in order to fully utilize the input pump power (which gets converted into both harmonics), setting up a fair comparison to a linear gyro under the same pump power. For a more conservative comparison, one may argue for using dual inputs and outputs in the linear case. Aside from the fact that having to use two different frequency lasers can be disadvantageous, a simple calculation readily shows that measuring two \textit{non-interacting} resonances in a linear gyro can offer only up to $\sqrt{2}\times$ improvement (under the same power budgets)---in fact, much less than $\sqrt{2}$ due to the smaller $Q_{i2}$---highlighting that nonlinear effects are indeed indispensable for significant sensitivity enhancements. Most importantly, the crucial insight we have drawn from our investigations is to realize that multiple resonances in a nonlinear resonator can be engineered to reinforce each other through nonlinear wave mixing, and can be used as powerful degrees of freedom to optimize sensitivities. This critical realization suggests an exciting future direction: to generalize our QONG from just two resonances to many more nonlinearly interacting resonances (see also Section \ref{sec:s&o}), which may lead to even better sensitivities and functionalities (approaching the ultimate Heisenberg limit).
\begin{table}
  \centering
  \includegraphics[width=0.5\textwidth]{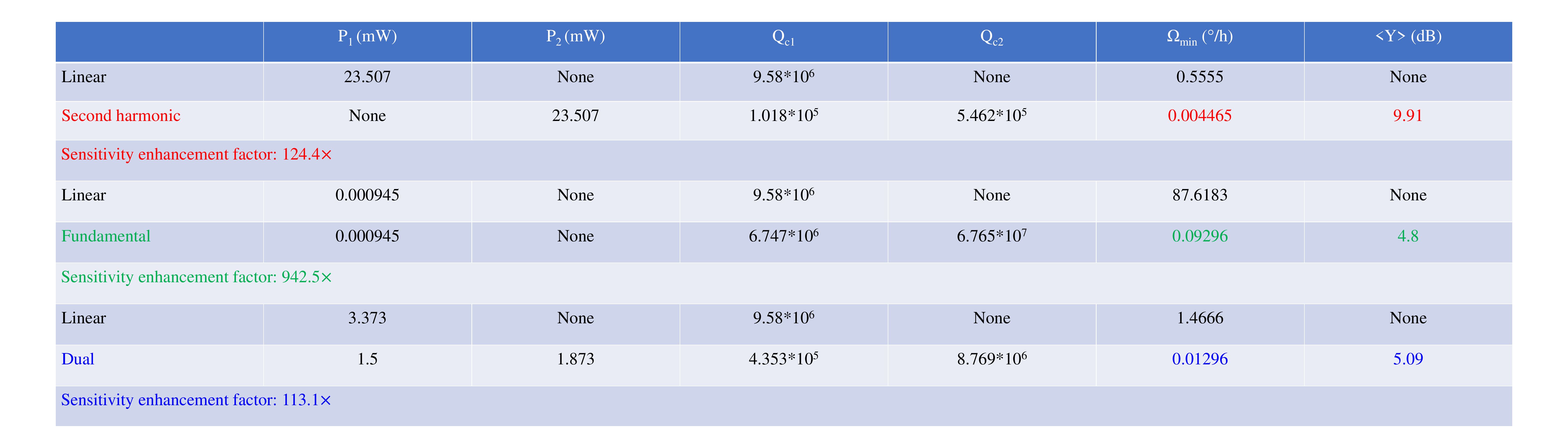}
  \caption{Optimal sensitivities of various injection schemes}
  \label{tab:1}
\end{table}
\section{Summary and Outlook}
\label{sec:s&o}
We have introduced a new type of quantum light gyroscopes based on nonlinear multi-resonant cavity quantum photonics (quantum-optic nonlinear gyro or QONG in short). Specifically, we analyzed and optimized the quantum-enhanced gyroscopic sensitivity of a doubly resonant $\chi^{(2)}$ cavity, revealing that, under quantum noise conditions, $\gtrsim 900\times$ enhancement is possible over the classical shot noise limit. \hl{A maximum sensitivity of 0.0044 $^\circ$/h has been achieved at an input power of 23.507 mW at the second harmonic injection, exhibiting comparable sensitivity performance and much lower power consumption than state-of-the-art FOGs (Boreas D90: 0.001 $^\circ$/h, 12 W,) and RLGs (Honeywell GG1320AN: 0.0035 $^\circ$/h, 1.6 W)} \citep{BoreasD90,GG1320N}. \hl{In addition, one of the main advantages of the current work is that the quantum states of light are generated on chip instead of being injected.} We highlight that our current design, which uses two resonances, represents only an elementary step and a relatively simple example of an QONG. In future works, we will develop a \textit{comprehensive} QONG inertial sensing paradigm, where a synergistic amalgamation of both quadratic $\chi^{(2)}$ and cubic $\chi^{(3)}$ nonlinearities, along with multiple intermixing resonances, mutually reinforced Sagnac shifts, co-arising quantum correlations, electro-optics dynamical control and geometry-induced anomalous dispersion effects, can unleash extraordinary complexities and freedoms, which can be fully exploited by state-of-the-art optimization techniques \citep{SaadHikmat,floudas2013deterministic,baydin2018automatic,Shakir} in order to identify unprecedented regimes for gyroscopic operation and sensitivities. A full incarnation of an QONG can be described by a Heisenberg-Langevin system of the form (or an equivalent density-operator Master equation \citep{Drummond_1980}):

\begin{align}
{d\hat{a}^\mu_j \over dt} &= \left(i\omega_j + i \delta_j^\mu(\Omega) - {\kappa_j \over 2}- {\gamma_j \over 2}\right) \hat{a}^\mu_j 
+ i \beta_j \hat{a}^\nu_j \notag \\
&+ \sum_{kl\alpha\beta} f^{(2)}_{kl\alpha\beta}\left( \hat{a}^\alpha_k, \hat{a}^\beta_l, (\hat{a}^\alpha_k)^\dagger, (\hat{a}^\beta_l)^\dagger \right ) \notag \\
&+ \sum_{klm\alpha\beta\theta} f^{(3)}_{klm\alpha\beta\theta}\left( \hat{a}^\alpha_k, \hat{a}^\beta_l, \hat{a}^\theta_m, 
(\hat{a}^\alpha_k)^\dagger, (\hat{a}^\beta_l)^\dagger, (\hat{a}^\theta_m)^\dagger \right ) \notag \\
&+ \sum_k \sqrt{\kappa^c_{jk}} \hat{a}_{jk,in}^\mu + \sum_k \sqrt{\gamma^r_{jk}} \hat{\eta}_{jk}^\mu \label{eq:a32}
\end{align}

for a selected set of carefully phase-matched and dispersion-engineered resonances $\{ \omega_j, j=1,...,N \}$. Here, $\delta^\mu (\Omega), \mu \in \{\text{cw,ccw}\}$, is the rotation-dependent Sagnac shift in the CW or CCW mode at each resonance. The functions $f^{(2)}$ and $f^{(3)}$ are polynomials of the annihilation and creation operators, representing all possible quantum-coherent three-wave mixing and four-wave mixing interactions between the selected resonances; these processes include sum and difference frequency generations of different orders and combinations as well as Kerr-variety self-phase and cross-phase modulation, and even cascaded processes \citep{Misoguti}. It is important to note that the strengths of different $f^{(2)}$ and $f^{(3)}$ terms are determined by nonlinear coupling factors \citep{Guo:16} which characterize the field concentration and nonlinear overlaps of the modes of the photonic resonator and can be computed from nanophotonic simulations. Therefore, on-chip structural parameters, ranging from a few simple shape parameters to entire permittivity distributions, can serve as design degrees of freedom \citep{Molesky2018}, by which we can engineer and optimize the different nonlinear processes (e.g. their relative contributions). The outputs of this multi-resonance system are collected by multiple waveguide ports and are set to passively interfere with each other and/or go through active electro-optics pulse processing (readily achievable on a TFLN platform \citep{Zhu:21}) before arriving at multiple photodetectors to yield multiple photocurrent signals $\mathbf{i} = \{\hat{i}_1,...,\hat{i}_M\}$. From these multi-variable (vector-valued) measurements, one can perform deep inferential analysis (such as advanced Bayesian computing \citep{csillery2010approximate}) to deduce the underlying non-inertial motion; the sensitivity of the entire process can be characterized by an end-to-end computation of Fisher Information, which will serve as an optimization figure of merit. We recognize tremendous opportunity in analyzing and optimizing such a system with increasing levels of mathematical and computational vigor, starting from steady-state analysis, small-signal modeling, classical stochastic simulations, to the non-perturbative quantum phase-space apparatus involving positive P-representations, Fokker-Planck equations and stochastic calculus \citep{Drummond2004,Drummond_1980}, from few-parameter deterministic global optimization \citep{horst2013global,floudas2013deterministic}, multi-parameter Bayesian optimization \citep{Shahriari} and evolutionary algorithms \citep{Back}, machine-learning assisted hybrid optimization \citep{Qi}, and Monte Carlo gradient computations \citep{Shakir} to billion-voxel topology optimization \citep{aage2017giga} and full end-to-end inverse design \citep{Zin2} of the entire workflow from the underlying resonator geometry to multi-variable inferential processes. Experimentally, thin film lithium niobate (TFLN) continues to offer the most suitable platform which features state-of-the-art on-chip frequency combs, pulse shaping, frequency shifting and ultra-fast signal processing capabilities \citep{McCutcheon:09,Wang2019,Yu2022,Lu2023}.

\begin{acknowledgments}
We thank Charles Roques-Carmes, Steven G. Johnson, Kiyoul Yang and Michael Larsen for informative discussions. Financial support was provided by generous gifts from the Virginia Tech Foundation. VK thanks the late Stavros Katsanevas for pointing out the new runs of VIRGO based on quantum mechanical states in November of 2018.
\end{acknowledgments}

\appendix
\setcounter{figure}{0} 
\renewcommand\thefigure{A\arabic{figure}}

\section{Steady state solutions}\label{sec:appendix1}

As discussed in Section \ref{sec:Noninearmodel}, the nonlinear coupled equations are solved by linearization. The classical scalar valued amplitudes $\alpha$ are obtained from the steady state analysis of Eqs.~\ref{eq:a15}-\ref{eq:a18}. To evaluate the steady state solutions,  the noise terms containing the quantum operators are omitted. The equations are thus simplified as follows:
\begin{align}
\begin{split}
    f_1 &= \left({\kappa_1 \over 2} + {\gamma_1 \over 2}-i\delta_1\right)a_\text{1,cw}-i\beta_1 a_\text{1,ccw} \\
    & -\chi a_\text{1,cw}^* a_\text{2,cw}-\sqrt{\kappa_1}b^\text{in}_\text{1,cw} 
\end{split}\label{eq:a33}\\
\begin{split}
    f_2 &= \left({\kappa_1 \over 2} + {\gamma_1 \over 2}+i\delta_1\right) a_\text{1,ccw}-i\beta_1 a_\text{1,cw} \\
    & -\chi a_\text{1,ccw}^* a_\text{2,ccw}-\sqrt{\kappa_1}b^\text{in}_\text{1,ccw} 
\end{split}\label{eq:a34}\\
\begin{split}
    f_3 &= \left({\kappa_2 \over 2} + {\gamma_2 \over 2}-i\delta_2\right)a_\text{2,cw}-i\beta_2 a_\text{2,ccw} \\
    & +{1 \over 2} \chi a_\text{1,cw}^2-\sqrt{\kappa_2}b^\text{in}_\text{2,cw}
\end{split}\label{eq:a35}\\
\begin{split}
    f_4 &= \left({\kappa_2 \over 2} + {\gamma_2 \over 2}+i\delta_2\right) a_\text{2,ccw}-i\beta_2 a_\text{2,cw} \\
    & +{1 \over 2} \chi a_\text{1,ccw}^2-\sqrt{\kappa_2}b^\text{in}_\text{2,ccw}
\end{split}\label{eq:a36}
\end{align}

Next, the steady state solutions are hence obtained by solving the equations $F(f_1,f_2,f_3,f_4) = 0$. Aside from the cavity modes $a_{n,cw/ccw}$, the input fields are also expressed as the steady states $b_{n,cw/ccw}$. These values are determined by the input power $P_n$ at both waveguide ports as discussed in Equation~\ref{eq:a3}. Here we fix $b_n$ as real values, which reduces the phase noise as reported by Dowling \citep{Dowling}. At different injection schemes, different arrangements of $b_n$ are employed. For example, $b_2=0$ at when the input light is injected at the fundamental frequency and $b_1=0$ at the second harmonic injection. Though the steady state solutions of a similar system has been studied by Drummond \citep{Drummond}, in which the analytical solutions are given at each injection scheme. In our system, however, the rotation-induced frequency shift $\delta_n$ and the Rayleigh back-scattering $\beta_n$ which introduces cross-coupling between the CW and the CCW modes make it impossible to calculate the analytical solution, hence we calculated the numerical solutions instead. Since these equations are nonlinear equations, multiple solutions are expected at each set of parameters. In order to discover the steady state solutions, linear stability analysis is performed \citep{erneux_glorieux_2010,GAVRIELIDES1997253}. By checking the eigen values of the Jacobian matrix \citep{2022Linear} associated with each set of solution we can determine the stability of these solutions. The Jacobian matrix J of Eqs~\ref{eq:a33}-\ref{eq:a36} is obtained by taking the gradient over a vector of the unknown variables $a_\text{n,cw/ccw}$:
\begin{align}
    J &= \nabla F |_{a_\text{n,cw/ccw}} \label{eq:a37}
\end{align}
Note that in Equation \ref{eq:a37} the function system $F$ and the variables $a_\text{n,cw/ccw}$ are both vectors, hence the gradient operator $\nabla$ generates a matrix $J$ instead of a vector. When the real part of the eigen values of the matrix is negative, the solution is stable and can be used for further calculation. 

\section{The algebra of the quantum operators}\label{sec:appendix2}

The system is assumed to be quantum-limited, meaning that the shot noise is considered as the main source of noise. To this end, it is necessary to investigate the statistical properties of the output light. As discussed in Section \ref{sec:Noninearmodel}, the output light are expressed by the operators $\hat{b}_\text{n,cw/ccw}^\text{out}$. As shown in Fig. \ref{fig:2}, the output light is measured by homodyne detection that $\hat{b}_\text{n,cw}^\text{out}$ and $\hat{b}_\text{n,ccw}^\text{out}$ are coupled with each other before being detected by two independent photodetectors:
\begin{align}
    \hat{b}_{1+} = \left( \hat{b}^\text{out}_\text{1,cw} e^{-i\phi_1}+i \hat{b}^\text{out}_\text{1,ccw} e^{i\phi_1}\right)/\sqrt{2} \label{eq:a38}\\
    \hat{b}_{1-} = \left( i \hat{b}^\text{out}_\text{1,cw} e^{-i\phi_1}+ \hat{b}^\text{out}_\text{1,ccw} e^{i\phi_1}\right)/\sqrt{2} \label{eq:a39}\\
    \hat{b}_{2+} = \left( \hat{b}^\text{out}_\text{2,cw} e^{-i\phi_2}+i \hat{b}^\text{out}_\text{2,ccw} e^{i\phi_2}\right)/\sqrt{2} \label{eq:a40}\\
    \hat{b}_{2-} = \left( i \hat{b}^\text{out}_\text{2,cw} e^{-i\phi_2}+ \hat{b}^\text{out}_\text{2,ccw} e^{i\phi_2}\right)/\sqrt{2} \label{eq:a41}
\end{align}
Here $\phi_1$ and $\phi_2$ are the propagation phase shifts of the output light at the sub and the second harmonics, which can be arbitrarily selected. Here we set them to zero. The resultant differential current current is given by:
\begin{align}
    \hat{i}_{1} = R \hbar \omega_1 \left( \hat{b}_{1+}^\dagger \hat{b}_{1+} - \hat{b}_{1-}^\dagger \hat{b}_{1-}\right) \label{eq:a42}\\
    \hat{i}_{2} = R \hbar \omega_2 \left( \hat{b}_{2+}^\dagger \hat{b}_{2+} - \hat{b}_{2-}^\dagger \hat{b}_{2-}\right) \label{eq:a43}
\end{align}
Here R is the responsivity of the photodetectors, set as $0.58~A/W$ in our analysis. Setting $A_1=R \hbar \omega_1$ and $A_2=R \hbar \omega_2$, Eqs.~\ref{eq:a42}-\ref{eq:a43} can be further simplified as:
\begin{align}
    \hat{i}_1 = i A_1 \left( \hat{b}^{\text{out} \dagger}_\text{1,cw} \hat{b}^\text{out}_\text{1,ccw} -\hat{b}^{\text{out} \dagger}_\text{1,ccw} \hat{b}^\text{out}_\text{1,cw} \right) \label{eq:a44}\\
    \hat{i}_2 = i A_2 \left( \hat{b}^{\text{out} \dagger}_\text{2,cw} \hat{b}^\text{out}_\text{2,ccw} -\hat{b}^{\text{out} \dagger}_\text{2,ccw} \hat{b}^\text{out}_\text{2,cw} \right) \label{eq:a45}
\end{align}
Following Maleki's approach \citep{MATSKO20182289}, the output operators are linearized as $\hat{b}_\text{n,cw/ccw}^\text{out} = b_\text{n,cw/ccw}^\text{out} + \hat{\delta b}_\text{n,cw/ccw}^\text{out}$. Here we analyze the perturbation terms, such that Eqs.~\ref{eq:a44}- \ref{eq:a45} are simplified as:
\begin{align}
\begin{split}
    \hat{\delta i}_1 = i A_1 \bigl(& b^{\text{out}}_\text{1,ccw} \hat{\delta b}^{\text{out} \dagger}_\text{1,cw} + b^{\text{out} *}_\text{1,cw} \hat{\delta b}^\text{out}_\text{1,ccw} \\
    &- b^{\text{out}^*}_\text{1,ccw} \hat{\delta b}^\text{out}_\text{1,cw} - b^{\text{out}}_\text{1,cw} \hat{\delta b}^{\text{out} \dagger}_\text{1,ccw} \bigr) 
\end{split}\label{eq:a46}\\
\begin{split}
    \hat{\delta i}_2 = i A_2 \bigl(& b^{\text{out}}_\text{2,ccw} \hat{\delta b}^{\text{out} \dagger}_\text{2,cw} + b^{\text{out} *}_\text{2,cw} \hat{\delta b}^\text{out}_\text{2,ccw} \\
    &- b^{\text{out}^*}_\text{2,ccw} \hat{\delta b}^\text{out}_\text{2,cw} - b^{\text{out}}_\text{2,cw} \hat{\delta b}^{\text{out} \dagger}_\text{2,ccw} \bigr)
\end{split}\label{eq:a47}
\end{align}
Then quadrature basis expansion is performed to separate the real and the imaginary parts ($X$ and $Y$) of the operators:
\begin{align}
    b_\text{n,cw/ccw}^\text{out} = X_\text{n,cw/ccw}^\text{out} + i Y_\text{n,cw/ccw}^\text{out} \label{eq:a48}\\
    b_\text{n,cw/ccw}^{\text{out} *} = X_\text{n,cw/ccw}^\text{out} - i Y_\text{n,cw/ccw}^\text{out} \label{eq:a49}\\
    \hat{\delta b}^\text{out}_\text{n,cw/ccw} = \hat{\delta X}_\text{n,cw/ccw}^\text{out} + i \hat{\delta Y}_\text{n,cw/ccw}^\text{out} \label{eq:a50}\\
    \hat{\delta b}^{\text{out} \dagger}_\text{n,cw/ccw} = \hat{\delta X}_\text{n,cw/ccw}^\text{out} - i \hat{\delta Y}_\text{n,cw/ccw}^\text{out} \label{eq:a51}
\end{align}
Hence, Eqs.~\ref{eq:a46}-\ref{eq:a47} are converted to:
\begin{align}
\begin{split}
    \hat{\delta i}_1 = 2 A_1 \bigl(& X^\text{out}_\text{1,ccw} \hat{\delta Y}^\text{out}_\text{1,cw} - Y^\text{out}_\text{1,ccw} \hat{\delta X}^\text{out}_\text{1,cw} \\
    &- X^\text{out}_\text{1,cw} \hat{\delta Y}^\text{out}_\text{1,ccw} + Y^\text{out}_\text{1,cw} \hat{\delta X}^\text{out}_\text{1,ccw} \bigr)
\end{split}\label{eq:a52}\\
\begin{split}
    \hat{\delta i}_2 = 2 A_2 \bigl(& X^\text{out}_\text{2,ccw} \hat{\delta Y}^\text{out}_\text{2,cw} - Y^\text{out}_\text{2,ccw} \hat{\delta X}^\text{out}_\text{2,cw} \\
    &- X^\text{out}_\text{2,cw} \hat{\delta Y}^\text{out}_\text{2,ccw} + Y^\text{out}_\text{2,cw} \hat{\delta X}^\text{out}_\text{2,ccw} \bigr)
\end{split}\label{eq:a53}
\end{align}
Next we need to determine the statistical properties of $\hat{\delta i}_1$ and $\hat{\delta i}_2$. Nevertheless, the output light is in complex quantum states (squeezed vacuum/squeezed coherent) which are difficult to calculate. On the other hand, these quadrature operators are nothing but linear combinations of the input light which is in relatively simple quantum states (vacuum/coherent). To this end, we calculate the mean values and the variances from the input states. 
Rewrite Eqs.~\ref{eq:a52} and \ref{eq:a53} in the form below:
\begin{align}
    {\hat{\delta i}_1} &= \sum_{n=1}^2 (b_\text{x,n,cw/ccw}^{(1)} \hat{b}_\text{X,n,cw/ccw}^\text{in} + b_\text{y,n,cw/ccw}^{(1)} \hat{b}_\text{Y,n,cw/ccw}^\text{in} \notag \\
&+ c_\text{x,n,cw/ccw}^{(1)} \hat{c}_\text{X,n,cw/ccw}^\text{in} + c_\text{y,n,cw/ccw}^{(1)} \hat{c}_\text{Y,n,cw/ccw}^\text{in} ) \label{eq:a54} \\
    {\hat{\delta i}_2} &= \sum_{n=1}^2 (b_\text{x,n,cw/ccw}^{(2)} \hat{b}_\text{X,n,cw/ccw}^\text{in} + b_\text{y,n,cw/ccw}^{(2)} \hat{b}_\text{Y,n,cw/ccw}^\text{in} \notag \\
&+ c_\text{x,n,cw/ccw}^{(2)} \hat{c}_\text{X,n,cw/ccw}^\text{in} + c_\text{y,n,cw/ccw}^{(2)} \hat{c}_\text{Y,n,cw/ccw}^\text{in} ) \label{eq:a55}
\end{align}
In Eqs.~\ref{eq:a54}-\ref{eq:a55}, $\hat{b}_\text{X,n,cw/ccw}^\text{in}$, $\hat{b}_\text{Y,n,cw/ccw}^\text{in}$, $\hat{c}_\text{X,n,cw/ccw}^\text{in}$ and $\hat{c}_\text{Y,n,cw/ccw}^\text{in}$ are the quadrature operators (real and imaginary parts) of the injection light light $\hat{b}_\text{n,cw/ccw}^\text{in}$ and the intrinsic loss channels $\hat{c}_\text{n,cw/ccw}^\text{in}$ for sub/second (n=1,2) harmonic light, where $\hat{b}_\text{X,n,cw/ccw}^\text{in}$, $\hat{b}_\text{Y,n,cw/ccw}^\text{in}$ are in coherent states and   $\hat{c}_\text{X,n,cw/ccw}^\text{in}$ and $\hat{c}_\text{Y,n,cw/ccw}^\text{in}$ are in vacuum states.  $b_\text{x,n,cw/ccw}^{(1,2)}$, $b_\text{y,n,cw/ccw}^{(1,2)}$, $c_\text{x,n,cw/ccw}^{(1,2)}$ and $c_\text{y,n,cw/ccw}^{(1,2)}$ are the corresponding coefficients of these operators at either sub or second harmonic. Assume $\psi_1$/$\psi_2$ and $N_1$/$N_2$ are the initial phases and the numbers of the injected photons of the input light $\hat{b}_\text{1,cw/ccw}$ and $\hat{b}_\text{2,cw/ccw}$, where $N_1 = |b_1^\text{in}|^2$ and $N_2 = |b_2^\text{in}|^2$.
With all these ingredients, now we can calculate mean values and the variances of both differential currents. For a coherent state $|\alpha>$, the mean values of both quadrature operators and their squares are given by~\citep{scully_zubairy_1997}:
\begin{align}
    <\alpha|\hat{X}|\alpha> = \sqrt{N} \cos{\psi} \label{eq:a56} \\
    <\alpha|\hat{Y}|\alpha> = \sqrt{N} \sin{\psi} \label{eq:a57} \\
    <\alpha|\hat{X}^2|\alpha> = {{4 N (\cos{\psi})^2 +1} \over 4} \label{eq:a58} \\
    <\alpha|\hat{Y}^2|\alpha> = {{4 N (\sin{\psi})^2 +1} \over 4} \label{eq:a59}
\end{align}
and the variances are defined as the mean values of the square minus the square of the mean values of the quadrature operators:
\begin{align}
    <\alpha|\Delta \hat{X}^2|\alpha> = <\alpha|\hat{X}^2|\alpha> - (<\alpha|\hat{X}|\alpha>)^2
    = {1 \over 4} \label{eq:a60} \\
    <\alpha|\Delta \hat{Y}^2|\alpha> = <\alpha|\hat{Y}^2|\alpha> - (<\alpha|\hat{Y}|\alpha>)^2
    = {1 \over 4} \label{eq:a61}
\end{align}
Note that when two quadrature operators do not share the same eigen vectors, the definitions are different. For example, the inner product of a real quadrature operator at the second harmonic and an imaginary quadrature operator at the fundamental frequency injection is given by:
\begin{align}
    <\alpha|\hat{X_2} \hat{Y_1}|\alpha> = \sqrt{N_1} \sqrt{N_2} \cos{\psi_2} \sin{\psi_1} \label{eq:a62} \\
    <\alpha|\hat{X_2} \hat{Y_1}|\alpha> - <\alpha|\hat{X_2} |\alpha> <\alpha|\hat{Y_1} |\alpha> = 0\label{eq:a63}
\end{align}
Now we can obtain the mean values and the variances of $\hat{\delta i}_1$ and $\hat{\delta i}_2$:
\begin{align}
\begin{split}
    <\hat{\delta i}_1> = \sum_{n=1}^2 \bigl(&b_\text{x,n,cw/ccw}^{(1)} \sqrt{N_n} \cos{\psi_n} \\
    &+ b_\text{y,n,cw/ccw}^{(1)} \sqrt{N_n} \sin{\psi_n} \bigr)
\end{split}\label{eq:a64} \\
\begin{split}
    <\hat{\delta i}_2> = \sum_{n=1}^2 \bigl(&b_\text{x,n,cw/ccw}^{(2)} \sqrt{N_n} \cos{\psi_n} \\
    &+ b_\text{y,n,cw/ccw}^{(2)} \sqrt{N_n} \sin{\psi_n} \bigr)
\end{split}\label{eq:a65} \\
\begin{split}
    {<\Delta \hat{\delta i}_1^2>} = {1 \over 4} \sum_{n=1}^2 \bigl([&(b_\text{x,n,cw/ccw}^{(1)})^2 + (b_\text{y,n,cw/ccw}^{(1)})^2 \\
    &+ (c_\text{x,n,cw/ccw}^{(1)})^2+(c_\text{y,n,cw/ccw}^{(1)})^2 ]\bigr)
\end{split}\label{eq:a66} \\
\begin{split}
    {<\Delta \hat{\delta i}_2^2>} = {1 \over 4} \sum_{n=1}^2 \bigl([&(b_\text{x,n,cw/ccw}^{(2)})^2 + (b_\text{y,n,cw/ccw}^{(2)})^2 \\
    &+ (c_\text{x,n,cw/ccw}^{(2)})^2+(c_\text{y,n,cw/ccw}^{(2)})^2 ]\bigr)
\end{split}\label{eq:a67} 
\end{align}
As discussed in Section \ref{sec:Noninearmodel}, in order to determine the covariance matrix, it is also necessary to calculate the correlation between $\hat{\delta i}_1$ and $\hat{\delta i}_2$ following the rule of operator calculation defined above:
\begin{align}
\begin{split}
    {< \hat{\delta i}_1 \hat{\delta i}_2> - <\hat{\delta i}_1> <\hat{\delta i}_2>} &= \sum_{n=1}^2 \bigl[b_\text{x,n,cw/ccw}^{(1)} b_\text{x,n,cw/ccw}^{(2)} \\
    &+ b_\text{y,n,cw/ccw}^{(1)} b_\text{y,n,cw/ccw}^{(2)} \\
    &+ c_\text{x,n,cw/ccw}^{(1)} c_\text{x,n,cw/ccw}^{(2)} \\
    &+c_\text{y,n,cw/ccw}^{(1)} c_\text{y,n,cw/ccw}^{(2)}\bigr]
\end{split}\label{eq:a68} \\
    {< \hat{\delta i}_2 \hat{\delta i}_1> - <\hat{\delta i}_2> <\hat{\delta i}_1>} &= {< \hat{\delta i}_1 \hat{\delta i}_2> - <\hat{\delta i}_1> <\hat{\delta i}_2>} \label{eq:a69}
\end{align}
With everything discussed in this section, particularly Eqs.~\ref{eq:a64}-\ref{eq:a69}, now we can calculate Eqs.~\ref{eq:a24}-\ref{eq:a26} to determine the Fisher information and the corresponding sensitivity of the system.

\section{Further analysis on sensitivity enhancement}\label{sec:appendix3}
\begin{figure}[htbp]
    \centering
    \includegraphics[width=\linewidth]{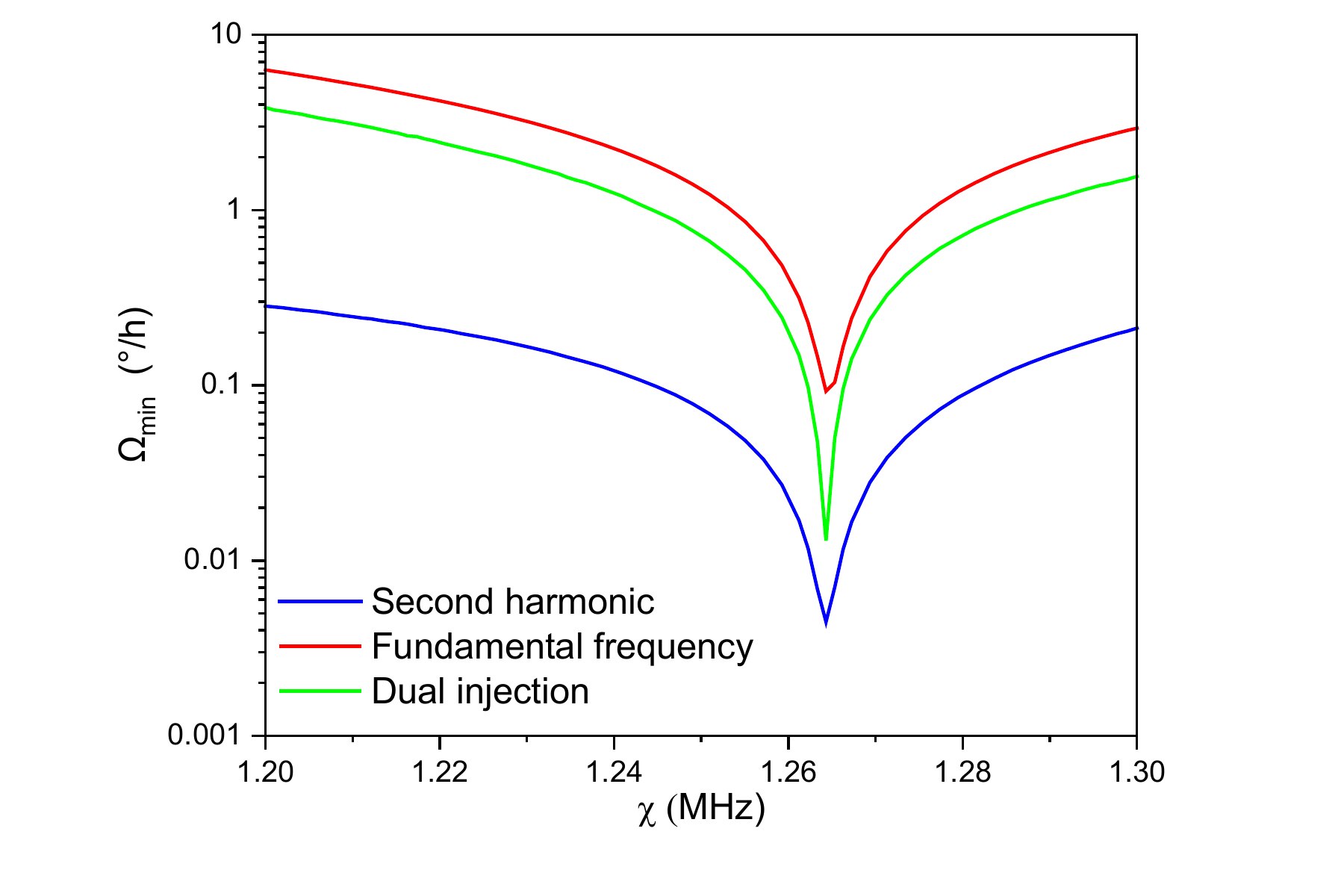}
    \caption{\hl{The sensitivities as functions of nonlinear coupling strength $\chi$ at various injection schemes.}}
    \label{fig:A2}
\end{figure}

\hl{To further investigate the mechanism for the sensitivity enhancements, we plotted the quadrature squeezing levels in all three injection schemes in the appendix} (Fig. \ref{fig:A1}). In Fig. \ref{fig:A1} (a) and (b), \hl{strong phase squeezing of $>$ 9 dB is observed for the output light at the fundamental frequency and increases at higher power, while negative amplitude squeezing is observed at the output signal at the second harmonic (0 to -3 dB), showing that phase squeezed light is generated by subharmonic generation.} In Fig. \ref{fig:A1} (c) and (d), \hl{relatively weaker phase squeezing is observed at the fundamental frequency injection (4.8 dB), while at the same time amplitude squeezing is also observed at the output at the second harmonics (6.03 dB). This observation is counter-intuitive because we expect up-conversion to be dominant at the fundamental frequency injection, which only generates amplitude-squeezed light at the second harmonic output. Nevertheless, since the phase-matching condition is satisfied, these high energy photons could be converted back to the fundamental frequency by down-conversion, leading to the phase squeezing. For the same reason, positive phase/amplitude squeezing is observed respectively at the output at the fundamental frequency/second harmonic at dual injection, as shown in} Fig. \ref{fig:A1} (e) and (f). \hl{Note that the sensitivity (MDR) monotonically increases with the level of phase squeezing, as shown in} Table \ref{tab:1}, \hl{showing that the squeezing of the phase noise indeed plays an important role in improving the sensitivity of the gyroscope. Furthermore, we examined the dependence of sensitivity (MDR) on the nonlinear coupling strength $\chi$ at all three injection schemes, as shown in} Fig. \ref{fig:A2}. \hl{The value of $\chi$ is 1.26 MHz, so here we scan over 1.2 – 1.3 MHz. Optimal sensitivities (minimum MDRs) are observed at 1.26 MHz in all three cases, suggesting that the gyroscope sensitivity is enhanced by the $\chi^{(2)}$ nonlinear coupling process, which is largely due to nonlinear eigenmode dispersion in the presence of critically sensitive $\chi^{(2)}$-mediated wave mixing processes, not dissimilar to enlarged Sagnac shifts reported in dispersive materials}~\citep{Dispersivegyro}.

\begin{figure}[htbp]
    \centering
    \includegraphics[width=\linewidth]{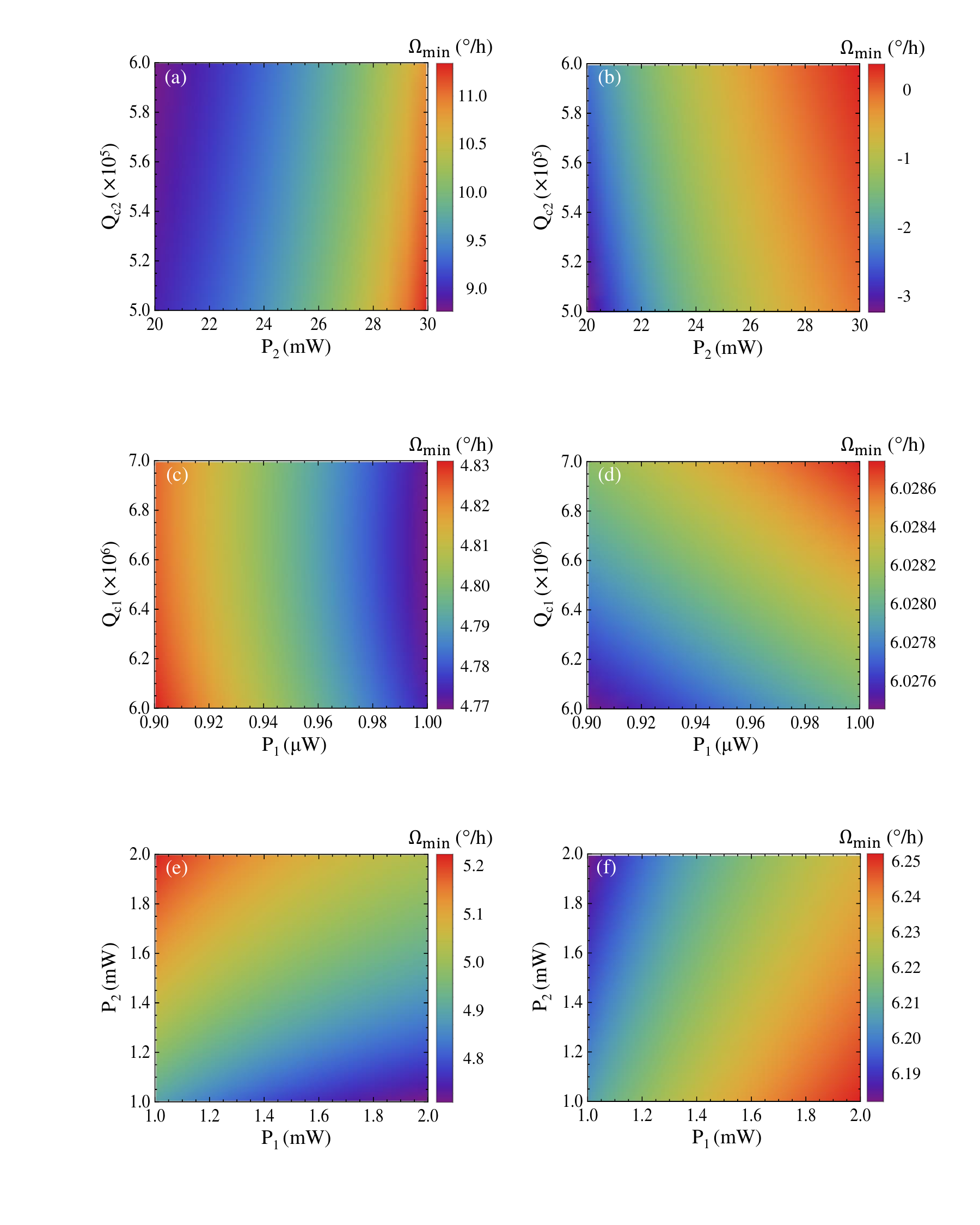}
    \caption{\hl{The squeezing levels of the phase and the amplitude quadrature at various injection schemes. (a)-(b): Second harmonic injection, (c)-(d): Fundamental frequency injection and (e)-(f): Dual injection.}}
    \label{fig:A1}
\end{figure}

\nocite{*}

\bibliography{apssamp}

\begin{thebibliography}{118}%
\makeatletter
\providecommand \@ifxundefined [1]{%
 \@ifx{#1\undefined}
}%
\providecommand \@ifnum [1]{%
 \ifnum #1\expandafter \@firstoftwo
 \else \expandafter \@secondoftwo
 \fi
}%
\providecommand \@ifx [1]{%
 \ifx #1\expandafter \@firstoftwo
 \else \expandafter \@secondoftwo
 \fi
}%
\providecommand \natexlab [1]{#1}%
\providecommand \enquote  [1]{``#1''}%
\providecommand \bibnamefont  [1]{#1}%
\providecommand \bibfnamefont [1]{#1}%
\providecommand \citenamefont [1]{#1}%
\providecommand \href@noop [0]{\@secondoftwo}%
\providecommand \href [0]{\begingroup \@sanitize@url \@href}%
\providecommand \@href[1]{\@@startlink{#1}\@@href}%
\providecommand \@@href[1]{\endgroup#1\@@endlink}%
\providecommand \@sanitize@url [0]{\catcode `\\12\catcode `\$12\catcode `\&12\catcode `\#12\catcode `\^12\catcode `\_12\catcode `\%12\relax}%
\providecommand \@@startlink[1]{}%
\providecommand \@@endlink[0]{}%
\providecommand \url  [0]{\begingroup\@sanitize@url \@url }%
\providecommand \@url [1]{\endgroup\@href {#1}{\urlprefix }}%
\providecommand \urlprefix  [0]{URL }%
\providecommand \Eprint [0]{\href }%
\providecommand \doibase [0]{https://doi.org/}%
\providecommand \selectlanguage [0]{\@gobble}%
\providecommand \bibinfo  [0]{\@secondoftwo}%
\providecommand \bibfield  [0]{\@secondoftwo}%
\providecommand \translation [1]{[#1]}%
\providecommand \BibitemOpen [0]{}%
\providecommand \bibitemStop [0]{}%
\providecommand \bibitemNoStop [0]{.\EOS\space}%
\providecommand \EOS [0]{\spacefactor3000\relax}%
\providecommand \BibitemShut  [1]{\csname bibitem#1\endcsname}%
\let\auto@bib@innerbib\@empty
\bibitem [{\citenamefont {Office}(2021)}]{doddoc}%
  \BibitemOpen
  \bibfield  {author} {\bibinfo {author} {\bibfnamefont {U.~S. G.~A.}\ \bibnamefont {Office}},\ }\href@noop {} {\bibinfo {title} {Defense navigation capabilities: Dod is developing positioning, navigation, and timing technologies to complement gps}} (\bibinfo {year} {2021})\BibitemShut {NoStop}%
\bibitem [{\citenamefont {Sagnac}(1913)}]{Sagnac}%
  \BibitemOpen
  \bibfield  {author} {\bibinfo {author} {\bibfnamefont {G.}~\bibnamefont {Sagnac}},\ }\bibfield  {title} {\bibinfo {title} {L'ether lumineux demontre par l'effet du vent relatif d'ether dans un interferometre en rotation uniforme},\ }\href@noop {} {\bibfield  {journal} {\bibinfo  {journal} {Proceedings of the USSR Academy of Sciences}\ }\textbf {\bibinfo {volume} {157}},\ \bibinfo {pages} {708} (\bibinfo {year} {1913})}\BibitemShut {NoStop}%
\bibitem [{\citenamefont {Post}(1967)}]{post1967sagnac}%
  \BibitemOpen
  \bibfield  {author} {\bibinfo {author} {\bibfnamefont {E.~J.}\ \bibnamefont {Post}},\ }\bibfield  {title} {\bibinfo {title} {Sagnac effect},\ }\href@noop {} {\bibfield  {journal} {\bibinfo  {journal} {Reviews of Modern Physics}\ }\textbf {\bibinfo {volume} {39}},\ \bibinfo {pages} {475} (\bibinfo {year} {1967})}\BibitemShut {NoStop}%
\bibitem [{\citenamefont {Arditty}\ and\ \citenamefont {Lef\`{e}vre}(1981)}]{Arditty}%
  \BibitemOpen
  \bibfield  {author} {\bibinfo {author} {\bibfnamefont {H.~J.}\ \bibnamefont {Arditty}}\ and\ \bibinfo {author} {\bibfnamefont {H.~C.}\ \bibnamefont {Lef\`{e}vre}},\ }\bibfield  {title} {\bibinfo {title} {Sagnac effect in fiber gyroscopes},\ }\href {https://doi.org/10.1364/OL.6.000401} {\bibfield  {journal} {\bibinfo  {journal} {Opt. Lett.}\ }\textbf {\bibinfo {volume} {6}},\ \bibinfo {pages} {401} (\bibinfo {year} {1981})}\BibitemShut {NoStop}%
\bibitem [{\citenamefont {Chow}\ \emph {et~al.}(1985)\citenamefont {Chow}, \citenamefont {Gea-Banacloche}, \citenamefont {Pedrotti}, \citenamefont {Sanders}, \citenamefont {Schleich},\ and\ \citenamefont {Scully}}]{Chow}%
  \BibitemOpen
  \bibfield  {author} {\bibinfo {author} {\bibfnamefont {W.~W.}\ \bibnamefont {Chow}}, \bibinfo {author} {\bibfnamefont {J.}~\bibnamefont {Gea-Banacloche}}, \bibinfo {author} {\bibfnamefont {L.~M.}\ \bibnamefont {Pedrotti}}, \bibinfo {author} {\bibfnamefont {V.~E.}\ \bibnamefont {Sanders}}, \bibinfo {author} {\bibfnamefont {W.}~\bibnamefont {Schleich}},\ and\ \bibinfo {author} {\bibfnamefont {M.~O.}\ \bibnamefont {Scully}},\ }\bibfield  {title} {\bibinfo {title} {The ring laser gyro},\ }\href {https://doi.org/10.1103/RevModPhys.57.61} {\bibfield  {journal} {\bibinfo  {journal} {Rev. Mod. Phys.}\ }\textbf {\bibinfo {volume} {57}},\ \bibinfo {pages} {61} (\bibinfo {year} {1985})}\BibitemShut {NoStop}%
\bibitem [{\citenamefont {Cresser}\ \emph {et~al.}(1982)\citenamefont {Cresser}, \citenamefont {Louisell}, \citenamefont {Meystre}, \citenamefont {Schleich},\ and\ \citenamefont {Scully}}]{Cresser}%
  \BibitemOpen
  \bibfield  {author} {\bibinfo {author} {\bibfnamefont {J.~D.}\ \bibnamefont {Cresser}}, \bibinfo {author} {\bibfnamefont {W.~H.}\ \bibnamefont {Louisell}}, \bibinfo {author} {\bibfnamefont {P.}~\bibnamefont {Meystre}}, \bibinfo {author} {\bibfnamefont {W.}~\bibnamefont {Schleich}},\ and\ \bibinfo {author} {\bibfnamefont {M.~O.}\ \bibnamefont {Scully}},\ }\bibfield  {title} {\bibinfo {title} {Quantum noise in ring-laser gyros. i. theoretical formulation of problem},\ }\href {https://doi.org/10.1103/PhysRevA.25.2214} {\bibfield  {journal} {\bibinfo  {journal} {Phys. Rev. A}\ }\textbf {\bibinfo {volume} {25}},\ \bibinfo {pages} {2214} (\bibinfo {year} {1982})}\BibitemShut {NoStop}%
\bibitem [{\citenamefont {Matsko}\ \emph {et~al.}(2018)\citenamefont {Matsko}, \citenamefont {Liang}, \citenamefont {Savchenkov}, \citenamefont {Ilchenko},\ and\ \citenamefont {Maleki}}]{MATSKO20182289}%
  \BibitemOpen
  \bibfield  {author} {\bibinfo {author} {\bibfnamefont {A.~B.}\ \bibnamefont {Matsko}}, \bibinfo {author} {\bibfnamefont {W.}~\bibnamefont {Liang}}, \bibinfo {author} {\bibfnamefont {A.~A.}\ \bibnamefont {Savchenkov}}, \bibinfo {author} {\bibfnamefont {V.~S.}\ \bibnamefont {Ilchenko}},\ and\ \bibinfo {author} {\bibfnamefont {L.}~\bibnamefont {Maleki}},\ }\bibfield  {title} {\bibinfo {title} {Fundamental limitations of sensitivity of whispering gallery mode gyroscopes},\ }\href {https://doi.org/https://doi.org/10.1016/j.physleta.2017.09.028} {\bibfield  {journal} {\bibinfo  {journal} {Physics Letters A}\ }\textbf {\bibinfo {volume} {382}},\ \bibinfo {pages} {2289} (\bibinfo {year} {2018})},\ \bibinfo {note} {special Issue in memory of Professor V.B. Braginsky}\BibitemShut {NoStop}%
\bibitem [{\citenamefont {Wang}\ \emph {et~al.}(2014)\citenamefont {Wang}, \citenamefont {Yang}, \citenamefont {Lu}, \citenamefont {Luo}, \citenamefont {Li}, \citenamefont {Zhao}, \citenamefont {Peng},\ and\ \citenamefont {Li}}]{Wang:14}%
  \BibitemOpen
  \bibfield  {author} {\bibinfo {author} {\bibfnamefont {Z.}~\bibnamefont {Wang}}, \bibinfo {author} {\bibfnamefont {Y.}~\bibnamefont {Yang}}, \bibinfo {author} {\bibfnamefont {P.}~\bibnamefont {Lu}}, \bibinfo {author} {\bibfnamefont {R.}~\bibnamefont {Luo}}, \bibinfo {author} {\bibfnamefont {Y.}~\bibnamefont {Li}}, \bibinfo {author} {\bibfnamefont {D.}~\bibnamefont {Zhao}}, \bibinfo {author} {\bibfnamefont {C.}~\bibnamefont {Peng}},\ and\ \bibinfo {author} {\bibfnamefont {Z.}~\bibnamefont {Li}},\ }\bibfield  {title} {\bibinfo {title} {Dual-polarization interferometric fiber-optic gyroscope with an ultra-simple configuration},\ }\href {https://doi.org/10.1364/OL.39.002463} {\bibfield  {journal} {\bibinfo  {journal} {Opt. Lett.}\ }\textbf {\bibinfo {volume} {39}},\ \bibinfo {pages} {2463} (\bibinfo {year} {2014})}\BibitemShut {NoStop}%
\bibitem [{\citenamefont {Ma}\ \emph {et~al.}(2013)\citenamefont {Ma}, \citenamefont {Wang}, \citenamefont {Ren},\ and\ \citenamefont {Jin}}]{Ma}%
  \BibitemOpen
  \bibfield  {author} {\bibinfo {author} {\bibfnamefont {H.}~\bibnamefont {Ma}}, \bibinfo {author} {\bibfnamefont {W.}~\bibnamefont {Wang}}, \bibinfo {author} {\bibfnamefont {Y.}~\bibnamefont {Ren}},\ and\ \bibinfo {author} {\bibfnamefont {Z.}~\bibnamefont {Jin}},\ }\bibfield  {title} {\bibinfo {title} {Low-noise low-delay digital signal processor for resonant micro optic gyro},\ }\href {https://doi.org/10.1109/LPT.2012.2233727} {\bibfield  {journal} {\bibinfo  {journal} {IEEE Photonics Technology Letters}\ }\textbf {\bibinfo {volume} {25}},\ \bibinfo {pages} {198} (\bibinfo {year} {2013})}\BibitemShut {NoStop}%
\bibitem [{\citenamefont {Sanders}\ \emph {et~al.}(2016)\citenamefont {Sanders}, \citenamefont {Sanders}, \citenamefont {Strandjord}, \citenamefont {Qiu}, \citenamefont {Wu}, \citenamefont {Smiciklas}, \citenamefont {Mead}, \citenamefont {Mosor}, \citenamefont {Arrizon}, \citenamefont {Ho},\ and\ \citenamefont {Salit}}]{Sanders}%
  \BibitemOpen
  \bibfield  {author} {\bibinfo {author} {\bibfnamefont {G.~A.}\ \bibnamefont {Sanders}}, \bibinfo {author} {\bibfnamefont {S.~J.}\ \bibnamefont {Sanders}}, \bibinfo {author} {\bibfnamefont {L.~K.}\ \bibnamefont {Strandjord}}, \bibinfo {author} {\bibfnamefont {T.}~\bibnamefont {Qiu}}, \bibinfo {author} {\bibfnamefont {J.}~\bibnamefont {Wu}}, \bibinfo {author} {\bibfnamefont {M.}~\bibnamefont {Smiciklas}}, \bibinfo {author} {\bibfnamefont {D.}~\bibnamefont {Mead}}, \bibinfo {author} {\bibfnamefont {S.}~\bibnamefont {Mosor}}, \bibinfo {author} {\bibfnamefont {A.}~\bibnamefont {Arrizon}}, \bibinfo {author} {\bibfnamefont {W.}~\bibnamefont {Ho}},\ and\ \bibinfo {author} {\bibfnamefont {M.}~\bibnamefont {Salit}},\ }\bibfield  {title} {\bibinfo {title} {{Fiber optic gyro development at Honeywell}},\ }in\ \href {https://doi.org/10.1117/12.2228893} {\emph {\bibinfo {booktitle} {Fiber Optic Sensors and Applications XIII}}},\ Vol.\ \bibinfo {volume} {9852},\ \bibinfo {editor} {edited by\ \bibinfo {editor}
  {\bibfnamefont {E.}~\bibnamefont {Udd}}, \bibinfo {editor} {\bibfnamefont {G.}~\bibnamefont {Pickrell}},\ and\ \bibinfo {editor} {\bibfnamefont {H.~H.}\ \bibnamefont {Du}}},\ \bibinfo {organization} {International Society for Optics and Photonics}\ (\bibinfo  {publisher} {SPIE},\ \bibinfo {year} {2016})\ p.\ \bibinfo {pages} {985207}\BibitemShut {NoStop}%
\bibitem [{\citenamefont {Korth}\ \emph {et~al.}(2016)\citenamefont {Korth}, \citenamefont {Heptonstall}, \citenamefont {Hall}, \citenamefont {Arai}, \citenamefont {Gustafson},\ and\ \citenamefont {Adhikari}}]{Korth}%
  \BibitemOpen
  \bibfield  {author} {\bibinfo {author} {\bibfnamefont {W.~Z.}\ \bibnamefont {Korth}}, \bibinfo {author} {\bibfnamefont {A.}~\bibnamefont {Heptonstall}}, \bibinfo {author} {\bibfnamefont {E.~D.}\ \bibnamefont {Hall}}, \bibinfo {author} {\bibfnamefont {K.}~\bibnamefont {Arai}}, \bibinfo {author} {\bibfnamefont {E.~K.}\ \bibnamefont {Gustafson}},\ and\ \bibinfo {author} {\bibfnamefont {R.~X.}\ \bibnamefont {Adhikari}},\ }\bibfield  {title} {\bibinfo {title} {Passive, free-space heterodyne laser gyroscope},\ }\href {https://doi.org/10.1088/0264-9381/33/3/035004} {\bibfield  {journal} {\bibinfo  {journal} {Classical and Quantum Gravity}\ }\textbf {\bibinfo {volume} {33}},\ \bibinfo {pages} {035004} (\bibinfo {year} {2016})}\BibitemShut {NoStop}%
\bibitem [{\citenamefont {Dowling}(1998)}]{Dowling}%
  \BibitemOpen
  \bibfield  {author} {\bibinfo {author} {\bibfnamefont {J.~P.}\ \bibnamefont {Dowling}},\ }\bibfield  {title} {\bibinfo {title} {Correlated input-port, matter-wave interferometer: Quantum-noise limits to the atom-laser gyroscope},\ }\href {https://doi.org/10.1103/PhysRevA.57.4736} {\bibfield  {journal} {\bibinfo  {journal} {Phys. Rev. A}\ }\textbf {\bibinfo {volume} {57}},\ \bibinfo {pages} {4736} (\bibinfo {year} {1998})}\BibitemShut {NoStop}%
\bibitem [{\citenamefont {Liang}\ \emph {et~al.}(2017)\citenamefont {Liang}, \citenamefont {Ilchenko}, \citenamefont {Savchenkov}, \citenamefont {Dale}, \citenamefont {Eliyahu}, \citenamefont {Matsko},\ and\ \citenamefont {Maleki}}]{Liang:17}%
  \BibitemOpen
  \bibfield  {author} {\bibinfo {author} {\bibfnamefont {W.}~\bibnamefont {Liang}}, \bibinfo {author} {\bibfnamefont {V.~S.}\ \bibnamefont {Ilchenko}}, \bibinfo {author} {\bibfnamefont {A.~A.}\ \bibnamefont {Savchenkov}}, \bibinfo {author} {\bibfnamefont {E.}~\bibnamefont {Dale}}, \bibinfo {author} {\bibfnamefont {D.}~\bibnamefont {Eliyahu}}, \bibinfo {author} {\bibfnamefont {A.~B.}\ \bibnamefont {Matsko}},\ and\ \bibinfo {author} {\bibfnamefont {L.}~\bibnamefont {Maleki}},\ }\bibfield  {title} {\bibinfo {title} {Resonant microphotonic gyroscope},\ }\href {https://doi.org/10.1364/OPTICA.4.000114} {\bibfield  {journal} {\bibinfo  {journal} {Optica}\ }\textbf {\bibinfo {volume} {4}},\ \bibinfo {pages} {114} (\bibinfo {year} {2017})}\BibitemShut {NoStop}%
\bibitem [{\citenamefont {Zhang}\ \emph {et~al.}(2017{\natexlab{a}})\citenamefont {Zhang}, \citenamefont {Ma}, \citenamefont {Li},\ and\ \citenamefont {Jin}}]{Zhang:17}%
  \BibitemOpen
  \bibfield  {author} {\bibinfo {author} {\bibfnamefont {J.}~\bibnamefont {Zhang}}, \bibinfo {author} {\bibfnamefont {H.}~\bibnamefont {Ma}}, \bibinfo {author} {\bibfnamefont {H.}~\bibnamefont {Li}},\ and\ \bibinfo {author} {\bibfnamefont {Z.}~\bibnamefont {Jin}},\ }\bibfield  {title} {\bibinfo {title} {Single-polarization fiber-pigtailed high-finesse silica waveguide ring resonator for a resonant micro-optic gyroscope},\ }\href {https://doi.org/10.1364/OL.42.003658} {\bibfield  {journal} {\bibinfo  {journal} {Opt. Lett.}\ }\textbf {\bibinfo {volume} {42}},\ \bibinfo {pages} {3658} (\bibinfo {year} {2017}{\natexlab{a}})}\BibitemShut {NoStop}%
\bibitem [{\citenamefont {Khial}\ \emph {et~al.}(2018)\citenamefont {Khial}, \citenamefont {White},\ and\ \citenamefont {Hajimiri}}]{Khial2018}%
  \BibitemOpen
  \bibfield  {author} {\bibinfo {author} {\bibfnamefont {P.~P.}\ \bibnamefont {Khial}}, \bibinfo {author} {\bibfnamefont {A.~D.}\ \bibnamefont {White}},\ and\ \bibinfo {author} {\bibfnamefont {A.}~\bibnamefont {Hajimiri}},\ }\bibfield  {title} {\bibinfo {title} {Nanophotonic optical gyroscope with reciprocal sensitivity enhancement},\ }\href {https://doi.org/10.1038/s41566-018-0266-5} {\bibfield  {journal} {\bibinfo  {journal} {Nature Photonics}\ }\textbf {\bibinfo {volume} {12}},\ \bibinfo {pages} {671} (\bibinfo {year} {2018})}\BibitemShut {NoStop}%
\bibitem [{\citenamefont {Lai}\ \emph {et~al.}(2020)\citenamefont {Lai}, \citenamefont {Suh}, \citenamefont {Lu}, \citenamefont {Shen}, \citenamefont {Yang}, \citenamefont {Wang}, \citenamefont {Li}, \citenamefont {Lee}, \citenamefont {Yang},\ and\ \citenamefont {Vahala}}]{vahala}%
  \BibitemOpen
  \bibfield  {author} {\bibinfo {author} {\bibfnamefont {Y.-H.}\ \bibnamefont {Lai}}, \bibinfo {author} {\bibfnamefont {M.-G.}\ \bibnamefont {Suh}}, \bibinfo {author} {\bibfnamefont {Y.-K.}\ \bibnamefont {Lu}}, \bibinfo {author} {\bibfnamefont {B.}~\bibnamefont {Shen}}, \bibinfo {author} {\bibfnamefont {Q.-F.}\ \bibnamefont {Yang}}, \bibinfo {author} {\bibfnamefont {H.}~\bibnamefont {Wang}}, \bibinfo {author} {\bibfnamefont {J.}~\bibnamefont {Li}}, \bibinfo {author} {\bibfnamefont {S.~H.}\ \bibnamefont {Lee}}, \bibinfo {author} {\bibfnamefont {K.~Y.}\ \bibnamefont {Yang}},\ and\ \bibinfo {author} {\bibfnamefont {K.}~\bibnamefont {Vahala}},\ }\bibfield  {title} {\bibinfo {title} {Earth rotation measured by a chip-scale ring laser gyroscope},\ }\href {https://doi.org/10.1038/s41566-020-0588-y} {\bibfield  {journal} {\bibinfo  {journal} {Nature Photonics}\ }\textbf {\bibinfo {volume} {14}},\ \bibinfo {pages} {345} (\bibinfo {year} {2020})}\BibitemShut {NoStop}%
\bibitem [{\citenamefont {Ge}\ \emph {et~al.}(2015)\citenamefont {Ge}, \citenamefont {Sarma},\ and\ \citenamefont {Cao}}]{Ge:15}%
  \BibitemOpen
  \bibfield  {author} {\bibinfo {author} {\bibfnamefont {L.}~\bibnamefont {Ge}}, \bibinfo {author} {\bibfnamefont {R.}~\bibnamefont {Sarma}},\ and\ \bibinfo {author} {\bibfnamefont {H.}~\bibnamefont {Cao}},\ }\bibfield  {title} {\bibinfo {title} {Rotation-induced evolution of far-field emission patterns of deformed microdisk cavities},\ }\href {https://doi.org/10.1364/OPTICA.2.000323} {\bibfield  {journal} {\bibinfo  {journal} {Optica}\ }\textbf {\bibinfo {volume} {2}},\ \bibinfo {pages} {323} (\bibinfo {year} {2015})}\BibitemShut {NoStop}%
\bibitem [{\citenamefont {Scheuer}(2007)}]{Scheuer:07}%
  \BibitemOpen
  \bibfield  {author} {\bibinfo {author} {\bibfnamefont {J.}~\bibnamefont {Scheuer}},\ }\bibfield  {title} {\bibinfo {title} {Direct rotation-induced intensity modulation in circular bragg micro-lasers},\ }\href {https://doi.org/10.1364/OE.15.015053} {\bibfield  {journal} {\bibinfo  {journal} {Opt. Express}\ }\textbf {\bibinfo {volume} {15}},\ \bibinfo {pages} {15053} (\bibinfo {year} {2007})}\BibitemShut {NoStop}%
\bibitem [{\citenamefont {Ciminelli}\ \emph {et~al.}(2016)\citenamefont {Ciminelli}, \citenamefont {D'Agostino}, \citenamefont {Carnicella}, \citenamefont {Dell'Olio}, \citenamefont {Conteduca}, \citenamefont {Ambrosius}, \citenamefont {Smit},\ and\ \citenamefont {Armenise}}]{Cimi}%
  \BibitemOpen
  \bibfield  {author} {\bibinfo {author} {\bibfnamefont {C.}~\bibnamefont {Ciminelli}}, \bibinfo {author} {\bibfnamefont {D.}~\bibnamefont {D'Agostino}}, \bibinfo {author} {\bibfnamefont {G.}~\bibnamefont {Carnicella}}, \bibinfo {author} {\bibfnamefont {F.}~\bibnamefont {Dell'Olio}}, \bibinfo {author} {\bibfnamefont {D.}~\bibnamefont {Conteduca}}, \bibinfo {author} {\bibfnamefont {H.~P. M.~M.}\ \bibnamefont {Ambrosius}}, \bibinfo {author} {\bibfnamefont {M.~K.}\ \bibnamefont {Smit}},\ and\ \bibinfo {author} {\bibfnamefont {M.~N.}\ \bibnamefont {Armenise}},\ }\bibfield  {title} {\bibinfo {title} {A high-q inp resonant angular velocity sensor for a monolithically integrated optical gyroscope},\ }\href {https://doi.org/10.1109/JPHOT.2015.2507549} {\bibfield  {journal} {\bibinfo  {journal} {IEEE Photonics Journal}\ }\textbf {\bibinfo {volume} {8}},\ \bibinfo {pages} {1} (\bibinfo {year} {2016})}\BibitemShut {NoStop}%
\bibitem [{\citenamefont {Grant}\ \emph {et~al.}(2022)\citenamefont {Grant}, \citenamefont {Vigneron}, \citenamefont {Feshali}, \citenamefont {Jin}, \citenamefont {Abrams}, \citenamefont {Paniccia},\ and\ \citenamefont {Digonnet}}]{MatthewGrant}%
  \BibitemOpen
  \bibfield  {author} {\bibinfo {author} {\bibfnamefont {M.~J.}\ \bibnamefont {Grant}}, \bibinfo {author} {\bibfnamefont {P.-B.}\ \bibnamefont {Vigneron}}, \bibinfo {author} {\bibfnamefont {A.}~\bibnamefont {Feshali}}, \bibinfo {author} {\bibfnamefont {W.}~\bibnamefont {Jin}}, \bibinfo {author} {\bibfnamefont {N.}~\bibnamefont {Abrams}}, \bibinfo {author} {\bibfnamefont {M.}~\bibnamefont {Paniccia}},\ and\ \bibinfo {author} {\bibfnamefont {M.}~\bibnamefont {Digonnet}},\ }\bibfield  {title} {\bibinfo {title} {{Chip-scale gyroscope using silicon-nitride waveguide resonator with a Q factor of 100 million}},\ }in\ \href {https://doi.org/10.1117/12.2617219} {\emph {\bibinfo {booktitle} {Optical and Quantum Sensing and Precision Metrology II}}},\ Vol.\ \bibinfo {volume} {12016},\ \bibinfo {editor} {edited by\ \bibinfo {editor} {\bibfnamefont {J.}~\bibnamefont {Scheuer}}\ and\ \bibinfo {editor} {\bibfnamefont {S.~M.}\ \bibnamefont {Shahriar}}},\ \bibinfo {organization} {International Society for Optics and
  Photonics}\ (\bibinfo  {publisher} {SPIE},\ \bibinfo {year} {2022})\ p.\ \bibinfo {pages} {120160S}\BibitemShut {NoStop}%
\bibitem [{\citenamefont {Ren}\ \emph {et~al.}(2017)\citenamefont {Ren}, \citenamefont {Hodaei}, \citenamefont {Harari}, \citenamefont {Hassan}, \citenamefont {Chow}, \citenamefont {Soltani}, \citenamefont {Christodoulides},\ and\ \citenamefont {Khajavikhan}}]{Ren:17}%
  \BibitemOpen
  \bibfield  {author} {\bibinfo {author} {\bibfnamefont {J.}~\bibnamefont {Ren}}, \bibinfo {author} {\bibfnamefont {H.}~\bibnamefont {Hodaei}}, \bibinfo {author} {\bibfnamefont {G.}~\bibnamefont {Harari}}, \bibinfo {author} {\bibfnamefont {A.~U.}\ \bibnamefont {Hassan}}, \bibinfo {author} {\bibfnamefont {W.}~\bibnamefont {Chow}}, \bibinfo {author} {\bibfnamefont {M.}~\bibnamefont {Soltani}}, \bibinfo {author} {\bibfnamefont {D.}~\bibnamefont {Christodoulides}},\ and\ \bibinfo {author} {\bibfnamefont {M.}~\bibnamefont {Khajavikhan}},\ }\bibfield  {title} {\bibinfo {title} {Ultrasensitive micro-scale parity-time-symmetric ring laser gyroscope},\ }\href {https://doi.org/10.1364/OL.42.001556} {\bibfield  {journal} {\bibinfo  {journal} {Opt. Lett.}\ }\textbf {\bibinfo {volume} {42}},\ \bibinfo {pages} {1556} (\bibinfo {year} {2017})}\BibitemShut {NoStop}%
\bibitem [{\citenamefont {Kononchuk}\ \emph {et~al.}(2022)\citenamefont {Kononchuk}, \citenamefont {Cai}, \citenamefont {Ellis}, \citenamefont {Thevamaran},\ and\ \citenamefont {Kottos}}]{Kononchuk2022}%
  \BibitemOpen
  \bibfield  {author} {\bibinfo {author} {\bibfnamefont {R.}~\bibnamefont {Kononchuk}}, \bibinfo {author} {\bibfnamefont {J.}~\bibnamefont {Cai}}, \bibinfo {author} {\bibfnamefont {F.}~\bibnamefont {Ellis}}, \bibinfo {author} {\bibfnamefont {R.}~\bibnamefont {Thevamaran}},\ and\ \bibinfo {author} {\bibfnamefont {T.}~\bibnamefont {Kottos}},\ }\bibfield  {title} {\bibinfo {title} {Exceptional-point-based accelerometers with enhanced signal-to-noise ratio},\ }\href {https://doi.org/10.1038/s41586-022-04904-w} {\bibfield  {journal} {\bibinfo  {journal} {Nature}\ }\textbf {\bibinfo {volume} {607}},\ \bibinfo {pages} {697} (\bibinfo {year} {2022})}\BibitemShut {NoStop}%
\bibitem [{\citenamefont {Scheuer}\ and\ \citenamefont {Yariv}(2006)}]{Yariv}%
  \BibitemOpen
  \bibfield  {author} {\bibinfo {author} {\bibfnamefont {J.}~\bibnamefont {Scheuer}}\ and\ \bibinfo {author} {\bibfnamefont {A.}~\bibnamefont {Yariv}},\ }\bibfield  {title} {\bibinfo {title} {Sagnac effect in coupled-resonator slow-light waveguide structures},\ }\href {https://doi.org/10.1103/PhysRevLett.96.053901} {\bibfield  {journal} {\bibinfo  {journal} {Phys. Rev. Lett.}\ }\textbf {\bibinfo {volume} {96}},\ \bibinfo {pages} {053901} (\bibinfo {year} {2006})}\BibitemShut {NoStop}%
\bibitem [{\citenamefont {Yan}\ \emph {et~al.}(2009)\citenamefont {Yan}, \citenamefont {Xiao}, \citenamefont {Guo},\ and\ \citenamefont {Huang}}]{Anping}%
  \BibitemOpen
  \bibfield  {author} {\bibinfo {author} {\bibfnamefont {L.}~\bibnamefont {Yan}}, \bibinfo {author} {\bibfnamefont {Z.}~\bibnamefont {Xiao}}, \bibinfo {author} {\bibfnamefont {X.}~\bibnamefont {Guo}},\ and\ \bibinfo {author} {\bibfnamefont {A.}~\bibnamefont {Huang}},\ }\bibfield  {title} {\bibinfo {title} {{Circle-coupled resonator waveguide with enhanced Sagnac phase-sensitivity for rotation sensing}},\ }\bibfield  {journal} {\bibinfo  {journal} {Applied Physics Letters}\ }\textbf {\bibinfo {volume} {95}},\ \href {https://doi.org/10.1063/1.3243456} {10.1063/1.3243456} (\bibinfo {year} {2009}),\ \bibinfo {note} {141104}\BibitemShut {NoStop}%
\bibitem [{\citenamefont {Smith}\ \emph {et~al.}(2008{\natexlab{a}})\citenamefont {Smith}, \citenamefont {Chang}, \citenamefont {Arissian},\ and\ \citenamefont {Diels}}]{Smith08}%
  \BibitemOpen
  \bibfield  {author} {\bibinfo {author} {\bibfnamefont {D.~D.}\ \bibnamefont {Smith}}, \bibinfo {author} {\bibfnamefont {H.}~\bibnamefont {Chang}}, \bibinfo {author} {\bibfnamefont {L.}~\bibnamefont {Arissian}},\ and\ \bibinfo {author} {\bibfnamefont {J.~C.}\ \bibnamefont {Diels}},\ }\bibfield  {title} {\bibinfo {title} {Dispersion-enhanced laser gyroscope},\ }\href {https://doi.org/10.1103/PhysRevA.78.053824} {\bibfield  {journal} {\bibinfo  {journal} {Phys. Rev. A}\ }\textbf {\bibinfo {volume} {78}},\ \bibinfo {pages} {053824} (\bibinfo {year} {2008}{\natexlab{a}})}\BibitemShut {NoStop}%
\bibitem [{\citenamefont {Shahriar}\ \emph {et~al.}(2007)\citenamefont {Shahriar}, \citenamefont {Pati}, \citenamefont {Tripathi}, \citenamefont {Gopal}, \citenamefont {Messall},\ and\ \citenamefont {Salit}}]{Shahriar}%
  \BibitemOpen
  \bibfield  {author} {\bibinfo {author} {\bibfnamefont {M.~S.}\ \bibnamefont {Shahriar}}, \bibinfo {author} {\bibfnamefont {G.~S.}\ \bibnamefont {Pati}}, \bibinfo {author} {\bibfnamefont {R.}~\bibnamefont {Tripathi}}, \bibinfo {author} {\bibfnamefont {V.}~\bibnamefont {Gopal}}, \bibinfo {author} {\bibfnamefont {M.}~\bibnamefont {Messall}},\ and\ \bibinfo {author} {\bibfnamefont {K.}~\bibnamefont {Salit}},\ }\bibfield  {title} {\bibinfo {title} {Ultrahigh enhancement in absolute and relative rotation sensing using fast and slow light},\ }\href {https://doi.org/10.1103/PhysRevA.75.053807} {\bibfield  {journal} {\bibinfo  {journal} {Phys. Rev. A}\ }\textbf {\bibinfo {volume} {75}},\ \bibinfo {pages} {053807} (\bibinfo {year} {2007})}\BibitemShut {NoStop}%
\bibitem [{\citenamefont {Digonnet}\ and\ \citenamefont {Chamoun}(2016)}]{Digonnet}%
  \BibitemOpen
  \bibfield  {author} {\bibinfo {author} {\bibfnamefont {M.~J.~F.}\ \bibnamefont {Digonnet}}\ and\ \bibinfo {author} {\bibfnamefont {J.~N.}\ \bibnamefont {Chamoun}},\ }\bibfield  {title} {\bibinfo {title} {{Recent developments in laser-driven and hollow-core fiber optic gyroscopes}},\ }in\ \href {https://doi.org/10.1117/12.2229080} {\emph {\bibinfo {booktitle} {Fiber Optic Sensors and Applications XIII}}},\ Vol.\ \bibinfo {volume} {9852},\ \bibinfo {editor} {edited by\ \bibinfo {editor} {\bibfnamefont {E.}~\bibnamefont {Udd}}, \bibinfo {editor} {\bibfnamefont {G.}~\bibnamefont {Pickrell}},\ and\ \bibinfo {editor} {\bibfnamefont {H.~H.}\ \bibnamefont {Du}}},\ \bibinfo {organization} {International Society for Optics and Photonics}\ (\bibinfo  {publisher} {SPIE},\ \bibinfo {year} {2016})\ p.\ \bibinfo {pages} {985204}\BibitemShut {NoStop}%
\bibitem [{\citenamefont {Larsen}\ and\ \citenamefont {Bulatowicz}(2014)}]{Larsen}%
  \BibitemOpen
  \bibfield  {author} {\bibinfo {author} {\bibfnamefont {M.}~\bibnamefont {Larsen}}\ and\ \bibinfo {author} {\bibfnamefont {M.}~\bibnamefont {Bulatowicz}},\ }\bibfield  {title} {\bibinfo {title} {Nuclear magnetic resonance gyroscope: For darpa's micro-technology for positioning, navigation and timing program},\ }in\ \href {https://doi.org/10.1109/ISISS.2014.6782506} {\emph {\bibinfo {booktitle} {2014 International Symposium on Inertial Sensors and Systems (ISISS)}}}\ (\bibinfo {year} {2014})\ pp.\ \bibinfo {pages} {1--5}\BibitemShut {NoStop}%
\bibitem [{\citenamefont {Tóth}\ and\ \citenamefont {Apellaniz}(2014)}]{Toth_2014}%
  \BibitemOpen
  \bibfield  {author} {\bibinfo {author} {\bibfnamefont {G.}~\bibnamefont {Tóth}}\ and\ \bibinfo {author} {\bibfnamefont {I.}~\bibnamefont {Apellaniz}},\ }\bibfield  {title} {\bibinfo {title} {Quantum metrology from a quantum information science perspective},\ }\href {https://doi.org/10.1088/1751-8113/47/42/424006} {\bibfield  {journal} {\bibinfo  {journal} {Journal of Physics A: Mathematical and Theoretical}\ }\textbf {\bibinfo {volume} {47}},\ \bibinfo {pages} {424006} (\bibinfo {year} {2014})}\BibitemShut {NoStop}%
\bibitem [{\citenamefont {Degen}\ \emph {et~al.}(2017)\citenamefont {Degen}, \citenamefont {Reinhard},\ and\ \citenamefont {Cappellaro}}]{Cappellaro}%
  \BibitemOpen
  \bibfield  {author} {\bibinfo {author} {\bibfnamefont {C.~L.}\ \bibnamefont {Degen}}, \bibinfo {author} {\bibfnamefont {F.}~\bibnamefont {Reinhard}},\ and\ \bibinfo {author} {\bibfnamefont {P.}~\bibnamefont {Cappellaro}},\ }\bibfield  {title} {\bibinfo {title} {Quantum sensing},\ }\href {https://doi.org/10.1103/RevModPhys.89.035002} {\bibfield  {journal} {\bibinfo  {journal} {Rev. Mod. Phys.}\ }\textbf {\bibinfo {volume} {89}},\ \bibinfo {pages} {035002} (\bibinfo {year} {2017})}\BibitemShut {NoStop}%
\bibitem [{\citenamefont {Roques-Carmes}\ \emph {et~al.}(2023)\citenamefont {Roques-Carmes}, \citenamefont {Salamin}, \citenamefont {Sloan}, \citenamefont {Choi}, \citenamefont {Velez}, \citenamefont {Koskas}, \citenamefont {Rivera}, \citenamefont {Kooi}, \citenamefont {Joannopoulos},\ and\ \citenamefont {Soljačić}}]{CRC}%
  \BibitemOpen
  \bibfield  {author} {\bibinfo {author} {\bibfnamefont {C.}~\bibnamefont {Roques-Carmes}}, \bibinfo {author} {\bibfnamefont {Y.}~\bibnamefont {Salamin}}, \bibinfo {author} {\bibfnamefont {J.}~\bibnamefont {Sloan}}, \bibinfo {author} {\bibfnamefont {S.}~\bibnamefont {Choi}}, \bibinfo {author} {\bibfnamefont {G.}~\bibnamefont {Velez}}, \bibinfo {author} {\bibfnamefont {E.}~\bibnamefont {Koskas}}, \bibinfo {author} {\bibfnamefont {N.}~\bibnamefont {Rivera}}, \bibinfo {author} {\bibfnamefont {S.~E.}\ \bibnamefont {Kooi}}, \bibinfo {author} {\bibfnamefont {J.~D.}\ \bibnamefont {Joannopoulos}},\ and\ \bibinfo {author} {\bibfnamefont {M.}~\bibnamefont {Soljačić}},\ }\bibfield  {title} {\bibinfo {title} {Biasing the quantum vacuum to control macroscopic probability distributions},\ }\href {https://doi.org/10.1126/science.adh4920} {\bibfield  {journal} {\bibinfo  {journal} {Science}\ }\textbf {\bibinfo {volume} {381}},\ \bibinfo {pages} {205} (\bibinfo {year} {2023})}\BibitemShut {NoStop}%
\bibitem [{\citenamefont {Caves}(1981)}]{Caves}%
  \BibitemOpen
  \bibfield  {author} {\bibinfo {author} {\bibfnamefont {C.~M.}\ \bibnamefont {Caves}},\ }\bibfield  {title} {\bibinfo {title} {Quantum-mechanical noise in an interferometer},\ }\href {https://doi.org/10.1103/PhysRevD.23.1693} {\bibfield  {journal} {\bibinfo  {journal} {Phys. Rev. D}\ }\textbf {\bibinfo {volume} {23}},\ \bibinfo {pages} {1693} (\bibinfo {year} {1981})}\BibitemShut {NoStop}%
\bibitem [{\citenamefont {Grace}\ \emph {et~al.}(2020)\citenamefont {Grace}, \citenamefont {Gagatsos}, \citenamefont {Zhuang},\ and\ \citenamefont {Guha}}]{Saikat}%
  \BibitemOpen
  \bibfield  {author} {\bibinfo {author} {\bibfnamefont {M.~R.}\ \bibnamefont {Grace}}, \bibinfo {author} {\bibfnamefont {C.~N.}\ \bibnamefont {Gagatsos}}, \bibinfo {author} {\bibfnamefont {Q.}~\bibnamefont {Zhuang}},\ and\ \bibinfo {author} {\bibfnamefont {S.}~\bibnamefont {Guha}},\ }\bibfield  {title} {\bibinfo {title} {Quantum-enhanced fiber-optic gyroscopes using quadrature squeezing and continuous-variable entanglement},\ }\href {https://doi.org/10.1103/PhysRevApplied.14.034065} {\bibfield  {journal} {\bibinfo  {journal} {Phys. Rev. Appl.}\ }\textbf {\bibinfo {volume} {14}},\ \bibinfo {pages} {034065} (\bibinfo {year} {2020})}\BibitemShut {NoStop}%
\bibitem [{\citenamefont {Fink}\ \emph {et~al.}(2019)\citenamefont {Fink}, \citenamefont {Steinlechner}, \citenamefont {Handsteiner}, \citenamefont {Dowling}, \citenamefont {Scheidl},\ and\ \citenamefont {Ursin}}]{Fink_2019}%
  \BibitemOpen
  \bibfield  {author} {\bibinfo {author} {\bibfnamefont {M.}~\bibnamefont {Fink}}, \bibinfo {author} {\bibfnamefont {F.}~\bibnamefont {Steinlechner}}, \bibinfo {author} {\bibfnamefont {J.}~\bibnamefont {Handsteiner}}, \bibinfo {author} {\bibfnamefont {J.~P.}\ \bibnamefont {Dowling}}, \bibinfo {author} {\bibfnamefont {T.}~\bibnamefont {Scheidl}},\ and\ \bibinfo {author} {\bibfnamefont {R.}~\bibnamefont {Ursin}},\ }\bibfield  {title} {\bibinfo {title} {Entanglement-enhanced optical gyroscope},\ }\href {https://doi.org/10.1088/1367-2630/ab1bb2} {\bibfield  {journal} {\bibinfo  {journal} {New Journal of Physics}\ }\textbf {\bibinfo {volume} {21}},\ \bibinfo {pages} {053010} (\bibinfo {year} {2019})}\BibitemShut {NoStop}%
\bibitem [{\citenamefont {et~al.}(2019{\natexlab{a}})}]{PRL2019}%
  \BibitemOpen
  \bibfield  {author} {\bibinfo {author} {\bibfnamefont {A.}~\bibnamefont {et~al.}} (\bibinfo {collaboration} {Virgo Collaboration}),\ }\bibfield  {title} {\bibinfo {title} {Increasing the astrophysical reach of the advanced virgo detector via the application of squeezed vacuum states of light},\ }\href {https://doi.org/10.1103/PhysRevLett.123.231108} {\bibfield  {journal} {\bibinfo  {journal} {Phys. Rev. Lett.}\ }\textbf {\bibinfo {volume} {123}},\ \bibinfo {pages} {231108} (\bibinfo {year} {2019}{\natexlab{a}})}\BibitemShut {NoStop}%
\bibitem [{\citenamefont {et~al.}(2019{\natexlab{b}})}]{PRL2}%
  \BibitemOpen
  \bibfield  {author} {\bibinfo {author} {\bibfnamefont {T.}~\bibnamefont {et~al.}},\ }\bibfield  {title} {\bibinfo {title} {Quantum-enhanced advanced ligo detectors in the era of gravitational-wave astronomy},\ }\href {https://doi.org/10.1103/PhysRevLett.123.231107} {\bibfield  {journal} {\bibinfo  {journal} {Phys. Rev. Lett.}\ }\textbf {\bibinfo {volume} {123}},\ \bibinfo {pages} {231107} (\bibinfo {year} {2019}{\natexlab{b}})}\BibitemShut {NoStop}%
\bibitem [{\citenamefont {Jin}\ \emph {et~al.}(2021)\citenamefont {Jin}, \citenamefont {Yang}, \citenamefont {Chang}, \citenamefont {Shen}, \citenamefont {Wang}, \citenamefont {Leal}, \citenamefont {Wu}, \citenamefont {Gao}, \citenamefont {Feshali}, \citenamefont {Paniccia}, \citenamefont {Vahala},\ and\ \citenamefont {Bowers}}]{Jin2021}%
  \BibitemOpen
  \bibfield  {author} {\bibinfo {author} {\bibfnamefont {W.}~\bibnamefont {Jin}}, \bibinfo {author} {\bibfnamefont {Q.-F.}\ \bibnamefont {Yang}}, \bibinfo {author} {\bibfnamefont {L.}~\bibnamefont {Chang}}, \bibinfo {author} {\bibfnamefont {B.}~\bibnamefont {Shen}}, \bibinfo {author} {\bibfnamefont {H.}~\bibnamefont {Wang}}, \bibinfo {author} {\bibfnamefont {M.~A.}\ \bibnamefont {Leal}}, \bibinfo {author} {\bibfnamefont {L.}~\bibnamefont {Wu}}, \bibinfo {author} {\bibfnamefont {M.}~\bibnamefont {Gao}}, \bibinfo {author} {\bibfnamefont {A.}~\bibnamefont {Feshali}}, \bibinfo {author} {\bibfnamefont {M.}~\bibnamefont {Paniccia}}, \bibinfo {author} {\bibfnamefont {K.~J.}\ \bibnamefont {Vahala}},\ and\ \bibinfo {author} {\bibfnamefont {J.~E.}\ \bibnamefont {Bowers}},\ }\bibfield  {title} {\bibinfo {title} {Hertz-linewidth semiconductor lasers using cmos-ready ultra-high-q microresonators},\ }\href {https://doi.org/10.1038/s41566-021-00761-7} {\bibfield  {journal} {\bibinfo  {journal} {Nature Photonics}\ }\textbf
  {\bibinfo {volume} {15}},\ \bibinfo {pages} {346} (\bibinfo {year} {2021})}\BibitemShut {NoStop}%
\bibitem [{\citenamefont {Krasnokutska}\ \emph {et~al.}(2018)\citenamefont {Krasnokutska}, \citenamefont {Tambasco}, \citenamefont {Li},\ and\ \citenamefont {Peruzzo}}]{Krasnokutska:18}%
  \BibitemOpen
  \bibfield  {author} {\bibinfo {author} {\bibfnamefont {I.}~\bibnamefont {Krasnokutska}}, \bibinfo {author} {\bibfnamefont {J.-L.~J.}\ \bibnamefont {Tambasco}}, \bibinfo {author} {\bibfnamefont {X.}~\bibnamefont {Li}},\ and\ \bibinfo {author} {\bibfnamefont {A.}~\bibnamefont {Peruzzo}},\ }\bibfield  {title} {\bibinfo {title} {Ultra-low loss photonic circuits in lithium niobate on insulator},\ }\href {https://doi.org/10.1364/OE.26.000897} {\bibfield  {journal} {\bibinfo  {journal} {Opt. Express}\ }\textbf {\bibinfo {volume} {26}},\ \bibinfo {pages} {897} (\bibinfo {year} {2018})}\BibitemShut {NoStop}%
\bibitem [{\citenamefont {Liu}\ \emph {et~al.}(2022)\citenamefont {Liu}, \citenamefont {Jin}, \citenamefont {Cheng}, \citenamefont {Chauhan}, \citenamefont {Puckett}, \citenamefont {Nelson}, \citenamefont {Behunin}, \citenamefont {Rakich},\ and\ \citenamefont {Blumenthal}}]{Liu:22}%
  \BibitemOpen
  \bibfield  {author} {\bibinfo {author} {\bibfnamefont {K.}~\bibnamefont {Liu}}, \bibinfo {author} {\bibfnamefont {N.}~\bibnamefont {Jin}}, \bibinfo {author} {\bibfnamefont {H.}~\bibnamefont {Cheng}}, \bibinfo {author} {\bibfnamefont {N.}~\bibnamefont {Chauhan}}, \bibinfo {author} {\bibfnamefont {M.~W.}\ \bibnamefont {Puckett}}, \bibinfo {author} {\bibfnamefont {K.~D.}\ \bibnamefont {Nelson}}, \bibinfo {author} {\bibfnamefont {R.~O.}\ \bibnamefont {Behunin}}, \bibinfo {author} {\bibfnamefont {P.~T.}\ \bibnamefont {Rakich}},\ and\ \bibinfo {author} {\bibfnamefont {D.~J.}\ \bibnamefont {Blumenthal}},\ }\bibfield  {title} {\bibinfo {title} {Ultralow 0.034 db/m loss wafer-scale integrated photonics realizing 720 million q and 380 $\mu$w threshold brillouin lasing},\ }\href {https://doi.org/10.1364/OL.454392} {\bibfield  {journal} {\bibinfo  {journal} {Opt. Lett.}\ }\textbf {\bibinfo {volume} {47}},\ \bibinfo {pages} {1855} (\bibinfo {year} {2022})}\BibitemShut {NoStop}%
\bibitem [{\citenamefont {Li}\ \emph {et~al.}(2018)\citenamefont {Li}, \citenamefont {Lin}, \citenamefont {Huang}, \citenamefont {Shiue}, \citenamefont {Yadav}, \citenamefont {Li}, \citenamefont {Michon}, \citenamefont {Englund}, \citenamefont {Richardson}, \citenamefont {Gu},\ and\ \citenamefont {Hu}}]{Li:18}%
  \BibitemOpen
  \bibfield  {author} {\bibinfo {author} {\bibfnamefont {L.}~\bibnamefont {Li}}, \bibinfo {author} {\bibfnamefont {H.}~\bibnamefont {Lin}}, \bibinfo {author} {\bibfnamefont {Y.}~\bibnamefont {Huang}}, \bibinfo {author} {\bibfnamefont {R.-J.}\ \bibnamefont {Shiue}}, \bibinfo {author} {\bibfnamefont {A.}~\bibnamefont {Yadav}}, \bibinfo {author} {\bibfnamefont {J.}~\bibnamefont {Li}}, \bibinfo {author} {\bibfnamefont {J.}~\bibnamefont {Michon}}, \bibinfo {author} {\bibfnamefont {D.}~\bibnamefont {Englund}}, \bibinfo {author} {\bibfnamefont {K.}~\bibnamefont {Richardson}}, \bibinfo {author} {\bibfnamefont {T.}~\bibnamefont {Gu}},\ and\ \bibinfo {author} {\bibfnamefont {J.}~\bibnamefont {Hu}},\ }\bibfield  {title} {\bibinfo {title} {High-performance flexible waveguide-integrated photodetectors},\ }\href {https://doi.org/10.1364/OPTICA.5.000044} {\bibfield  {journal} {\bibinfo  {journal} {Optica}\ }\textbf {\bibinfo {volume} {5}},\ \bibinfo {pages} {44} (\bibinfo {year} {2018})}\BibitemShut {NoStop}%
\bibitem [{\citenamefont {Zhang}\ \emph {et~al.}(2019{\natexlab{a}})\citenamefont {Zhang}, \citenamefont {Sweeney}, \citenamefont {Hsu}, \citenamefont {Yang}, \citenamefont {Stone},\ and\ \citenamefont {Jiang}}]{Zhang2019}%
  \BibitemOpen
  \bibfield  {author} {\bibinfo {author} {\bibfnamefont {M.}~\bibnamefont {Zhang}}, \bibinfo {author} {\bibfnamefont {W.}~\bibnamefont {Sweeney}}, \bibinfo {author} {\bibfnamefont {C.~W.}\ \bibnamefont {Hsu}}, \bibinfo {author} {\bibfnamefont {L.}~\bibnamefont {Yang}}, \bibinfo {author} {\bibfnamefont {A.~D.}\ \bibnamefont {Stone}},\ and\ \bibinfo {author} {\bibfnamefont {L.}~\bibnamefont {Jiang}},\ }\bibfield  {title} {\bibinfo {title} {Quantum noise theory of exceptional point amplifying sensors},\ }\href {https://doi.org/10.1103/PhysRevLett.123.180501} {\bibfield  {journal} {\bibinfo  {journal} {Phys. Rev. Lett.}\ }\textbf {\bibinfo {volume} {123}},\ \bibinfo {pages} {180501} (\bibinfo {year} {2019}{\natexlab{a}})}\BibitemShut {NoStop}%
\bibitem [{\citenamefont {Anderson}\ \emph {et~al.}(2023)\citenamefont {Anderson}, \citenamefont {Shah},\ and\ \citenamefont {Fan}}]{Anderson2023}%
  \BibitemOpen
  \bibfield  {author} {\bibinfo {author} {\bibfnamefont {D.}~\bibnamefont {Anderson}}, \bibinfo {author} {\bibfnamefont {M.}~\bibnamefont {Shah}},\ and\ \bibinfo {author} {\bibfnamefont {L.}~\bibnamefont {Fan}},\ }\bibfield  {title} {\bibinfo {title} {Clarification of the exceptional-point contribution to photonic sensing},\ }\href {https://doi.org/10.1103/PhysRevApplied.19.034059} {\bibfield  {journal} {\bibinfo  {journal} {Phys. Rev. Appl.}\ }\textbf {\bibinfo {volume} {19}},\ \bibinfo {pages} {034059} (\bibinfo {year} {2023})}\BibitemShut {NoStop}%
\bibitem [{\citenamefont {Shahriari}\ \emph {et~al.}(2016)\citenamefont {Shahriari}, \citenamefont {Swersky}, \citenamefont {Wang}, \citenamefont {Adams},\ and\ \citenamefont {de~Freitas}}]{Shahriari}%
  \BibitemOpen
  \bibfield  {author} {\bibinfo {author} {\bibfnamefont {B.}~\bibnamefont {Shahriari}}, \bibinfo {author} {\bibfnamefont {K.}~\bibnamefont {Swersky}}, \bibinfo {author} {\bibfnamefont {Z.}~\bibnamefont {Wang}}, \bibinfo {author} {\bibfnamefont {R.~P.}\ \bibnamefont {Adams}},\ and\ \bibinfo {author} {\bibfnamefont {N.}~\bibnamefont {de~Freitas}},\ }\bibfield  {title} {\bibinfo {title} {Taking the human out of the loop: A review of bayesian optimization},\ }\href {https://doi.org/10.1109/JPROC.2015.2494218} {\bibfield  {journal} {\bibinfo  {journal} {Proceedings of the IEEE}\ }\textbf {\bibinfo {volume} {104}},\ \bibinfo {pages} {148} (\bibinfo {year} {2016})}\BibitemShut {NoStop}%
\bibitem [{\citenamefont {Zhang}\ \emph {et~al.}(2019{\natexlab{b}})\citenamefont {Zhang}, \citenamefont {Sweeney}, \citenamefont {Hsu}, \citenamefont {Yang}, \citenamefont {Stone},\ and\ \citenamefont {Jiang}}]{zhang2019quantum}%
  \BibitemOpen
  \bibfield  {author} {\bibinfo {author} {\bibfnamefont {M.}~\bibnamefont {Zhang}}, \bibinfo {author} {\bibfnamefont {W.}~\bibnamefont {Sweeney}}, \bibinfo {author} {\bibfnamefont {C.~W.}\ \bibnamefont {Hsu}}, \bibinfo {author} {\bibfnamefont {L.}~\bibnamefont {Yang}}, \bibinfo {author} {\bibfnamefont {A.}~\bibnamefont {Stone}},\ and\ \bibinfo {author} {\bibfnamefont {L.}~\bibnamefont {Jiang}},\ }\bibfield  {title} {\bibinfo {title} {Quantum noise theory of exceptional point amplifying sensors},\ }\href@noop {} {\bibfield  {journal} {\bibinfo  {journal} {Physical review letters}\ }\textbf {\bibinfo {volume} {123}},\ \bibinfo {pages} {180501} (\bibinfo {year} {2019}{\natexlab{b}})}\BibitemShut {NoStop}%
\bibitem [{\citenamefont {Lee}\ \emph {et~al.}(2013)\citenamefont {Lee}, \citenamefont {Suh}, \citenamefont {Chen}, \citenamefont {Li}, \citenamefont {Diddams},\ and\ \citenamefont {Vahala}}]{Lee2013}%
  \BibitemOpen
  \bibfield  {author} {\bibinfo {author} {\bibfnamefont {H.}~\bibnamefont {Lee}}, \bibinfo {author} {\bibfnamefont {M.-G.}\ \bibnamefont {Suh}}, \bibinfo {author} {\bibfnamefont {T.}~\bibnamefont {Chen}}, \bibinfo {author} {\bibfnamefont {J.}~\bibnamefont {Li}}, \bibinfo {author} {\bibfnamefont {S.~A.}\ \bibnamefont {Diddams}},\ and\ \bibinfo {author} {\bibfnamefont {K.~J.}\ \bibnamefont {Vahala}},\ }\bibfield  {title} {\bibinfo {title} {Spiral resonators for on-chip laser frequency stabilization},\ }\href {https://doi.org/10.1038/ncomms3468} {\bibfield  {journal} {\bibinfo  {journal} {Nature Communications}\ }\textbf {\bibinfo {volume} {4}},\ \bibinfo {pages} {2468} (\bibinfo {year} {2013})}\BibitemShut {NoStop}%
\bibitem [{\citenamefont {Ciminelli}\ \emph {et~al.}(2012)\citenamefont {Ciminelli}, \citenamefont {Dell'Olio},\ and\ \citenamefont {Armenise}}]{Ciminelli}%
  \BibitemOpen
  \bibfield  {author} {\bibinfo {author} {\bibfnamefont {C.}~\bibnamefont {Ciminelli}}, \bibinfo {author} {\bibfnamefont {F.}~\bibnamefont {Dell'Olio}},\ and\ \bibinfo {author} {\bibfnamefont {M.~N.}\ \bibnamefont {Armenise}},\ }\bibfield  {title} {\bibinfo {title} {High-q spiral resonator for optical gyroscope applications: Numerical and experimental investigation},\ }\href {https://doi.org/10.1109/JPHOT.2012.2218098} {\bibfield  {journal} {\bibinfo  {journal} {IEEE Photonics Journal}\ }\textbf {\bibinfo {volume} {4}},\ \bibinfo {pages} {1844} (\bibinfo {year} {2012})}\BibitemShut {NoStop}%
\bibitem [{\citenamefont {Kondratiev}\ \emph {et~al.}(2023)\citenamefont {Kondratiev}, \citenamefont {Lobanov}, \citenamefont {Shitikov}, \citenamefont {Galiev}, \citenamefont {Chermoshentsev}, \citenamefont {Dmitriev}, \citenamefont {Danilin}, \citenamefont {Lonshakov}, \citenamefont {Min'kov}, \citenamefont {Sokol}, \citenamefont {Cordette}, \citenamefont {Luo}, \citenamefont {Liang}, \citenamefont {Liu},\ and\ \citenamefont {Bilenko}}]{Kondratiev2023}%
  \BibitemOpen
  \bibfield  {author} {\bibinfo {author} {\bibfnamefont {N.~M.}\ \bibnamefont {Kondratiev}}, \bibinfo {author} {\bibfnamefont {V.~E.}\ \bibnamefont {Lobanov}}, \bibinfo {author} {\bibfnamefont {A.~E.}\ \bibnamefont {Shitikov}}, \bibinfo {author} {\bibfnamefont {R.~R.}\ \bibnamefont {Galiev}}, \bibinfo {author} {\bibfnamefont {D.~A.}\ \bibnamefont {Chermoshentsev}}, \bibinfo {author} {\bibfnamefont {N.~Y.}\ \bibnamefont {Dmitriev}}, \bibinfo {author} {\bibfnamefont {A.~N.}\ \bibnamefont {Danilin}}, \bibinfo {author} {\bibfnamefont {E.~A.}\ \bibnamefont {Lonshakov}}, \bibinfo {author} {\bibfnamefont {K.~N.}\ \bibnamefont {Min'kov}}, \bibinfo {author} {\bibfnamefont {D.~M.}\ \bibnamefont {Sokol}}, \bibinfo {author} {\bibfnamefont {S.~J.}\ \bibnamefont {Cordette}}, \bibinfo {author} {\bibfnamefont {Y.-H.}\ \bibnamefont {Luo}}, \bibinfo {author} {\bibfnamefont {W.}~\bibnamefont {Liang}}, \bibinfo {author} {\bibfnamefont {J.}~\bibnamefont {Liu}},\ and\ \bibinfo {author} {\bibfnamefont {I.~A.}\ \bibnamefont
  {Bilenko}},\ }\bibfield  {title} {\bibinfo {title} {Recent advances in laser self-injection locking to high-q microresonators},\ }\href {https://doi.org/10.1007/s11467-022-1245-3} {\bibfield  {journal} {\bibinfo  {journal} {Frontiers of Physics}\ }\textbf {\bibinfo {volume} {18}},\ \bibinfo {pages} {21305} (\bibinfo {year} {2023})}\BibitemShut {NoStop}%
\bibitem [{\citenamefont {Drummond}(2004)}]{Drummond2004}%
  \BibitemOpen
  \bibfield  {author} {\bibinfo {author} {\bibfnamefont {P.~D.}\ \bibnamefont {Drummond}},\ }\bibinfo {title} {Squeezing with nonlinear optics},\ in\ \href {https://doi.org/10.1007/978-3-662-09645-1_4} {\emph {\bibinfo {booktitle} {Quantum Squeezing}}}\ (\bibinfo  {publisher} {Springer Berlin Heidelberg},\ \bibinfo {address} {Berlin, Heidelberg},\ \bibinfo {year} {2004})\ pp.\ \bibinfo {pages} {99--139}\BibitemShut {NoStop}%
\bibitem [{\citenamefont {Wang}\ \emph {et~al.}(2020)\citenamefont {Wang}, \citenamefont {Lai}, \citenamefont {Yuan}, \citenamefont {Suh},\ and\ \citenamefont {Vahala}}]{Wang2020}%
  \BibitemOpen
  \bibfield  {author} {\bibinfo {author} {\bibfnamefont {H.}~\bibnamefont {Wang}}, \bibinfo {author} {\bibfnamefont {Y.-H.}\ \bibnamefont {Lai}}, \bibinfo {author} {\bibfnamefont {Z.}~\bibnamefont {Yuan}}, \bibinfo {author} {\bibfnamefont {M.-G.}\ \bibnamefont {Suh}},\ and\ \bibinfo {author} {\bibfnamefont {K.}~\bibnamefont {Vahala}},\ }\bibfield  {title} {\bibinfo {title} {Petermann-factor sensitivity limit near an exceptional point in a brillouin ring laser gyroscope},\ }\href {https://doi.org/10.1038/s41467-020-15341-6} {\bibfield  {journal} {\bibinfo  {journal} {Nature Communications}\ }\textbf {\bibinfo {volume} {11}},\ \bibinfo {pages} {1610} (\bibinfo {year} {2020})}\BibitemShut {NoStop}%
\bibitem [{\citenamefont {Zhao}\ \emph {et~al.}(2020{\natexlab{a}})\citenamefont {Zhao}, \citenamefont {Okawachi}, \citenamefont {Jang}, \citenamefont {Ji}, \citenamefont {Lipson},\ and\ \citenamefont {Gaeta}}]{GaetaandLipson}%
  \BibitemOpen
  \bibfield  {author} {\bibinfo {author} {\bibfnamefont {Y.}~\bibnamefont {Zhao}}, \bibinfo {author} {\bibfnamefont {Y.}~\bibnamefont {Okawachi}}, \bibinfo {author} {\bibfnamefont {J.~K.}\ \bibnamefont {Jang}}, \bibinfo {author} {\bibfnamefont {X.}~\bibnamefont {Ji}}, \bibinfo {author} {\bibfnamefont {M.}~\bibnamefont {Lipson}},\ and\ \bibinfo {author} {\bibfnamefont {A.~L.}\ \bibnamefont {Gaeta}},\ }\bibfield  {title} {\bibinfo {title} {Near-degenerate quadrature-squeezed vacuum generation on a silicon-nitride chip},\ }\href {https://doi.org/10.1103/PhysRevLett.124.193601} {\bibfield  {journal} {\bibinfo  {journal} {Phys. Rev. Lett.}\ }\textbf {\bibinfo {volume} {124}},\ \bibinfo {pages} {193601} (\bibinfo {year} {2020}{\natexlab{a}})}\BibitemShut {NoStop}%
\bibitem [{\citenamefont {Masada}\ \emph {et~al.}(2015)\citenamefont {Masada}, \citenamefont {Miyata}, \citenamefont {Politi}, \citenamefont {Hashimoto}, \citenamefont {O'Brien},\ and\ \citenamefont {Furusawa}}]{Masada2015}%
  \BibitemOpen
  \bibfield  {author} {\bibinfo {author} {\bibfnamefont {G.}~\bibnamefont {Masada}}, \bibinfo {author} {\bibfnamefont {K.}~\bibnamefont {Miyata}}, \bibinfo {author} {\bibfnamefont {A.}~\bibnamefont {Politi}}, \bibinfo {author} {\bibfnamefont {T.}~\bibnamefont {Hashimoto}}, \bibinfo {author} {\bibfnamefont {J.~L.}\ \bibnamefont {O'Brien}},\ and\ \bibinfo {author} {\bibfnamefont {A.}~\bibnamefont {Furusawa}},\ }\bibfield  {title} {\bibinfo {title} {Continuous-variable entanglement on a chip},\ }\href {https://doi.org/10.1038/nphoton.2015.42} {\bibfield  {journal} {\bibinfo  {journal} {Nature Photonics}\ }\textbf {\bibinfo {volume} {9}},\ \bibinfo {pages} {316} (\bibinfo {year} {2015})}\BibitemShut {NoStop}%
\bibitem [{\citenamefont {Couteau}(2018)}]{Couteau}%
  \BibitemOpen
  \bibfield  {author} {\bibinfo {author} {\bibfnamefont {C.}~\bibnamefont {Couteau}},\ }\bibfield  {title} {\bibinfo {title} {Spontaneous parametric down-conversion},\ }\href {https://doi.org/10.1080/00107514.2018.1488463} {\bibfield  {journal} {\bibinfo  {journal} {Contemporary Physics}\ }\textbf {\bibinfo {volume} {59}},\ \bibinfo {pages} {291} (\bibinfo {year} {2018})}\BibitemShut {NoStop}%
\bibitem [{\citenamefont {Kleinman}(1962)}]{Kleinman}%
  \BibitemOpen
  \bibfield  {author} {\bibinfo {author} {\bibfnamefont {D.~A.}\ \bibnamefont {Kleinman}},\ }\bibfield  {title} {\bibinfo {title} {Theory of second harmonic generation of light},\ }\href {https://doi.org/10.1103/PhysRev.128.1761} {\bibfield  {journal} {\bibinfo  {journal} {Phys. Rev.}\ }\textbf {\bibinfo {volume} {128}},\ \bibinfo {pages} {1761} (\bibinfo {year} {1962})}\BibitemShut {NoStop}%
\bibitem [{\citenamefont {Breunig}(2016)}]{Breunig}%
  \BibitemOpen
  \bibfield  {author} {\bibinfo {author} {\bibfnamefont {I.}~\bibnamefont {Breunig}},\ }\bibfield  {title} {\bibinfo {title} {Three-wave mixing in whispering gallery resonators},\ }\href {https://doi.org/https://doi.org/10.1002/lpor.201600038} {\bibfield  {journal} {\bibinfo  {journal} {Laser and Photonics Reviews}\ }\textbf {\bibinfo {volume} {10}},\ \bibinfo {pages} {569} (\bibinfo {year} {2016})}\BibitemShut {NoStop}%
\bibitem [{\citenamefont {Sarma}\ \emph {et~al.}(2015)\citenamefont {Sarma}, \citenamefont {Ge},\ and\ \citenamefont {Cao}}]{Sarma:15}%
  \BibitemOpen
  \bibfield  {author} {\bibinfo {author} {\bibfnamefont {R.}~\bibnamefont {Sarma}}, \bibinfo {author} {\bibfnamefont {L.}~\bibnamefont {Ge}},\ and\ \bibinfo {author} {\bibfnamefont {H.}~\bibnamefont {Cao}},\ }\bibfield  {title} {\bibinfo {title} {Optical resonances in rotating dielectric microcavities of deformed shape},\ }\href {https://doi.org/10.1364/JOSAB.32.001736} {\bibfield  {journal} {\bibinfo  {journal} {J. Opt. Soc. Am. B}\ }\textbf {\bibinfo {volume} {32}},\ \bibinfo {pages} {1736} (\bibinfo {year} {2015})}\BibitemShut {NoStop}%
\bibitem [{\citenamefont {Lu}\ \emph {et~al.}(2019)\citenamefont {Lu}, \citenamefont {Surya}, \citenamefont {Liu}, \citenamefont {Bruch}, \citenamefont {Gong}, \citenamefont {Xu},\ and\ \citenamefont {Tang}}]{Lu:19}%
  \BibitemOpen
  \bibfield  {author} {\bibinfo {author} {\bibfnamefont {J.}~\bibnamefont {Lu}}, \bibinfo {author} {\bibfnamefont {J.~B.}\ \bibnamefont {Surya}}, \bibinfo {author} {\bibfnamefont {X.}~\bibnamefont {Liu}}, \bibinfo {author} {\bibfnamefont {A.~W.}\ \bibnamefont {Bruch}}, \bibinfo {author} {\bibfnamefont {Z.}~\bibnamefont {Gong}}, \bibinfo {author} {\bibfnamefont {Y.}~\bibnamefont {Xu}},\ and\ \bibinfo {author} {\bibfnamefont {H.~X.}\ \bibnamefont {Tang}},\ }\bibfield  {title} {\bibinfo {title} {Periodically poled thin-film lithium niobate microring resonators with a second-harmonic generation efficiency of 250,000\%/w},\ }\href {https://doi.org/10.1364/OPTICA.6.001455} {\bibfield  {journal} {\bibinfo  {journal} {Optica}\ }\textbf {\bibinfo {volume} {6}},\ \bibinfo {pages} {1455} (\bibinfo {year} {2019})}\BibitemShut {NoStop}%
\bibitem [{\citenamefont {Guo}\ \emph {et~al.}(2016)\citenamefont {Guo}, \citenamefont {Zou},\ and\ \citenamefont {Tang}}]{Guo:16}%
  \BibitemOpen
  \bibfield  {author} {\bibinfo {author} {\bibfnamefont {X.}~\bibnamefont {Guo}}, \bibinfo {author} {\bibfnamefont {C.-L.}\ \bibnamefont {Zou}},\ and\ \bibinfo {author} {\bibfnamefont {H.~X.}\ \bibnamefont {Tang}},\ }\bibfield  {title} {\bibinfo {title} {Second-harmonic generation in aluminum nitride microrings with 2500\%/w conversion efficiency},\ }\href {https://doi.org/10.1364/OPTICA.3.001126} {\bibfield  {journal} {\bibinfo  {journal} {Optica}\ }\textbf {\bibinfo {volume} {3}},\ \bibinfo {pages} {1126} (\bibinfo {year} {2016})}\BibitemShut {NoStop}%
\bibitem [{\citenamefont {Drummond}\ \emph {et~al.}(1980)\citenamefont {Drummond}, \citenamefont {McNeil},\ and\ \citenamefont {Walls}}]{Drummond}%
  \BibitemOpen
  \bibfield  {author} {\bibinfo {author} {\bibfnamefont {P.}~\bibnamefont {Drummond}}, \bibinfo {author} {\bibfnamefont {K.}~\bibnamefont {McNeil}},\ and\ \bibinfo {author} {\bibfnamefont {D.}~\bibnamefont {Walls}},\ }\bibfield  {title} {\bibinfo {title} {Non-equilibrium transitions in sub/second harmonic generation},\ }\href {https://doi.org/10.1080/713820226} {\bibfield  {journal} {\bibinfo  {journal} {Optica Acta: International Journal of Optics}\ }\textbf {\bibinfo {volume} {27}},\ \bibinfo {pages} {321} (\bibinfo {year} {1980})}\BibitemShut {NoStop}%
\bibitem [{\citenamefont {Johansson}\ \emph {et~al.}(2012)\citenamefont {Johansson}, \citenamefont {Nation},\ and\ \citenamefont {Nori}}]{johansson2012qutip}%
  \BibitemOpen
  \bibfield  {author} {\bibinfo {author} {\bibfnamefont {J.~R.}\ \bibnamefont {Johansson}}, \bibinfo {author} {\bibfnamefont {P.~D.}\ \bibnamefont {Nation}},\ and\ \bibinfo {author} {\bibfnamefont {F.}~\bibnamefont {Nori}},\ }\bibfield  {title} {\bibinfo {title} {Qutip: An open-source python framework for the dynamics of open quantum systems},\ }\href@noop {} {\bibfield  {journal} {\bibinfo  {journal} {Computer Physics Communications}\ }\textbf {\bibinfo {volume} {183}},\ \bibinfo {pages} {1760} (\bibinfo {year} {2012})}\BibitemShut {NoStop}%
\bibitem [{\citenamefont {Tang}\ \emph {et~al.}(1993)\citenamefont {Tang}, \citenamefont {Lee},\ and\ \citenamefont {Ridley}}]{Tang}%
  \BibitemOpen
  \bibfield  {author} {\bibinfo {author} {\bibfnamefont {W.}~\bibnamefont {Tang}}, \bibinfo {author} {\bibfnamefont {F.}~\bibnamefont {Lee}},\ and\ \bibinfo {author} {\bibfnamefont {R.}~\bibnamefont {Ridley}},\ }\bibfield  {title} {\bibinfo {title} {Small-signal modeling of average current-mode control},\ }\href {https://doi.org/10.1109/63.223961} {\bibfield  {journal} {\bibinfo  {journal} {IEEE Transactions on Power Electronics}\ }\textbf {\bibinfo {volume} {8}},\ \bibinfo {pages} {112} (\bibinfo {year} {1993})}\BibitemShut {NoStop}%
\bibitem [{\citenamefont {Gillner}\ \emph {et~al.}(1990)\citenamefont {Gillner}, \citenamefont {Bj\"ork},\ and\ \citenamefont {Yamamoto}}]{Gillner}%
  \BibitemOpen
  \bibfield  {author} {\bibinfo {author} {\bibfnamefont {L.}~\bibnamefont {Gillner}}, \bibinfo {author} {\bibfnamefont {G.}~\bibnamefont {Bj\"ork}},\ and\ \bibinfo {author} {\bibfnamefont {Y.}~\bibnamefont {Yamamoto}},\ }\bibfield  {title} {\bibinfo {title} {Quantum noise properties of an injection-locked laser oscillator with pump-noise suppression and squeezed injection},\ }\href {https://doi.org/10.1103/PhysRevA.41.5053} {\bibfield  {journal} {\bibinfo  {journal} {Phys. Rev. A}\ }\textbf {\bibinfo {volume} {41}},\ \bibinfo {pages} {5053} (\bibinfo {year} {1990})}\BibitemShut {NoStop}%
\bibitem [{\citenamefont {Machida}\ \emph {et~al.}(1987)\citenamefont {Machida}, \citenamefont {Yamamoto},\ and\ \citenamefont {Itaya}}]{Yamamoto1}%
  \BibitemOpen
  \bibfield  {author} {\bibinfo {author} {\bibfnamefont {S.}~\bibnamefont {Machida}}, \bibinfo {author} {\bibfnamefont {Y.}~\bibnamefont {Yamamoto}},\ and\ \bibinfo {author} {\bibfnamefont {Y.}~\bibnamefont {Itaya}},\ }\bibfield  {title} {\bibinfo {title} {Observation of amplitude squeezing in a constant-current--driven semiconductor laser},\ }\href {https://doi.org/10.1103/PhysRevLett.58.1000} {\bibfield  {journal} {\bibinfo  {journal} {Phys. Rev. Lett.}\ }\textbf {\bibinfo {volume} {58}},\ \bibinfo {pages} {1000} (\bibinfo {year} {1987})}\BibitemShut {NoStop}%
\bibitem [{\citenamefont {Chembo}(2016)}]{chembo2016quantum}%
  \BibitemOpen
  \bibfield  {author} {\bibinfo {author} {\bibfnamefont {Y.~K.}\ \bibnamefont {Chembo}},\ }\bibfield  {title} {\bibinfo {title} {Quantum dynamics of kerr optical frequency combs below and above threshold: Spontaneous four-wave mixing, entanglement, and squeezed states of light},\ }\href@noop {} {\bibfield  {journal} {\bibinfo  {journal} {Physical Review A}\ }\textbf {\bibinfo {volume} {93}},\ \bibinfo {pages} {033820} (\bibinfo {year} {2016})}\BibitemShut {NoStop}%
\bibitem [{\citenamefont {Pontula}\ \emph {et~al.}(2022)\citenamefont {Pontula}, \citenamefont {Sloan}, \citenamefont {Rivera},\ and\ \citenamefont {Soljacic}}]{pontula2022strong}%
  \BibitemOpen
  \bibfield  {author} {\bibinfo {author} {\bibfnamefont {S.}~\bibnamefont {Pontula}}, \bibinfo {author} {\bibfnamefont {J.}~\bibnamefont {Sloan}}, \bibinfo {author} {\bibfnamefont {N.}~\bibnamefont {Rivera}},\ and\ \bibinfo {author} {\bibfnamefont {M.}~\bibnamefont {Soljacic}},\ }\bibfield  {title} {\bibinfo {title} {Strong intensity noise condensation using nonlinear dispersive loss in semiconductor lasers},\ }\href@noop {} {\bibfield  {journal} {\bibinfo  {journal} {arXiv preprint arXiv:2212.07300}\ } (\bibinfo {year} {2022})}\BibitemShut {NoStop}%
\bibitem [{\citenamefont {Costa}(2021)}]{Costa2021}%
  \BibitemOpen
  \bibfield  {author} {\bibinfo {author} {\bibfnamefont {P.~J.}\ \bibnamefont {Costa}},\ }\bibinfo {title} {The hartman--grobman theorem},\ in\ \href {https://doi.org/10.1007/978-3-031-02434-4_9} {\emph {\bibinfo {booktitle} {Select Ideas in Partial Differential Equations}}}\ (\bibinfo  {publisher} {Springer International Publishing},\ \bibinfo {address} {Cham},\ \bibinfo {year} {2021})\ pp.\ \bibinfo {pages} {173--198}\BibitemShut {NoStop}%
\bibitem [{\citenamefont {Drummond}\ and\ \citenamefont {Gardiner}(1980)}]{Drummond_1980}%
  \BibitemOpen
  \bibfield  {author} {\bibinfo {author} {\bibfnamefont {P.~D.}\ \bibnamefont {Drummond}}\ and\ \bibinfo {author} {\bibfnamefont {C.~W.}\ \bibnamefont {Gardiner}},\ }\bibfield  {title} {\bibinfo {title} {Generalised p-representations in quantum optics},\ }\href {https://doi.org/10.1088/0305-4470/13/7/018} {\bibfield  {journal} {\bibinfo  {journal} {Journal of Physics A: Mathematical and General}\ }\textbf {\bibinfo {volume} {13}},\ \bibinfo {pages} {2353} (\bibinfo {year} {1980})}\BibitemShut {NoStop}%
\bibitem [{\citenamefont {Braunstein}\ and\ \citenamefont {Caves}(1994)}]{Braunstein}%
  \BibitemOpen
  \bibfield  {author} {\bibinfo {author} {\bibfnamefont {S.~L.}\ \bibnamefont {Braunstein}}\ and\ \bibinfo {author} {\bibfnamefont {C.~M.}\ \bibnamefont {Caves}},\ }\bibfield  {title} {\bibinfo {title} {Statistical distance and the geometry of quantum states},\ }\href {https://doi.org/10.1103/PhysRevLett.72.3439} {\bibfield  {journal} {\bibinfo  {journal} {Phys. Rev. Lett.}\ }\textbf {\bibinfo {volume} {72}},\ \bibinfo {pages} {3439} (\bibinfo {year} {1994})}\BibitemShut {NoStop}%
\bibitem [{\citenamefont {Kapral}(2015)}]{Kapral_2015}%
  \BibitemOpen
  \bibfield  {author} {\bibinfo {author} {\bibfnamefont {R.}~\bibnamefont {Kapral}},\ }\bibfield  {title} {\bibinfo {title} {Quantum dynamics in open quantum-classical systems},\ }\href {https://doi.org/10.1088/0953-8984/27/7/073201} {\bibfield  {journal} {\bibinfo  {journal} {Journal of Physics: Condensed Matter}\ }\textbf {\bibinfo {volume} {27}},\ \bibinfo {pages} {073201} (\bibinfo {year} {2015})}\BibitemShut {NoStop}%
\bibitem [{\citenamefont {Drummond}\ \emph {et~al.}(1981)\citenamefont {Drummond}, \citenamefont {McNeil},\ and\ \citenamefont {Walls}}]{Drummond1}%
  \BibitemOpen
  \bibfield  {author} {\bibinfo {author} {\bibfnamefont {P.}~\bibnamefont {Drummond}}, \bibinfo {author} {\bibfnamefont {K.}~\bibnamefont {McNeil}},\ and\ \bibinfo {author} {\bibfnamefont {D.}~\bibnamefont {Walls}},\ }\bibfield  {title} {\bibinfo {title} {Non-equilibrium transitions in sub/second harmonic generation},\ }\href {https://doi.org/10.1080/713820531} {\bibfield  {journal} {\bibinfo  {journal} {Optica Acta: International Journal of Optics}\ }\textbf {\bibinfo {volume} {28}},\ \bibinfo {pages} {211} (\bibinfo {year} {1981})}\BibitemShut {NoStop}%
\bibitem [{\citenamefont {Giglio}\ \emph {et~al.}(2016)\citenamefont {Giglio}, \citenamefont {Patimisco}, \citenamefont {Sampaolo}, \citenamefont {Scamarcio}, \citenamefont {Tittel},\ and\ \citenamefont {Spagnolo}}]{Giglio}%
  \BibitemOpen
  \bibfield  {author} {\bibinfo {author} {\bibfnamefont {M.}~\bibnamefont {Giglio}}, \bibinfo {author} {\bibfnamefont {P.}~\bibnamefont {Patimisco}}, \bibinfo {author} {\bibfnamefont {A.}~\bibnamefont {Sampaolo}}, \bibinfo {author} {\bibfnamefont {G.}~\bibnamefont {Scamarcio}}, \bibinfo {author} {\bibfnamefont {F.~K.}\ \bibnamefont {Tittel}},\ and\ \bibinfo {author} {\bibfnamefont {V.}~\bibnamefont {Spagnolo}},\ }\bibfield  {title} {\bibinfo {title} {Allan deviation plot as a tool for quartz-enhanced photoacoustic sensors noise analysis},\ }\href {https://doi.org/10.1109/TUFFC.2015.2495013} {\bibfield  {journal} {\bibinfo  {journal} {IEEE Transactions on Ultrasonics, Ferroelectrics, and Frequency Control}\ }\textbf {\bibinfo {volume} {63}},\ \bibinfo {pages} {555} (\bibinfo {year} {2016})}\BibitemShut {NoStop}%
\bibitem [{\citenamefont {Shapiro}(2009)}]{Shapiro}%
  \BibitemOpen
  \bibfield  {author} {\bibinfo {author} {\bibfnamefont {J.~H.}\ \bibnamefont {Shapiro}},\ }\bibfield  {title} {\bibinfo {title} {The quantum theory of optical communications},\ }\href {https://doi.org/10.1109/JSTQE.2009.2024959} {\bibfield  {journal} {\bibinfo  {journal} {IEEE Journal of Selected Topics in Quantum Electronics}\ }\textbf {\bibinfo {volume} {15}},\ \bibinfo {pages} {1547} (\bibinfo {year} {2009})}\BibitemShut {NoStop}%
\bibitem [{\citenamefont {Carmichael}(1999)}]{Carmichael1999}%
  \BibitemOpen
  \bibfield  {author} {\bibinfo {author} {\bibfnamefont {H.~J.}\ \bibnamefont {Carmichael}},\ }\bibinfo {title} {Dissipation in quantum mechanics: The master equation approach},\ in\ \href {https://doi.org/10.1007/978-3-662-03875-8_1} {\emph {\bibinfo {booktitle} {Statistical Methods in Quantum Optics 1: Master Equations and Fokker-Planck Equations}}}\ (\bibinfo  {publisher} {Springer Berlin Heidelberg},\ \bibinfo {address} {Berlin, Heidelberg},\ \bibinfo {year} {1999})\ pp.\ \bibinfo {pages} {1--28}\BibitemShut {NoStop}%
\bibitem [{\citenamefont {Zhu}\ \emph {et~al.}(2021)\citenamefont {Zhu}, \citenamefont {Shao}, \citenamefont {Yu}, \citenamefont {Cheng}, \citenamefont {Desiatov}, \citenamefont {Xin}, \citenamefont {Hu}, \citenamefont {Holzgrafe}, \citenamefont {Ghosh}, \citenamefont {Shams-Ansari}, \citenamefont {Puma}, \citenamefont {Sinclair}, \citenamefont {Reimer}, \citenamefont {Zhang},\ and\ \citenamefont {Lon\v{c}ar}}]{Zhu:21}%
  \BibitemOpen
  \bibfield  {author} {\bibinfo {author} {\bibfnamefont {D.}~\bibnamefont {Zhu}}, \bibinfo {author} {\bibfnamefont {L.}~\bibnamefont {Shao}}, \bibinfo {author} {\bibfnamefont {M.}~\bibnamefont {Yu}}, \bibinfo {author} {\bibfnamefont {R.}~\bibnamefont {Cheng}}, \bibinfo {author} {\bibfnamefont {B.}~\bibnamefont {Desiatov}}, \bibinfo {author} {\bibfnamefont {C.~J.}\ \bibnamefont {Xin}}, \bibinfo {author} {\bibfnamefont {Y.}~\bibnamefont {Hu}}, \bibinfo {author} {\bibfnamefont {J.}~\bibnamefont {Holzgrafe}}, \bibinfo {author} {\bibfnamefont {S.}~\bibnamefont {Ghosh}}, \bibinfo {author} {\bibfnamefont {A.}~\bibnamefont {Shams-Ansari}}, \bibinfo {author} {\bibfnamefont {E.}~\bibnamefont {Puma}}, \bibinfo {author} {\bibfnamefont {N.}~\bibnamefont {Sinclair}}, \bibinfo {author} {\bibfnamefont {C.}~\bibnamefont {Reimer}}, \bibinfo {author} {\bibfnamefont {M.}~\bibnamefont {Zhang}},\ and\ \bibinfo {author} {\bibfnamefont {M.}~\bibnamefont {Lon\v{c}ar}},\ }\bibfield  {title} {\bibinfo {title} {Integrated photonics on
  thin-film lithium niobate},\ }\href {https://doi.org/10.1364/AOP.411024} {\bibfield  {journal} {\bibinfo  {journal} {Adv. Opt. Photon.}\ }\textbf {\bibinfo {volume} {13}},\ \bibinfo {pages} {242} (\bibinfo {year} {2021})}\BibitemShut {NoStop}%
\bibitem [{\citenamefont {Cui}\ \emph {et~al.}(2022)\citenamefont {Cui}, \citenamefont {Zhang},\ and\ \citenamefont {Fan}}]{Cui2022}%
  \BibitemOpen
  \bibfield  {author} {\bibinfo {author} {\bibfnamefont {C.}~\bibnamefont {Cui}}, \bibinfo {author} {\bibfnamefont {L.}~\bibnamefont {Zhang}},\ and\ \bibinfo {author} {\bibfnamefont {L.}~\bibnamefont {Fan}},\ }\bibfield  {title} {\bibinfo {title} {In situ control of effective kerr nonlinearity with pockels integrated photonics},\ }\href {https://doi.org/10.1038/s41567-022-01542-x} {\bibfield  {journal} {\bibinfo  {journal} {Nature Physics}\ }\textbf {\bibinfo {volume} {18}},\ \bibinfo {pages} {497} (\bibinfo {year} {2022})}\BibitemShut {NoStop}%
\bibitem [{\citenamefont {Silver}\ \emph {et~al.}(2021)\citenamefont {Silver}, \citenamefont {Del~Bino}, \citenamefont {Woodley}, \citenamefont {Ghalanos}, \citenamefont {Svela}, \citenamefont {Moroney}, \citenamefont {Zhang}, \citenamefont {Grattan},\ and\ \citenamefont {Del’Haye}}]{silver2021nonlinear}%
  \BibitemOpen
  \bibfield  {author} {\bibinfo {author} {\bibfnamefont {J.~M.}\ \bibnamefont {Silver}}, \bibinfo {author} {\bibfnamefont {L.}~\bibnamefont {Del~Bino}}, \bibinfo {author} {\bibfnamefont {M.~T.}\ \bibnamefont {Woodley}}, \bibinfo {author} {\bibfnamefont {G.~N.}\ \bibnamefont {Ghalanos}}, \bibinfo {author} {\bibfnamefont {A.~{\O}.}\ \bibnamefont {Svela}}, \bibinfo {author} {\bibfnamefont {N.}~\bibnamefont {Moroney}}, \bibinfo {author} {\bibfnamefont {S.}~\bibnamefont {Zhang}}, \bibinfo {author} {\bibfnamefont {K.~T.}\ \bibnamefont {Grattan}},\ and\ \bibinfo {author} {\bibfnamefont {P.}~\bibnamefont {Del’Haye}},\ }\bibfield  {title} {\bibinfo {title} {Nonlinear enhanced microresonator gyroscope},\ }\href@noop {} {\bibfield  {journal} {\bibinfo  {journal} {Optica}\ }\textbf {\bibinfo {volume} {8}},\ \bibinfo {pages} {1219} (\bibinfo {year} {2021})}\BibitemShut {NoStop}%
\bibitem [{\citenamefont {Peters}\ and\ \citenamefont {Rodriguez}(2022)}]{peters2022exceptional}%
  \BibitemOpen
  \bibfield  {author} {\bibinfo {author} {\bibfnamefont {K.~J.}\ \bibnamefont {Peters}}\ and\ \bibinfo {author} {\bibfnamefont {S.~R.}\ \bibnamefont {Rodriguez}},\ }\bibfield  {title} {\bibinfo {title} {Exceptional precision of a nonlinear optical sensor at a square-root singularity},\ }\href@noop {} {\bibfield  {journal} {\bibinfo  {journal} {Physical Review Letters}\ }\textbf {\bibinfo {volume} {129}},\ \bibinfo {pages} {013901} (\bibinfo {year} {2022})}\BibitemShut {NoStop}%
\bibitem [{\citenamefont {Pastrňák}\ and\ \citenamefont {Roskovcová}(1966)}]{ALN}%
  \BibitemOpen
  \bibfield  {author} {\bibinfo {author} {\bibfnamefont {J.}~\bibnamefont {Pastrňák}}\ and\ \bibinfo {author} {\bibfnamefont {L.}~\bibnamefont {Roskovcová}},\ }\bibfield  {title} {\bibinfo {title} {Refraction index measurements on aln single crystals},\ }\href {https://doi.org/https://doi.org/10.1002/pssb.19660140127} {\bibfield  {journal} {\bibinfo  {journal} {physica status solidi (b)}\ }\textbf {\bibinfo {volume} {14}},\ \bibinfo {pages} {K5} (\bibinfo {year} {1966})}\BibitemShut {NoStop}%
\bibitem [{\citenamefont {Singh}\ \emph {et~al.}(2003)\citenamefont {Singh}, \citenamefont {Potopowicz}, \citenamefont {Van~Uitert},\ and\ \citenamefont {Wemple}}]{Sic}%
  \BibitemOpen
  \bibfield  {author} {\bibinfo {author} {\bibfnamefont {S.}~\bibnamefont {Singh}}, \bibinfo {author} {\bibfnamefont {J.~R.}\ \bibnamefont {Potopowicz}}, \bibinfo {author} {\bibfnamefont {L.~G.}\ \bibnamefont {Van~Uitert}},\ and\ \bibinfo {author} {\bibfnamefont {S.~H.}\ \bibnamefont {Wemple}},\ }\bibfield  {title} {\bibinfo {title} {{Nonlinear Optical Properties of Hexagonal Silicon Carbide}},\ }\href {https://doi.org/10.1063/1.1653819} {\bibfield  {journal} {\bibinfo  {journal} {Applied Physics Letters}\ }\textbf {\bibinfo {volume} {19}},\ \bibinfo {pages} {53} (\bibinfo {year} {2003})}\BibitemShut {NoStop}%
\bibitem [{\citenamefont {Aspnes}\ \emph {et~al.}(1986)\citenamefont {Aspnes}, \citenamefont {Kelso}, \citenamefont {Logan},\ and\ \citenamefont {Bhat}}]{GaAs}%
  \BibitemOpen
  \bibfield  {author} {\bibinfo {author} {\bibfnamefont {D.~E.}\ \bibnamefont {Aspnes}}, \bibinfo {author} {\bibfnamefont {S.~M.}\ \bibnamefont {Kelso}}, \bibinfo {author} {\bibfnamefont {R.~A.}\ \bibnamefont {Logan}},\ and\ \bibinfo {author} {\bibfnamefont {R.}~\bibnamefont {Bhat}},\ }\bibfield  {title} {\bibinfo {title} {Optical properties of alxga1x as},\ }\href {https://doi.org/10.1063/1.337426} {\bibfield  {journal} {\bibinfo  {journal} {Journal of Applied Physics}\ }\textbf {\bibinfo {volume} {60}},\ \bibinfo {pages} {754} (\bibinfo {year} {1986})}\BibitemShut {NoStop}%
\bibitem [{\citenamefont {Nehra}\ \emph {et~al.}(2022)\citenamefont {Nehra}, \citenamefont {Sekine}, \citenamefont {Ledezma}, \citenamefont {Guo}, \citenamefont {Gray}, \citenamefont {Roy},\ and\ \citenamefont {Marandi}}]{Alireza}%
  \BibitemOpen
  \bibfield  {author} {\bibinfo {author} {\bibfnamefont {R.}~\bibnamefont {Nehra}}, \bibinfo {author} {\bibfnamefont {R.}~\bibnamefont {Sekine}}, \bibinfo {author} {\bibfnamefont {L.}~\bibnamefont {Ledezma}}, \bibinfo {author} {\bibfnamefont {Q.}~\bibnamefont {Guo}}, \bibinfo {author} {\bibfnamefont {R.~M.}\ \bibnamefont {Gray}}, \bibinfo {author} {\bibfnamefont {A.}~\bibnamefont {Roy}},\ and\ \bibinfo {author} {\bibfnamefont {A.}~\bibnamefont {Marandi}},\ }\bibfield  {title} {\bibinfo {title} {Few-cycle vacuum squeezing in nanophotonics},\ }\href {https://doi.org/10.1126/science.abo6213} {\bibfield  {journal} {\bibinfo  {journal} {Science}\ }\textbf {\bibinfo {volume} {377}},\ \bibinfo {pages} {1333} (\bibinfo {year} {2022})}\BibitemShut {NoStop}%
\bibitem [{\citenamefont {McCutcheon}\ \emph {et~al.}(2009)\citenamefont {McCutcheon}, \citenamefont {Chang}, \citenamefont {Zhang}, \citenamefont {Lukin},\ and\ \citenamefont {Lon\v{c}ar}}]{McCutcheon:09}%
  \BibitemOpen
  \bibfield  {author} {\bibinfo {author} {\bibfnamefont {M.~W.}\ \bibnamefont {McCutcheon}}, \bibinfo {author} {\bibfnamefont {D.~E.}\ \bibnamefont {Chang}}, \bibinfo {author} {\bibfnamefont {Y.}~\bibnamefont {Zhang}}, \bibinfo {author} {\bibfnamefont {M.~D.}\ \bibnamefont {Lukin}},\ and\ \bibinfo {author} {\bibfnamefont {M.}~\bibnamefont {Lon\v{c}ar}},\ }\bibfield  {title} {\bibinfo {title} {Broadband frequency conversion and shaping of single photons emitted from a nonlinear cavity},\ }\href {https://doi.org/10.1364/OE.17.022689} {\bibfield  {journal} {\bibinfo  {journal} {Opt. Express}\ }\textbf {\bibinfo {volume} {17}},\ \bibinfo {pages} {22689} (\bibinfo {year} {2009})}\BibitemShut {NoStop}%
\bibitem [{\citenamefont {Chen}\ \emph {et~al.}(2022)\citenamefont {Chen}, \citenamefont {Briggs}, \citenamefont {Hou},\ and\ \citenamefont {Fan}}]{Chen:22}%
  \BibitemOpen
  \bibfield  {author} {\bibinfo {author} {\bibfnamefont {P.-K.}\ \bibnamefont {Chen}}, \bibinfo {author} {\bibfnamefont {I.}~\bibnamefont {Briggs}}, \bibinfo {author} {\bibfnamefont {S.}~\bibnamefont {Hou}},\ and\ \bibinfo {author} {\bibfnamefont {L.}~\bibnamefont {Fan}},\ }\bibfield  {title} {\bibinfo {title} {Ultra-broadband quadrature squeezing with thin-film lithium niobate nanophotonics},\ }\href {https://doi.org/10.1364/OL.447695} {\bibfield  {journal} {\bibinfo  {journal} {Opt. Lett.}\ }\textbf {\bibinfo {volume} {47}},\ \bibinfo {pages} {1506} (\bibinfo {year} {2022})}\BibitemShut {NoStop}%
\bibitem [{\citenamefont {Zhao}\ \emph {et~al.}(2020{\natexlab{b}})\citenamefont {Zhao}, \citenamefont {Ma}, \citenamefont {R\"using},\ and\ \citenamefont {Mookherjea}}]{Zhao}%
  \BibitemOpen
  \bibfield  {author} {\bibinfo {author} {\bibfnamefont {J.}~\bibnamefont {Zhao}}, \bibinfo {author} {\bibfnamefont {C.}~\bibnamefont {Ma}}, \bibinfo {author} {\bibfnamefont {M.}~\bibnamefont {R\"using}},\ and\ \bibinfo {author} {\bibfnamefont {S.}~\bibnamefont {Mookherjea}},\ }\bibfield  {title} {\bibinfo {title} {High quality entangled photon pair generation in periodically poled thin-film lithium niobate waveguides},\ }\href {https://doi.org/10.1103/PhysRevLett.124.163603} {\bibfield  {journal} {\bibinfo  {journal} {Phys. Rev. Lett.}\ }\textbf {\bibinfo {volume} {124}},\ \bibinfo {pages} {163603} (\bibinfo {year} {2020}{\natexlab{b}})}\BibitemShut {NoStop}%
\bibitem [{\citenamefont {Zhang}\ \emph {et~al.}(2017{\natexlab{b}})\citenamefont {Zhang}, \citenamefont {Wang}, \citenamefont {Cheng}, \citenamefont {Shams-Ansari},\ and\ \citenamefont {Lon\v{c}ar}}]{MianZhang}%
  \BibitemOpen
  \bibfield  {author} {\bibinfo {author} {\bibfnamefont {M.}~\bibnamefont {Zhang}}, \bibinfo {author} {\bibfnamefont {C.}~\bibnamefont {Wang}}, \bibinfo {author} {\bibfnamefont {R.}~\bibnamefont {Cheng}}, \bibinfo {author} {\bibfnamefont {A.}~\bibnamefont {Shams-Ansari}},\ and\ \bibinfo {author} {\bibfnamefont {M.}~\bibnamefont {Lon\v{c}ar}},\ }\bibfield  {title} {\bibinfo {title} {Monolithic ultra-high-q lithium niobate microring resonator},\ }\href {https://doi.org/10.1364/OPTICA.4.001536} {\bibfield  {journal} {\bibinfo  {journal} {Optica}\ }\textbf {\bibinfo {volume} {4}},\ \bibinfo {pages} {1536} (\bibinfo {year} {2017}{\natexlab{b}})}\BibitemShut {NoStop}%
\bibitem [{\citenamefont {Bi}\ \emph {et~al.}(2012)\citenamefont {Bi}, \citenamefont {Rodriguez}, \citenamefont {Hashemi}, \citenamefont {Duchesne}, \citenamefont {Loncar}, \citenamefont {Wang},\ and\ \citenamefont {Johnson}}]{Bi:12}%
  \BibitemOpen
  \bibfield  {author} {\bibinfo {author} {\bibfnamefont {Z.-F.}\ \bibnamefont {Bi}}, \bibinfo {author} {\bibfnamefont {A.~W.}\ \bibnamefont {Rodriguez}}, \bibinfo {author} {\bibfnamefont {H.}~\bibnamefont {Hashemi}}, \bibinfo {author} {\bibfnamefont {D.}~\bibnamefont {Duchesne}}, \bibinfo {author} {\bibfnamefont {M.}~\bibnamefont {Loncar}}, \bibinfo {author} {\bibfnamefont {K.-M.}\ \bibnamefont {Wang}},\ and\ \bibinfo {author} {\bibfnamefont {S.~G.}\ \bibnamefont {Johnson}},\ }\bibfield  {title} {\bibinfo {title} {High-efficiency second-harmonic generation in doubly-resonant $\chi$(2) microring resonators},\ }\href {https://doi.org/10.1364/OE.20.007526} {\bibfield  {journal} {\bibinfo  {journal} {Opt. Express}\ }\textbf {\bibinfo {volume} {20}},\ \bibinfo {pages} {7526} (\bibinfo {year} {2012})}\BibitemShut {NoStop}%
\bibitem [{\citenamefont {Shams-Ansari}\ \emph {et~al.}(2022{\natexlab{a}})\citenamefont {Shams-Ansari}, \citenamefont {Yu}, \citenamefont {Chen}, \citenamefont {Reimer}, \citenamefont {Zhang}, \citenamefont {Picqu{\'e}},\ and\ \citenamefont {Lon{\v{c}}ar}}]{Shams-Ansari2022}%
  \BibitemOpen
  \bibfield  {author} {\bibinfo {author} {\bibfnamefont {A.}~\bibnamefont {Shams-Ansari}}, \bibinfo {author} {\bibfnamefont {M.}~\bibnamefont {Yu}}, \bibinfo {author} {\bibfnamefont {Z.}~\bibnamefont {Chen}}, \bibinfo {author} {\bibfnamefont {C.}~\bibnamefont {Reimer}}, \bibinfo {author} {\bibfnamefont {M.}~\bibnamefont {Zhang}}, \bibinfo {author} {\bibfnamefont {N.}~\bibnamefont {Picqu{\'e}}},\ and\ \bibinfo {author} {\bibfnamefont {M.}~\bibnamefont {Lon{\v{c}}ar}},\ }\bibfield  {title} {\bibinfo {title} {Thin-film lithium-niobate electro-optic platform for spectrally tailored dual-comb spectroscopy},\ }\href {https://doi.org/10.1038/s42005-022-00865-8} {\bibfield  {journal} {\bibinfo  {journal} {Communications Physics}\ }\textbf {\bibinfo {volume} {5}},\ \bibinfo {pages} {88} (\bibinfo {year} {2022}{\natexlab{a}})}\BibitemShut {NoStop}%
\bibitem [{\citenamefont {Nakamura}\ \emph {et~al.}(2002)\citenamefont {Nakamura}, \citenamefont {Kurz}, \citenamefont {Parameswaran},\ and\ \citenamefont {Fejer}}]{Fejer}%
  \BibitemOpen
  \bibfield  {author} {\bibinfo {author} {\bibfnamefont {K.}~\bibnamefont {Nakamura}}, \bibinfo {author} {\bibfnamefont {J.}~\bibnamefont {Kurz}}, \bibinfo {author} {\bibfnamefont {K.}~\bibnamefont {Parameswaran}},\ and\ \bibinfo {author} {\bibfnamefont {M.~M.}\ \bibnamefont {Fejer}},\ }\bibfield  {title} {\bibinfo {title} {Periodic poling of magnesium-oxide-doped lithium niobate},\ }\href {https://doi.org/10.1063/1.1456965} {\bibfield  {journal} {\bibinfo  {journal} {Journal of Applied Physics}\ }\textbf {\bibinfo {volume} {91}},\ \bibinfo {pages} {4528} (\bibinfo {year} {2002})}\BibitemShut {NoStop}%
\bibitem [{\citenamefont {Rantamaki}\ \emph {et~al.}(2015)\citenamefont {Rantamaki}, \citenamefont {Saarinen}, \citenamefont {Lyytik{\"a}inen}, \citenamefont {Kontio}, \citenamefont {Heikkinen}, \citenamefont {Lahtonen}, \citenamefont {Valden},\ and\ \citenamefont {Okhotnikov}}]{Okhotnikov}%
  \BibitemOpen
  \bibfield  {author} {\bibinfo {author} {\bibfnamefont {A.}~\bibnamefont {Rantamaki}}, \bibinfo {author} {\bibfnamefont {E.}~\bibnamefont {Saarinen}}, \bibinfo {author} {\bibfnamefont {J.}~\bibnamefont {Lyytik{\"a}inen}}, \bibinfo {author} {\bibfnamefont {J.}~\bibnamefont {Kontio}}, \bibinfo {author} {\bibfnamefont {J.}~\bibnamefont {Heikkinen}}, \bibinfo {author} {\bibfnamefont {K.}~\bibnamefont {Lahtonen}}, \bibinfo {author} {\bibfnamefont {M.}~\bibnamefont {Valden}},\ and\ \bibinfo {author} {\bibfnamefont {O.}~\bibnamefont {Okhotnikov}},\ }\bibfield  {title} {\bibinfo {title} {Towards high power flip-chip long-wavelength semiconductor disk lasers},\ }in\ \href {https://doi.org/10.1117/12.2076795} {\emph {\bibinfo {booktitle} {Vertical External Cavity Surface Emitting Lasers (VECSELs) V}}},\ Vol.\ \bibinfo {volume} {9349},\ \bibinfo {editor} {edited by\ \bibinfo {editor} {\bibfnamefont {M.}~\bibnamefont {Guina}}},\ \bibinfo {organization} {International Society for Optics and Photonics}\ (\bibinfo
  {publisher} {SPIE},\ \bibinfo {year} {2015})\ p.\ \bibinfo {pages} {934908}\BibitemShut {NoStop}%
\bibitem [{\citenamefont {Guo}\ \emph {et~al.}(2022)\citenamefont {Guo}, \citenamefont {Shao}, \citenamefont {He}, \citenamefont {Luke}, \citenamefont {Morgan}, \citenamefont {Sun}, \citenamefont {Gao}, \citenamefont {Tzu}, \citenamefont {Shen}, \citenamefont {Chen}, \citenamefont {Guo}, \citenamefont {Yu}, \citenamefont {Yu}, \citenamefont {Jafari}, \citenamefont {Lon\v{c}ar}, \citenamefont {Zhang},\ and\ \citenamefont {Beling}}]{Guo:22}%
  \BibitemOpen
  \bibfield  {author} {\bibinfo {author} {\bibfnamefont {X.}~\bibnamefont {Guo}}, \bibinfo {author} {\bibfnamefont {L.}~\bibnamefont {Shao}}, \bibinfo {author} {\bibfnamefont {L.}~\bibnamefont {He}}, \bibinfo {author} {\bibfnamefont {K.}~\bibnamefont {Luke}}, \bibinfo {author} {\bibfnamefont {J.}~\bibnamefont {Morgan}}, \bibinfo {author} {\bibfnamefont {K.}~\bibnamefont {Sun}}, \bibinfo {author} {\bibfnamefont {J.}~\bibnamefont {Gao}}, \bibinfo {author} {\bibfnamefont {T.-C.}\ \bibnamefont {Tzu}}, \bibinfo {author} {\bibfnamefont {Y.}~\bibnamefont {Shen}}, \bibinfo {author} {\bibfnamefont {D.}~\bibnamefont {Chen}}, \bibinfo {author} {\bibfnamefont {B.}~\bibnamefont {Guo}}, \bibinfo {author} {\bibfnamefont {F.}~\bibnamefont {Yu}}, \bibinfo {author} {\bibfnamefont {Q.}~\bibnamefont {Yu}}, \bibinfo {author} {\bibfnamefont {M.}~\bibnamefont {Jafari}}, \bibinfo {author} {\bibfnamefont {M.}~\bibnamefont {Lon\v{c}ar}}, \bibinfo {author} {\bibfnamefont {M.}~\bibnamefont {Zhang}},\ and\ \bibinfo {author}
  {\bibfnamefont {A.}~\bibnamefont {Beling}},\ }\bibfield  {title} {\bibinfo {title} {High-performance modified uni-traveling carrier photodiode integrated on a thin-film lithium niobate platform},\ }\href {https://doi.org/10.1364/PRJ.455969} {\bibfield  {journal} {\bibinfo  {journal} {Photon. Res.}\ }\textbf {\bibinfo {volume} {10}},\ \bibinfo {pages} {1338} (\bibinfo {year} {2022})}\BibitemShut {NoStop}%
\bibitem [{\citenamefont {Guarino}\ \emph {et~al.}(2007)\citenamefont {Guarino}, \citenamefont {Poberaj}, \citenamefont {Rezzonico}, \citenamefont {Degl'Innocenti},\ and\ \citenamefont {G{\"u}nter}}]{Guarino2007}%
  \BibitemOpen
  \bibfield  {author} {\bibinfo {author} {\bibfnamefont {A.}~\bibnamefont {Guarino}}, \bibinfo {author} {\bibfnamefont {G.}~\bibnamefont {Poberaj}}, \bibinfo {author} {\bibfnamefont {D.}~\bibnamefont {Rezzonico}}, \bibinfo {author} {\bibfnamefont {R.}~\bibnamefont {Degl'Innocenti}},\ and\ \bibinfo {author} {\bibfnamefont {P.}~\bibnamefont {G{\"u}nter}},\ }\bibfield  {title} {\bibinfo {title} {Electro--optically tunable microring resonators in lithium niobate},\ }\href {https://doi.org/10.1038/nphoton.2007.93} {\bibfield  {journal} {\bibinfo  {journal} {Nature Photonics}\ }\textbf {\bibinfo {volume} {1}},\ \bibinfo {pages} {407} (\bibinfo {year} {2007})}\BibitemShut {NoStop}%
\bibitem [{\citenamefont {Wang}\ \emph {et~al.}(2022)\citenamefont {Wang}, \citenamefont {Li}, \citenamefont {Yao}, \citenamefont {Li}, \citenamefont {Wu}, \citenamefont {Chiang},\ and\ \citenamefont {Chen}}]{Wang:22}%
  \BibitemOpen
  \bibfield  {author} {\bibinfo {author} {\bibfnamefont {M.}~\bibnamefont {Wang}}, \bibinfo {author} {\bibfnamefont {J.}~\bibnamefont {Li}}, \bibinfo {author} {\bibfnamefont {H.}~\bibnamefont {Yao}}, \bibinfo {author} {\bibfnamefont {X.}~\bibnamefont {Li}}, \bibinfo {author} {\bibfnamefont {J.}~\bibnamefont {Wu}}, \bibinfo {author} {\bibfnamefont {K.~S.}\ \bibnamefont {Chiang}},\ and\ \bibinfo {author} {\bibfnamefont {K.}~\bibnamefont {Chen}},\ }\bibfield  {title} {\bibinfo {title} {Thin-film lithium-niobate modulator with a combined passive bias and thermo-optic bias},\ }\href {https://doi.org/10.1364/OE.474594} {\bibfield  {journal} {\bibinfo  {journal} {Opt. Express}\ }\textbf {\bibinfo {volume} {30}},\ \bibinfo {pages} {39706} (\bibinfo {year} {2022})}\BibitemShut {NoStop}%
\bibitem [{\citenamefont {Lu}\ \emph {et~al.}(2020)\citenamefont {Lu}, \citenamefont {Yang}, \citenamefont {Li},\ and\ \citenamefont {Gong}}]{8979176}%
  \BibitemOpen
  \bibfield  {author} {\bibinfo {author} {\bibfnamefont {R.}~\bibnamefont {Lu}}, \bibinfo {author} {\bibfnamefont {Y.}~\bibnamefont {Yang}}, \bibinfo {author} {\bibfnamefont {M.-H.}\ \bibnamefont {Li}},\ and\ \bibinfo {author} {\bibfnamefont {S.}~\bibnamefont {Gong}},\ }\bibfield  {title} {\bibinfo {title} {Ghz low-loss acoustic rf couplers in lithium niobate thin film},\ }\href {https://doi.org/10.1109/TUFFC.2020.2971196} {\bibfield  {journal} {\bibinfo  {journal} {IEEE Transactions on Ultrasonics, Ferroelectrics, and Frequency Control}\ }\textbf {\bibinfo {volume} {67}},\ \bibinfo {pages} {1448} (\bibinfo {year} {2020})}\BibitemShut {NoStop}%
\bibitem [{\citenamefont {Shams-Ansari}\ \emph {et~al.}(2022{\natexlab{b}})\citenamefont {Shams-Ansari}, \citenamefont {Huang}, \citenamefont {He}, \citenamefont {Li}, \citenamefont {Holzgrafe}, \citenamefont {Jankowski}, \citenamefont {Churaev}, \citenamefont {Kharel}, \citenamefont {Cheng}, \citenamefont {Zhu}, \citenamefont {Sinclair}, \citenamefont {Desiatov}, \citenamefont {Zhang}, \citenamefont {Kippenberg},\ and\ \citenamefont {Lončar}}]{Tobias}%
  \BibitemOpen
  \bibfield  {author} {\bibinfo {author} {\bibfnamefont {A.}~\bibnamefont {Shams-Ansari}}, \bibinfo {author} {\bibfnamefont {G.}~\bibnamefont {Huang}}, \bibinfo {author} {\bibfnamefont {L.}~\bibnamefont {He}}, \bibinfo {author} {\bibfnamefont {Z.}~\bibnamefont {Li}}, \bibinfo {author} {\bibfnamefont {J.}~\bibnamefont {Holzgrafe}}, \bibinfo {author} {\bibfnamefont {M.}~\bibnamefont {Jankowski}}, \bibinfo {author} {\bibfnamefont {M.}~\bibnamefont {Churaev}}, \bibinfo {author} {\bibfnamefont {P.}~\bibnamefont {Kharel}}, \bibinfo {author} {\bibfnamefont {R.}~\bibnamefont {Cheng}}, \bibinfo {author} {\bibfnamefont {D.}~\bibnamefont {Zhu}}, \bibinfo {author} {\bibfnamefont {N.}~\bibnamefont {Sinclair}}, \bibinfo {author} {\bibfnamefont {B.}~\bibnamefont {Desiatov}}, \bibinfo {author} {\bibfnamefont {M.}~\bibnamefont {Zhang}}, \bibinfo {author} {\bibfnamefont {T.~J.}\ \bibnamefont {Kippenberg}},\ and\ \bibinfo {author} {\bibfnamefont {M.}~\bibnamefont {Lončar}},\ }\bibfield  {title} {\bibinfo {title} {{Reduced
  material loss in thin-film lithium niobate waveguides}},\ }\href {https://doi.org/10.1063/5.0095146} {\bibfield  {journal} {\bibinfo  {journal} {APL Photonics}\ }\textbf {\bibinfo {volume} {7}},\ \bibinfo {pages} {081301} (\bibinfo {year} {2022}{\natexlab{b}})}\BibitemShut {NoStop}%
\bibitem [{\citenamefont {Bertsimas}\ and\ \citenamefont {Tsitsiklis}(1993)}]{SimulatedAnnealing}%
  \BibitemOpen
  \bibfield  {author} {\bibinfo {author} {\bibfnamefont {D.}~\bibnamefont {Bertsimas}}\ and\ \bibinfo {author} {\bibfnamefont {J.}~\bibnamefont {Tsitsiklis}},\ }\bibfield  {title} {\bibinfo {title} {{Simulated Annealing}},\ }\href {https://doi.org/10.1214/ss/1177011077} {\bibfield  {journal} {\bibinfo  {journal} {Statistical Science}\ }\textbf {\bibinfo {volume} {8}},\ \bibinfo {pages} {10 } (\bibinfo {year} {1993})}\BibitemShut {NoStop}%
\bibitem [{\citenamefont {Eiben}\ \emph {et~al.}(1999)\citenamefont {Eiben}, \citenamefont {Hinterding},\ and\ \citenamefont {Michalewicz}}]{Eiben}%
  \BibitemOpen
  \bibfield  {author} {\bibinfo {author} {\bibfnamefont {A.}~\bibnamefont {Eiben}}, \bibinfo {author} {\bibfnamefont {R.}~\bibnamefont {Hinterding}},\ and\ \bibinfo {author} {\bibfnamefont {Z.}~\bibnamefont {Michalewicz}},\ }\bibfield  {title} {\bibinfo {title} {Parameter control in evolutionary algorithms},\ }\href {https://doi.org/10.1109/4235.771166} {\bibfield  {journal} {\bibinfo  {journal} {IEEE Transactions on Evolutionary Computation}\ }\textbf {\bibinfo {volume} {3}},\ \bibinfo {pages} {124} (\bibinfo {year} {1999})}\BibitemShut {NoStop}%
\bibitem [{Bor(2021)}]{BoreasD90}%
  \BibitemOpen
  \href@noop {} {\bibinfo {title} {Boreas d90 gnss/ins (advanced navigation, sydney, austratia)}},\ \bibinfo {howpublished} {Product information} (\bibinfo {year} {2021}),\ \bibinfo {note} {uRL: \url{https://www.advancednavigation.com/inertial-navigation-systems/fog-gnss-ins/boreas/}}\BibitemShut {NoStop}%
\bibitem [{GG1(2015)}]{GG1320N}%
  \BibitemOpen
  \href@noop {} {\bibinfo {title} {Gg1320an digital ring laser gyroscope (honeywell, charllote, usa)}},\ \bibinfo {howpublished} {Product information} (\bibinfo {year} {2015}),\ \bibinfo {note} {uRL: \url{https://aerospace.honeywell.com/us/en/products-and-services/product/hardware-and-systems/sensors/gg1320an-digital-ring-laser-gyroscope}}\BibitemShut {NoStop}%
\bibitem [{\citenamefont {Haji}\ and\ \citenamefont {Abdulazeez}(2021)}]{SaadHikmat}%
  \BibitemOpen
  \bibfield  {author} {\bibinfo {author} {\bibfnamefont {S.~H.}\ \bibnamefont {Haji}}\ and\ \bibinfo {author} {\bibfnamefont {A.~M.}\ \bibnamefont {Abdulazeez}},\ }\bibfield  {title} {\bibinfo {title} {Comparison of optimization techniques based on gradient descent algorithm: A review},\ }\href {https://archives.palarch.nl/index.php/jae/article/view/6705} {\bibfield  {journal} {\bibinfo  {journal} {PalArch’s Journal of Archaeology of Egypt / Egyptology}\ }\textbf {\bibinfo {volume} {18}},\ \bibinfo {pages} {2715} (\bibinfo {year} {2021})}\BibitemShut {NoStop}%
\bibitem [{\citenamefont {Floudas}(2013)}]{floudas2013deterministic}%
  \BibitemOpen
  \bibfield  {author} {\bibinfo {author} {\bibfnamefont {C.~A.}\ \bibnamefont {Floudas}},\ }\href@noop {} {\emph {\bibinfo {title} {Deterministic global optimization: theory, methods and applications}}},\ Vol.~\bibinfo {volume} {37}\ (\bibinfo  {publisher} {Springer Science and Business Media},\ \bibinfo {year} {2013})\BibitemShut {NoStop}%
\bibitem [{\citenamefont {Baydin}\ \emph {et~al.}(2018)\citenamefont {Baydin}, \citenamefont {Pearlmutter}, \citenamefont {Radul},\ and\ \citenamefont {Siskind}}]{baydin2018automatic}%
  \BibitemOpen
  \bibfield  {author} {\bibinfo {author} {\bibfnamefont {A.~G.}\ \bibnamefont {Baydin}}, \bibinfo {author} {\bibfnamefont {B.~A.}\ \bibnamefont {Pearlmutter}}, \bibinfo {author} {\bibfnamefont {A.~A.}\ \bibnamefont {Radul}},\ and\ \bibinfo {author} {\bibfnamefont {J.~M.}\ \bibnamefont {Siskind}},\ }\bibfield  {title} {\bibinfo {title} {Automatic differentiation in machine learning: a survey},\ }\href@noop {} {\bibfield  {journal} {\bibinfo  {journal} {Journal of Marchine Learning Research}\ }\textbf {\bibinfo {volume} {18}},\ \bibinfo {pages} {1} (\bibinfo {year} {2018})}\BibitemShut {NoStop}%
\bibitem [{\citenamefont {Mohamed}\ \emph {et~al.}(2020)\citenamefont {Mohamed}, \citenamefont {Rosca}, \citenamefont {Figurnov},\ and\ \citenamefont {Mnih}}]{Shakir}%
  \BibitemOpen
  \bibfield  {author} {\bibinfo {author} {\bibfnamefont {S.}~\bibnamefont {Mohamed}}, \bibinfo {author} {\bibfnamefont {M.}~\bibnamefont {Rosca}}, \bibinfo {author} {\bibfnamefont {M.}~\bibnamefont {Figurnov}},\ and\ \bibinfo {author} {\bibfnamefont {A.}~\bibnamefont {Mnih}},\ }\bibfield  {title} {\bibinfo {title} {Monte carlo gradient estimation in machine learning},\ }\href@noop {} {\bibfield  {journal} {\bibinfo  {journal} {Journal of Machine Learning Research}\ }\textbf {\bibinfo {volume} {21}} (\bibinfo {year} {2020})}\BibitemShut {NoStop}%
\bibitem [{\citenamefont {Misoguti}\ \emph {et~al.}(2001)\citenamefont {Misoguti}, \citenamefont {Backus}, \citenamefont {Durfee}, \citenamefont {Bartels}, \citenamefont {Murnane},\ and\ \citenamefont {Kapteyn}}]{Misoguti}%
  \BibitemOpen
  \bibfield  {author} {\bibinfo {author} {\bibfnamefont {L.}~\bibnamefont {Misoguti}}, \bibinfo {author} {\bibfnamefont {S.}~\bibnamefont {Backus}}, \bibinfo {author} {\bibfnamefont {C.~G.}\ \bibnamefont {Durfee}}, \bibinfo {author} {\bibfnamefont {R.}~\bibnamefont {Bartels}}, \bibinfo {author} {\bibfnamefont {M.~M.}\ \bibnamefont {Murnane}},\ and\ \bibinfo {author} {\bibfnamefont {H.~C.}\ \bibnamefont {Kapteyn}},\ }\bibfield  {title} {\bibinfo {title} {Generation of broadband vuv light using third-order cascaded processes},\ }\href {https://doi.org/10.1103/PhysRevLett.87.013601} {\bibfield  {journal} {\bibinfo  {journal} {Phys. Rev. Lett.}\ }\textbf {\bibinfo {volume} {87}},\ \bibinfo {pages} {013601} (\bibinfo {year} {2001})}\BibitemShut {NoStop}%
\bibitem [{\citenamefont {Molesky}\ \emph {et~al.}(2018)\citenamefont {Molesky}, \citenamefont {Lin}, \citenamefont {Piggott}, \citenamefont {Jin}, \citenamefont {Vuckovi{\'{c}}},\ and\ \citenamefont {Rodriguez}}]{Molesky2018}%
  \BibitemOpen
  \bibfield  {author} {\bibinfo {author} {\bibfnamefont {S.}~\bibnamefont {Molesky}}, \bibinfo {author} {\bibfnamefont {Z.}~\bibnamefont {Lin}}, \bibinfo {author} {\bibfnamefont {A.~Y.}\ \bibnamefont {Piggott}}, \bibinfo {author} {\bibfnamefont {W.}~\bibnamefont {Jin}}, \bibinfo {author} {\bibfnamefont {J.}~\bibnamefont {Vuckovi{\'{c}}}},\ and\ \bibinfo {author} {\bibfnamefont {A.~W.}\ \bibnamefont {Rodriguez}},\ }\bibfield  {title} {\bibinfo {title} {Inverse design in nanophotonics},\ }\href {https://doi.org/10.1038/s41566-018-0246-9} {\bibfield  {journal} {\bibinfo  {journal} {Nature Photonics}\ }\textbf {\bibinfo {volume} {12}},\ \bibinfo {pages} {659} (\bibinfo {year} {2018})}\BibitemShut {NoStop}%
\bibitem [{\citenamefont {Csill{\'e}ry}\ \emph {et~al.}(2010)\citenamefont {Csill{\'e}ry}, \citenamefont {Blum}, \citenamefont {Gaggiotti},\ and\ \citenamefont {Fran{\c{c}}ois}}]{csillery2010approximate}%
  \BibitemOpen
  \bibfield  {author} {\bibinfo {author} {\bibfnamefont {K.}~\bibnamefont {Csill{\'e}ry}}, \bibinfo {author} {\bibfnamefont {M.~G.}\ \bibnamefont {Blum}}, \bibinfo {author} {\bibfnamefont {O.~E.}\ \bibnamefont {Gaggiotti}},\ and\ \bibinfo {author} {\bibfnamefont {O.}~\bibnamefont {Fran{\c{c}}ois}},\ }\bibfield  {title} {\bibinfo {title} {Approximate bayesian computation (abc) in practice},\ }\href@noop {} {\bibfield  {journal} {\bibinfo  {journal} {Trends in ecology and evolution}\ }\textbf {\bibinfo {volume} {25}},\ \bibinfo {pages} {410} (\bibinfo {year} {2010})}\BibitemShut {NoStop}%
\bibitem [{\citenamefont {Horst}\ and\ \citenamefont {Tuy}(2013)}]{horst2013global}%
  \BibitemOpen
  \bibfield  {author} {\bibinfo {author} {\bibfnamefont {R.}~\bibnamefont {Horst}}\ and\ \bibinfo {author} {\bibfnamefont {H.}~\bibnamefont {Tuy}},\ }\href {https://books.google.com/books?id=Pe\_1CAAAQBAJ} {\emph {\bibinfo {title} {Global Optimization: Deterministic Approaches}}}\ (\bibinfo  {publisher} {Springer Berlin Heidelberg},\ \bibinfo {year} {2013})\BibitemShut {NoStop}%
\bibitem [{\citenamefont {Bäck}\ and\ \citenamefont {Schwefel}(1993)}]{Back}%
  \BibitemOpen
  \bibfield  {author} {\bibinfo {author} {\bibfnamefont {T.}~\bibnamefont {Bäck}}\ and\ \bibinfo {author} {\bibfnamefont {H.-P.}\ \bibnamefont {Schwefel}},\ }\bibfield  {title} {\bibinfo {title} {An overview of evolutionary algorithms for parameter optimization},\ }\href {https://doi.org/10.1162/evco.1993.1.1.1} {\bibfield  {journal} {\bibinfo  {journal} {Evolutionary Computation}\ }\textbf {\bibinfo {volume} {1}},\ \bibinfo {pages} {1} (\bibinfo {year} {1993})}\BibitemShut {NoStop}%
\bibitem [{\citenamefont {Wu}\ \emph {et~al.}(2020)\citenamefont {Wu}, \citenamefont {Cao}, \citenamefont {Wang},\ and\ \citenamefont {Hong}}]{Qi}%
  \BibitemOpen
  \bibfield  {author} {\bibinfo {author} {\bibfnamefont {Q.}~\bibnamefont {Wu}}, \bibinfo {author} {\bibfnamefont {Y.}~\bibnamefont {Cao}}, \bibinfo {author} {\bibfnamefont {H.}~\bibnamefont {Wang}},\ and\ \bibinfo {author} {\bibfnamefont {W.}~\bibnamefont {Hong}},\ }\bibfield  {title} {\bibinfo {title} {Machine-learning-assisted optimization and its application to antenna designs: Opportunities and challenges},\ }\href {https://doi.org/10.23919/JCC.2020.04.014} {\bibfield  {journal} {\bibinfo  {journal} {China Communications}\ }\textbf {\bibinfo {volume} {17}},\ \bibinfo {pages} {152} (\bibinfo {year} {2020})}\BibitemShut {NoStop}%
\bibitem [{\citenamefont {Aage}\ \emph {et~al.}(2017)\citenamefont {Aage}, \citenamefont {Andreassen}, \citenamefont {Lazarov},\ and\ \citenamefont {Sigmund}}]{aage2017giga}%
  \BibitemOpen
  \bibfield  {author} {\bibinfo {author} {\bibfnamefont {N.}~\bibnamefont {Aage}}, \bibinfo {author} {\bibfnamefont {E.}~\bibnamefont {Andreassen}}, \bibinfo {author} {\bibfnamefont {B.~S.}\ \bibnamefont {Lazarov}},\ and\ \bibinfo {author} {\bibfnamefont {O.}~\bibnamefont {Sigmund}},\ }\bibfield  {title} {\bibinfo {title} {Giga-voxel computational morphogenesis for structural design},\ }\href@noop {} {\bibfield  {journal} {\bibinfo  {journal} {Nature}\ }\textbf {\bibinfo {volume} {550}},\ \bibinfo {pages} {84} (\bibinfo {year} {2017})}\BibitemShut {NoStop}%
\bibitem [{\citenamefont {Lin}\ \emph {et~al.}(2022)\citenamefont {Lin}, \citenamefont {Pestourie}, \citenamefont {Roques-Carmes}, \citenamefont {Li}, \citenamefont {Capasso}, \citenamefont {Solja\v{c}i\'{c}},\ and\ \citenamefont {Johnson}}]{Zin2}%
  \BibitemOpen
  \bibfield  {author} {\bibinfo {author} {\bibfnamefont {Z.}~\bibnamefont {Lin}}, \bibinfo {author} {\bibfnamefont {R.}~\bibnamefont {Pestourie}}, \bibinfo {author} {\bibfnamefont {C.}~\bibnamefont {Roques-Carmes}}, \bibinfo {author} {\bibfnamefont {Z.}~\bibnamefont {Li}}, \bibinfo {author} {\bibfnamefont {F.}~\bibnamefont {Capasso}}, \bibinfo {author} {\bibfnamefont {M.}~\bibnamefont {Solja\v{c}i\'{c}}},\ and\ \bibinfo {author} {\bibfnamefont {S.~G.}\ \bibnamefont {Johnson}},\ }\bibfield  {title} {\bibinfo {title} {End-to-end metasurface inverse design for single-shot multi-channel imaging},\ }\href {https://doi.org/10.1364/OE.449985} {\bibfield  {journal} {\bibinfo  {journal} {Opt. Express}\ }\textbf {\bibinfo {volume} {30}},\ \bibinfo {pages} {28358} (\bibinfo {year} {2022})}\BibitemShut {NoStop}%
\bibitem [{\citenamefont {Wang}\ \emph {et~al.}(2019)\citenamefont {Wang}, \citenamefont {Zhang}, \citenamefont {Yu}, \citenamefont {Zhu}, \citenamefont {Hu},\ and\ \citenamefont {Loncar}}]{Wang2019}%
  \BibitemOpen
  \bibfield  {author} {\bibinfo {author} {\bibfnamefont {C.}~\bibnamefont {Wang}}, \bibinfo {author} {\bibfnamefont {M.}~\bibnamefont {Zhang}}, \bibinfo {author} {\bibfnamefont {M.}~\bibnamefont {Yu}}, \bibinfo {author} {\bibfnamefont {R.}~\bibnamefont {Zhu}}, \bibinfo {author} {\bibfnamefont {H.}~\bibnamefont {Hu}},\ and\ \bibinfo {author} {\bibfnamefont {M.}~\bibnamefont {Loncar}},\ }\bibfield  {title} {\bibinfo {title} {Monolithic lithium niobate photonic circuits for kerr frequency comb generation and modulation},\ }\href {https://doi.org/10.1038/s41467-019-08969-6} {\bibfield  {journal} {\bibinfo  {journal} {Nature Communications}\ }\textbf {\bibinfo {volume} {10}},\ \bibinfo {pages} {978} (\bibinfo {year} {2019})}\BibitemShut {NoStop}%
\bibitem [{\citenamefont {Yu}\ \emph {et~al.}(2022)\citenamefont {Yu}, \citenamefont {Barton~III}, \citenamefont {Cheng}, \citenamefont {Reimer}, \citenamefont {Kharel}, \citenamefont {He}, \citenamefont {Shao}, \citenamefont {Zhu}, \citenamefont {Hu}, \citenamefont {Grant}, \citenamefont {Johansson}, \citenamefont {Okawachi}, \citenamefont {Gaeta}, \citenamefont {Zhang},\ and\ \citenamefont {Lon{\v{c}}ar}}]{Yu2022}%
  \BibitemOpen
  \bibfield  {author} {\bibinfo {author} {\bibfnamefont {M.}~\bibnamefont {Yu}}, \bibinfo {author} {\bibfnamefont {D.}~\bibnamefont {Barton~III}}, \bibinfo {author} {\bibfnamefont {R.}~\bibnamefont {Cheng}}, \bibinfo {author} {\bibfnamefont {C.}~\bibnamefont {Reimer}}, \bibinfo {author} {\bibfnamefont {P.}~\bibnamefont {Kharel}}, \bibinfo {author} {\bibfnamefont {L.}~\bibnamefont {He}}, \bibinfo {author} {\bibfnamefont {L.}~\bibnamefont {Shao}}, \bibinfo {author} {\bibfnamefont {D.}~\bibnamefont {Zhu}}, \bibinfo {author} {\bibfnamefont {Y.}~\bibnamefont {Hu}}, \bibinfo {author} {\bibfnamefont {H.~R.}\ \bibnamefont {Grant}}, \bibinfo {author} {\bibfnamefont {L.}~\bibnamefont {Johansson}}, \bibinfo {author} {\bibfnamefont {Y.}~\bibnamefont {Okawachi}}, \bibinfo {author} {\bibfnamefont {A.~L.}\ \bibnamefont {Gaeta}}, \bibinfo {author} {\bibfnamefont {M.}~\bibnamefont {Zhang}},\ and\ \bibinfo {author} {\bibfnamefont {M.}~\bibnamefont {Lon{\v{c}}ar}},\ }\bibfield  {title} {\bibinfo {title} {Integrated femtosecond
  pulse generator on thin-film lithium niobate},\ }\href {https://doi.org/10.1038/s41586-022-05345-1} {\bibfield  {journal} {\bibinfo  {journal} {Nature}\ }\textbf {\bibinfo {volume} {612}},\ \bibinfo {pages} {252} (\bibinfo {year} {2022})}\BibitemShut {NoStop}%
\bibitem [{\citenamefont {Lu}\ \emph {et~al.}(2023)\citenamefont {Lu}, \citenamefont {Puzyrev}, \citenamefont {Pankratov}, \citenamefont {Skryabin}, \citenamefont {Yang}, \citenamefont {Gong}, \citenamefont {Surya},\ and\ \citenamefont {Tang}}]{Lu2023}%
  \BibitemOpen
  \bibfield  {author} {\bibinfo {author} {\bibfnamefont {J.}~\bibnamefont {Lu}}, \bibinfo {author} {\bibfnamefont {D.~N.}\ \bibnamefont {Puzyrev}}, \bibinfo {author} {\bibfnamefont {V.~V.}\ \bibnamefont {Pankratov}}, \bibinfo {author} {\bibfnamefont {D.~V.}\ \bibnamefont {Skryabin}}, \bibinfo {author} {\bibfnamefont {F.}~\bibnamefont {Yang}}, \bibinfo {author} {\bibfnamefont {Z.}~\bibnamefont {Gong}}, \bibinfo {author} {\bibfnamefont {J.~B.}\ \bibnamefont {Surya}},\ and\ \bibinfo {author} {\bibfnamefont {H.~X.}\ \bibnamefont {Tang}},\ }\bibfield  {title} {\bibinfo {title} {Two-colour dissipative solitons and breathers in microresonator second-harmonic generation},\ }\href {https://doi.org/10.1038/s41467-023-38412-w} {\bibfield  {journal} {\bibinfo  {journal} {Nature Communications}\ }\textbf {\bibinfo {volume} {14}},\ \bibinfo {pages} {2798} (\bibinfo {year} {2023})}\BibitemShut {NoStop}%
\bibitem [{\citenamefont {Erneux}\ and\ \citenamefont {Glorieux}(2010)}]{erneux_glorieux_2010}%
  \BibitemOpen
  \bibfield  {author} {\bibinfo {author} {\bibfnamefont {T.}~\bibnamefont {Erneux}}\ and\ \bibinfo {author} {\bibfnamefont {P.}~\bibnamefont {Glorieux}},\ }\href {https://doi.org/10.1017/CBO9780511776908} {\emph {\bibinfo {title} {Laser Dynamics}}}\ (\bibinfo  {publisher} {Cambridge University Press},\ \bibinfo {year} {2010})\BibitemShut {NoStop}%
\bibitem [{\citenamefont {Gavrielides}\ \emph {et~al.}(1997)\citenamefont {Gavrielides}, \citenamefont {Kovanis},\ and\ \citenamefont {Erneux}}]{GAVRIELIDES1997253}%
  \BibitemOpen
  \bibfield  {author} {\bibinfo {author} {\bibfnamefont {A.}~\bibnamefont {Gavrielides}}, \bibinfo {author} {\bibfnamefont {V.}~\bibnamefont {Kovanis}},\ and\ \bibinfo {author} {\bibfnamefont {T.}~\bibnamefont {Erneux}},\ }\bibfield  {title} {\bibinfo {title} {Analytical stability boundaries for a semiconductor laser subject to optical injection},\ }\href {https://doi.org/https://doi.org/10.1016/S0030-4018(96)00705-5} {\bibfield  {journal} {\bibinfo  {journal} {Optics Communications}\ }\textbf {\bibinfo {volume} {136}},\ \bibinfo {pages} {253} (\bibinfo {year} {1997})}\BibitemShut {NoStop}%
\bibitem [{\citenamefont {Sayama}(2015)}]{2022Linear}%
  \BibitemOpen
  \bibfield  {author} {\bibinfo {author} {\bibfnamefont {H.}~\bibnamefont {Sayama}},\ }\href@noop {} {\bibinfo {title} {Introduction to the modeling and analysis of complex systems}} (\bibinfo {year} {2015})\BibitemShut {NoStop}%
\bibitem [{\citenamefont {Scully}\ and\ \citenamefont {Zubairy}(1997)}]{scully_zubairy_1997}%
  \BibitemOpen
  \bibfield  {author} {\bibinfo {author} {\bibfnamefont {M.~O.}\ \bibnamefont {Scully}}\ and\ \bibinfo {author} {\bibfnamefont {M.~S.}\ \bibnamefont {Zubairy}},\ }\href {https://doi.org/10.1017/CBO9780511813993} {\emph {\bibinfo {title} {Quantum Optics}}}\ (\bibinfo  {publisher} {Cambridge University Press},\ \bibinfo {year} {1997})\BibitemShut {NoStop}%
\bibitem [{\citenamefont {Smith}\ \emph {et~al.}(2008{\natexlab{b}})\citenamefont {Smith}, \citenamefont {Chang}, \citenamefont {Arissian},\ and\ \citenamefont {Diels}}]{Dispersivegyro}%
  \BibitemOpen
  \bibfield  {author} {\bibinfo {author} {\bibfnamefont {D.~D.}\ \bibnamefont {Smith}}, \bibinfo {author} {\bibfnamefont {H.}~\bibnamefont {Chang}}, \bibinfo {author} {\bibfnamefont {L.}~\bibnamefont {Arissian}},\ and\ \bibinfo {author} {\bibfnamefont {J.~C.}\ \bibnamefont {Diels}},\ }\bibfield  {title} {\bibinfo {title} {Dispersion-enhanced laser gyroscope},\ }\href {https://doi.org/10.1103/PhysRevA.78.053824} {\bibfield  {journal} {\bibinfo  {journal} {Phys. Rev. A}\ }\textbf {\bibinfo {volume} {78}},\ \bibinfo {pages} {053824} (\bibinfo {year} {2008}{\natexlab{b}})}\BibitemShut {NoStop}%
\bibitem [{\citenamefont {Stefszky}\ \emph {et~al.}(2012)\citenamefont {Stefszky}, \citenamefont {Mow-Lowry}, \citenamefont {Chua}, \citenamefont {Shaddock}, \citenamefont {Buchler}, \citenamefont {Vahlbruch}, \citenamefont {Khalaidovski}, \citenamefont {Schnabel}, \citenamefont {Lam},\ and\ \citenamefont {McClelland}}]{Stefszky_2012}%
  \BibitemOpen
  \bibfield  {author} {\bibinfo {author} {\bibfnamefont {M.~S.}\ \bibnamefont {Stefszky}}, \bibinfo {author} {\bibfnamefont {C.~M.}\ \bibnamefont {Mow-Lowry}}, \bibinfo {author} {\bibfnamefont {S.~S.~Y.}\ \bibnamefont {Chua}}, \bibinfo {author} {\bibfnamefont {D.~A.}\ \bibnamefont {Shaddock}}, \bibinfo {author} {\bibfnamefont {B.~C.}\ \bibnamefont {Buchler}}, \bibinfo {author} {\bibfnamefont {H.}~\bibnamefont {Vahlbruch}}, \bibinfo {author} {\bibfnamefont {A.}~\bibnamefont {Khalaidovski}}, \bibinfo {author} {\bibfnamefont {R.}~\bibnamefont {Schnabel}}, \bibinfo {author} {\bibfnamefont {P.~K.}\ \bibnamefont {Lam}},\ and\ \bibinfo {author} {\bibfnamefont {D.~E.}\ \bibnamefont {McClelland}},\ }\bibfield  {title} {\bibinfo {title} {Balanced homodyne detection of optical quantum states at audio-band frequencies and below},\ }\href {https://doi.org/10.1088/0264-9381/29/14/145015} {\bibfield  {journal} {\bibinfo  {journal} {Classical and Quantum Gravity}\ }\textbf {\bibinfo {volume} {29}},\ \bibinfo {pages} {145015}
  (\bibinfo {year} {2012})}\BibitemShut {NoStop}%
\end{thebibliography}%

\end{document}